\documentclass[preprint,12pt,sort&compress]{elsarticle}

\usepackage{amssymb}
\usepackage{amsmath}
\usepackage{tabularx}
\usepackage{amsthm}
\usepackage{lineno, color}
\usepackage{hyperref}
\usepackage{xcolor}
\usepackage{setspace}
\usepackage{array}
\usepackage{booktabs}
\usepackage{rotating}

\journal{Int. J. Heat Mass Transf.}
\begin{document}
\begin{frontmatter}

\title{Interface-resolved simulations of boiling heat transfer}

\author[label1,label2]{Alessio Roccon}
\affiliation[label1]{organization={Polytechnic Department of Engineering and Architecture, University of Udine},
             addressline={Via delle Scienze 206},
            city={Udine},
            postcode={33100},
            country={Italy}}
            
 \affiliation[label2]{organization={Institute of Fluid Mechanics and Heat Transfer, TU Wien},
             addressline={Getreidemarkt 9},
            city={Vienna},
             postcode={1060},
             country={Austria}}
             
\author[label3]{Luca Brandt}
\affiliation[label3]{organization={Department of Environment, Land and Infrastructure Engineering, Politecnico di Torino},
             addressline={Corso Duca degli Abruzzi, 24},
            city={Torino},
            postcode={10129},
            country={Italy}}

\begin{abstract}
Boiling heat transfer underpins many applications where large heat fluxes must be dissipated.
Boiling is characterized by a wide range of coupled multiscale transport phenomena, making its accurate numerical prediction a long-standing challenge. 
In this context, interface-resolved simulation approaches -- where the liquid-vapor interface is either explicitly or implicitly described -- have undergone significant advancements, driven by improved numerical formulations, more complete thermodynamic and compressibility modeling, and the growing availability of computational resources.
Here, we provide a comprehensive overview of the state of the art in interface-resolved simulations of boiling flows. 
We examine the different formulations governing the flow and temperature field, strategies for coupling heat and mass transfer at the interface, and current modeling approaches for small-scale physics such as nucleation, microlayer evaporation, and contact-line dynamics.
By analyzing insights from recent high-fidelity simulations, a central observation is that interface-resolved simulations of boiling are hydrodynamically resolved but microscopically modeled: their predictive accuracy depends on the sub-grid closures for nucleation, microlayer, and contact-line physics no less than on the fidelity of the hydrodynamic solver.
We discuss the strengths and limitations of available methods on this basis, together with the open challenges in the accurate description of boiling heat transfer.
\end{abstract}



\end{frontmatter}
\thispagestyle{pprintTitle}

\section{Introduction}

Boiling heat transfer is fundamental to a wide range of thermal management technologies, from nuclear energy systems to advanced cooling applications~\cite{chen2018numerical,silvi2021understanding,LEONG2017}. 
Owing to its ability to dissipate high heat fluxes, because latent heat is typically much larger than sensible heat, boiling is particularly attractive for efficient heat removal \cite{dhir1998boiling,nukiyama1966maximum,moghaddam2009physical1,moghaddam2009physical2}.
The thermal performance of devices employing boiling is commonly characterized by two fundamental metrics: the heat transfer coefficient (HTC) and the critical heat flux (CHF). 
The HTC quantifies the efficiency of heat removal, relating the wall heat flux to the temperature difference between the heated surface and the bulk fluid. 
In contrast, the CHF defines the upper limit of safe operation and represents the maximum heat flux that can be sustained in the nucleate boiling regime before transition to less efficient regimes -- namely transitional and film boiling \cite{mudawar1999ultra,ganesan2021universal}.
These latter regimes (see also Fig.~\ref{fig:curve}) are characterized by the disruption of the liquid supply to the heated surface, partial or complete dryout, and consequently a sharp increase in surface temperature.

From a practical point of view, a deeper understanding of boiling heat transfer is highly desirable in order to enable systems to operate as close as possible to the CHF limit while maintaining safe and stable conditions \cite{dhir1998boiling,kam2020heat}.
In this context, recent advances in micro- and nano-fabrication have enabled the design of engineered surfaces that control bubble dynamics and significantly enhance both HTC and CHF~\cite{chu2012structured,kim2015enhanced,kim2015review,liang2019review,dhillon2015critical}, while porous and foam-based structures offer a complementary route to enhancement at the macroscale~\cite{righetti2020water,calati2021water}.
However, despite significant progress, our predictive capabilities remain limited, particularly outside the range of conditions for which models or correlations were developed \cite{cheung2014modelingp1,cheung2014modelingp2,darges2022assessment}.
Closing this predictive gap would have significant industrial impact, from nuclear reactor safety analysis to the design of the next-generation cooling systems~\cite{yadigaroglu2014cmfd,mudawar2002assessment}.

\begin{figure}[!t]
\setlength{\unitlength}{0.0025\columnwidth}
\begin{picture}(255,170)
\put(-2,-10){\includegraphics[width=1.00\columnwidth, keepaspectratio]{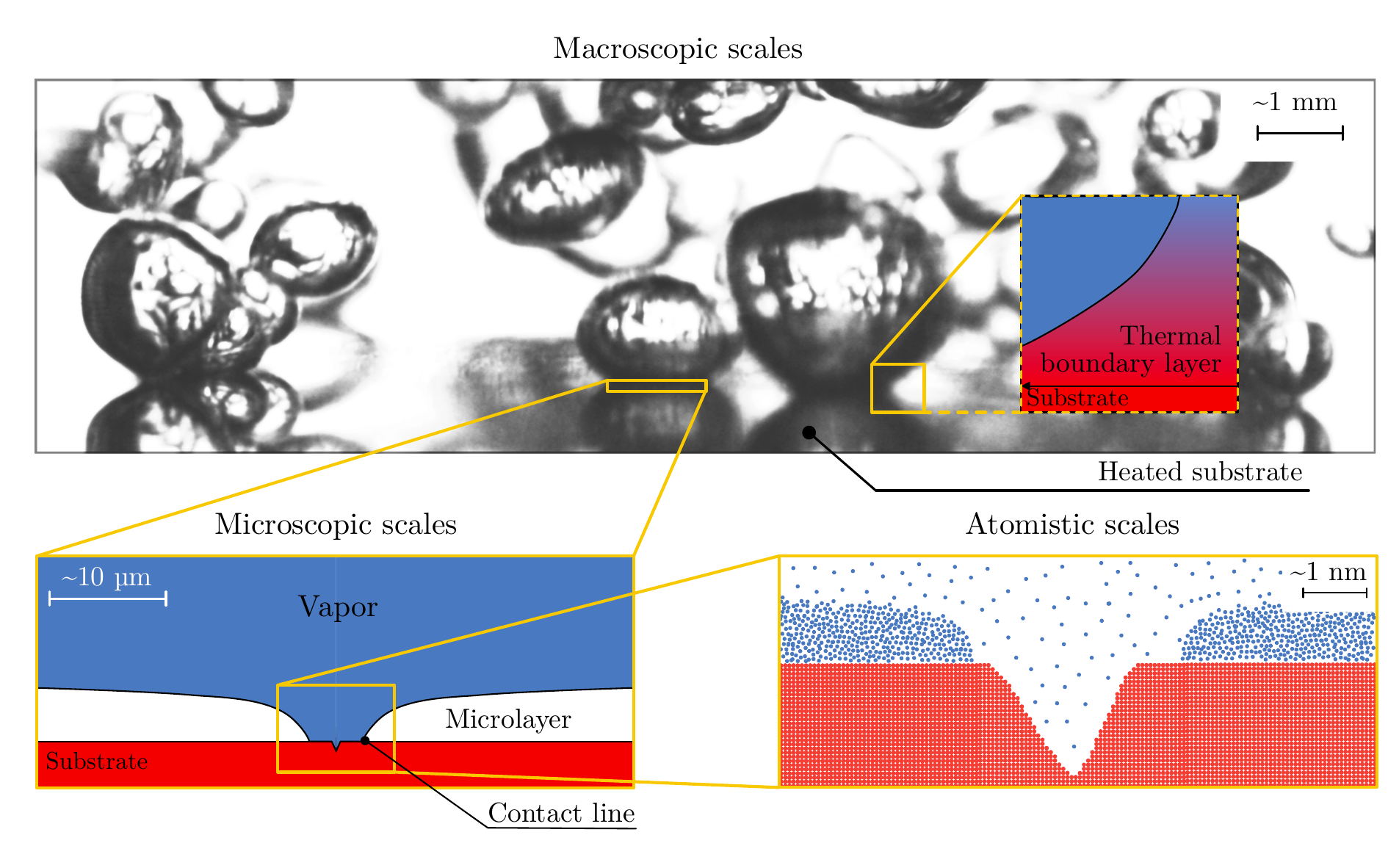}}
\put(10,225){$(a)$}
\put(10,90){$(b)$}
\put(220,90){$(c)$}
\end{picture}
\caption{Qualitative overview of the spectrum of spatial scales involved in boiling heat transfer.
At the macroscopic level (panel~$a$, adapted from Hao {\em et al.}~\cite{hao2025experimental}), flow structures induced by turbulence and buoyancy influence bubble dynamics and determine global mixing.
Moving to smaller scales (panel~$b$), microlayer and contact-line evaporation are controlled by the microphysics of thin liquid films.
Finally, nucleation and interface dynamics are controlled by the small-scale physics that occur at the atomistic scales (panel~$c$).}
\label{fig1} 
\end{figure}
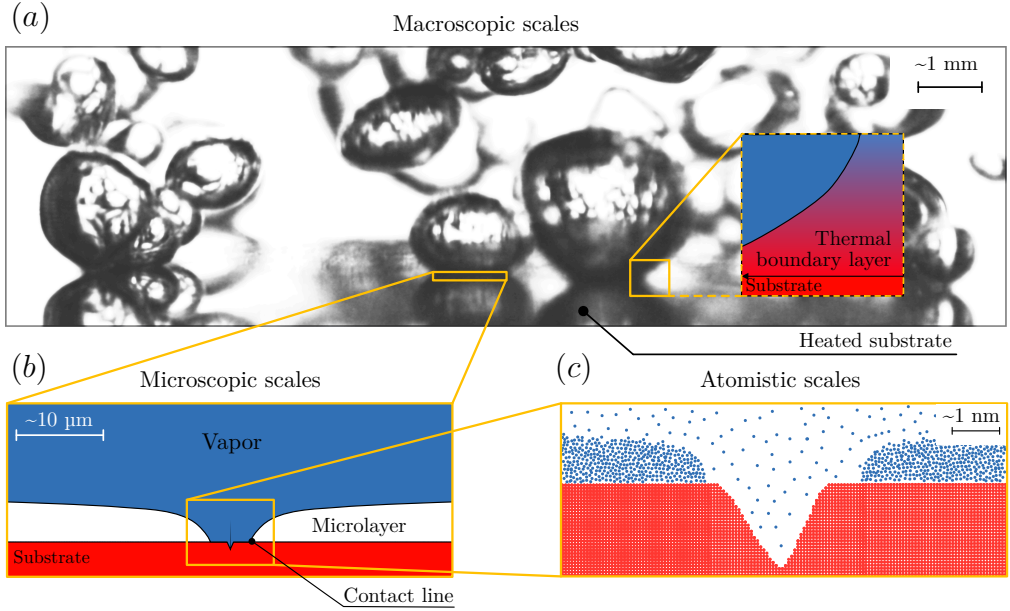

The current limitations stem from the inherently non-equilibrium and multiscale character of boiling, where interfacial transport, phase change, and hydrodynamic instabilities are strongly coupled across a wide spectrum of spatial and temporal scales \cite{chen2018numerical,soligo2021turbulent}.
Capturing this complex interplay remains a major challenge for numerical modeling, because we would like numerical simulations to resolve the interfacial dynamics while maintaining computational efficiency and robustness \cite{kharangate2017review,chen2024review}.
The phenomena and the corresponding time and length scales involved in boiling heat transfer are illustrated in Fig.~\ref{fig1}. 
At the macroscopic scale, Fig.~\ref{fig1}$a$, large-scale hydrodynamic effects such as turbulence and buoyancy-driven flow structures govern mixing and bubble departure and coalescence \cite{mathai2020bubbly,demou2022pressure,Kossolapov2024,garbin2025bubbles} as well as the behavior of momentum and thermal boundary layers (see inset).
These macroscopic scales range from the scale of the system down to the smallest hydrodynamic and thermal length scales, whose magnitudes depend on the specific configuration (e.g., laminar or turbulent conditions, pool or flow boiling).
When moving to smaller scales, Fig.~\ref{fig1}$b$, microscopic processes that distinctly affect the overall system performance emerge.
Specifically, microlayer formation and contact-line dynamics are clear examples of phenomena that have an important effect on heat transfer and bubble growth, and can account for a considerable amount of the total heat flux \cite{kim2009review,jung2014experimental,urbano2018direct,chen2020measurement,chen2023modeling,bures2024coarse,garbin2025bubbles,torres2024coupling}. 
At the smallest atomistic scales, Fig.~\ref{fig1}$c$, molecular interactions at the liquid-vapor interface control bubble nucleation and interaction with the wall surface (e.g biphilic surfaces and nanostructures) \cite{su2020investigation,zupanvcivc2015enhanced,mozze2020laser,vachhani2025numerical,garbin2025bubbles}.
Crucially, these phenomena do not evolve independently; rather, they are strongly coupled, with each playing an important role in the overall boiling process.

In this context, numerical simulations represent an essential tool for advancing our understanding of the complex physics of boiling heat transfer and improving our ability to manipulate it.
By providing space- and time-resolved information on velocity, temperature, and interface evolution, simulations enable systematic investigations of individual physical mechanisms that are difficult to isolate experimentally.
An example is the investigation of gravity effects, which can be enabled or disabled in simulations, whereas experimentally low- or zero-gravity conditions are attainable only during parabolic flights or aboard the International Space Station~\cite{garivalis2021critical,raj2012pool,dhir2012nucleate,warrier2015nucleate,sielaff2022multiscale,mudawar2023heat}.
However, because boiling heat transfer spans such a wide range of scales, the numerical methods used to simulate it are diverse, each targeting a different portion of that range.
Simulating the entire range of scales remains infeasible for two primary reasons: i) the computational cost far exceeds current resource capabilities, and ii) a unified set of governing equations valid across both atomistic and continuum regimes has yet to be established.

\begin{figure}[!t]
\setlength{\unitlength}{0.0025\columnwidth}
\begin{picture}(400,130)
\put(0,-10){\includegraphics[width=1.00\columnwidth, keepaspectratio]{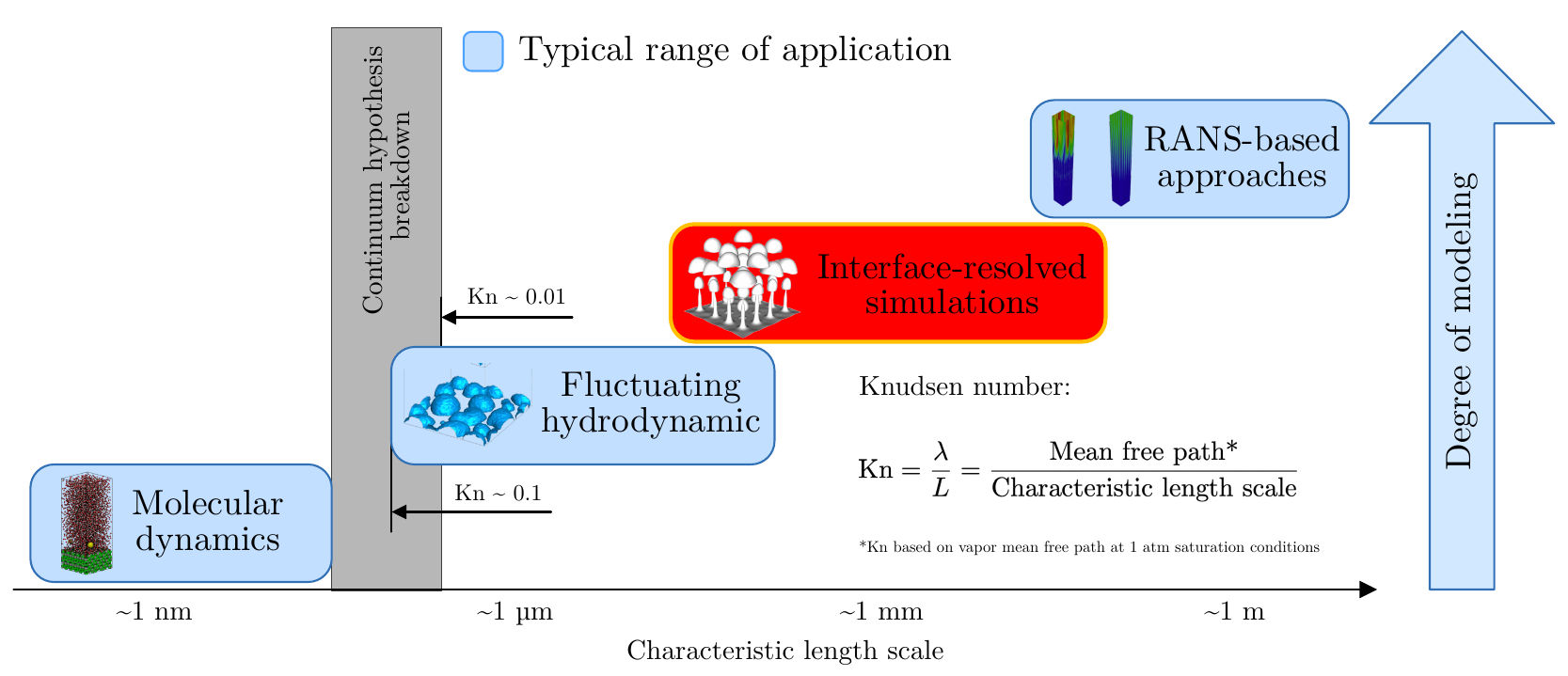}}
\end{picture}
\caption{Overview of numerical approaches for boiling flow simulations. 
The horizontal axis represents the characteristic length scale, while the vertical axis indicates the degree of modeling abstraction. 
Molecular dynamics approaches operate at the discrete atomistic level where $\text{Kn}>1$.
Fluctuating hydrodynamics extends continuum formulations toward smaller scales by incorporating thermal fluctuations. 
Interface-resolved simulations operate in the continuum regime and resolve the liquid–vapor interface. 
At larger, engineering-relevant scales, RANS-based approaches introduce additional modeling closures to represent unresolved turbulence and interfacial dynamics.
Images reproduced from Karalis {\em et al.} \cite{karalis2021deciphering}, Magaletti {\em et al.}~\cite{magaletti2020unraveling}, Gu {\em et al.}~\cite{gu2024three} and Wang {\em et al.}~\cite{wang2020cfd}.}
\label{fig2} 
\end{figure}

Given these limitations, existing approaches focus on a selected range of scales, and can be broadly classified into atomistic methods, mesoscopic formulations, and macroscopic approaches~\cite{soligo2021turbulent,chen2024review,garcia2025numerical}, as summarized in Fig.~\ref{fig2}.
On the far left of the scale spectrum, molecular dynamics (MD) methods operate at the atomistic level to capture fundamental interfacial physics.
These methods model the behavior of individual atoms and molecules by solving Newton equations of motion.
Because each atom is tracked individually and the required time step is typically a fraction of a femtosecond, MD simulations are restricted to extremely small spatial and temporal scales, of the order of nanometers and nanoseconds~\cite{shahmardi2021effects,lavino2021surface,wu2020molecular,lin2024recent}.
Moving to larger scales, fluctuating hydrodynamics (FH) bridges the gap between the atomistic and continuum regimes.
By augmenting the continuum equations with stochastic fluxes that obey the fluctuation-dissipation theorem, FH extends classical hydrodynamic formulations to scales at which thermal fluctuations are no longer negligible thus enabling the description of rare, thermally activated events, such as the spontaneous nucleation of vapor bubbles in metastable liquids \cite{gallo2018thermally,gallo2020nucleation,magaletti2020unraveling}.
Moving to even larger scales, we reach macroscopic approaches, which operate at conventional hydrodynamic scales. 
At the lower end of these hydrodynamic scales, a first class of methods is represented by interface-resolved simulation methods (the focus of this review), in which the liquid-vapor interface is resolved \cite{cmmf2009,kharangate2017review,mirjalili2017interface,soligo2021turbulent,chen2024review}.
Interface-resolved simulations mainly rely on direct numerical simulations (DNS) for the flow field description, where the definition of DNS is here suitably adapted to account for the multiscale character of multiphase turbulence \cite{tryggvason2013multiscale,soligo2021turbulent}.
At engineering-relevant scales, Reynolds-averaged Navier-Stokes (RANS) formulations are adopted, where the unresolved turbulence and interfacial dynamics are modeled \cite{wang2020cfd,gilman2017self,yuan2024assessment}.

This review focuses on interface-resolved simulations of boiling heat transfer, with primary emphasis on pool boiling.
For comprehensive discussions of other aspects of boiling heat transfer, readers are referred to existing review articles covering fundamental physics and boiling mechanisms \cite{dhir1998boiling,thome2004boiling,kim2009review}, enhancement techniques and applications \cite{LEONG2017,darshan2024numerical,shang2025comprehensive}, predictive methods \cite{liang2018pool1,liang2018pool2} and numerical approaches at other scales~\cite{kharangate2017review,li2023recent,jiang2023review,chen2024review}.
A theme that recurs throughout is that these methods resolve the hydrodynamics of the liquid-vapor interface directly but must represent a number of small-scale physics phenomena -- nucleation, microlayer evaporation, contact-line motion, thin-film forces -- through sub-grid closures.
Interface-resolved simulations of boiling are, in this sense, hydrodynamically resolved but microscopically modeled, a point we shall return this when assessing what we can learn from these simulations.
The review is organized as follows.
Section~\ref{sec:jump} discusses the interfacial jump conditions in the presence of phase change while Section~\ref{sec:interface} briefly recalls interface tracking and capturing methods. 
Section~\ref{sec:fluid} details the numerical approaches available to impose the interfacial jump conditions. Section~\ref{sec:eos} discusses the treatment of compressibility effects, and the modeling of small-scale physical phenomena is addressed in Section~\ref{sec:sss}. 
Finally, Section~\ref{sec:insights} summarizes the physical insights gained from interface-resolved simulations and Section~\ref{sec:final} concludes with a summary on open challenges and future research directions.

\section{Interfacial conditions}
\label{sec:jump}

The primary challenge in interface-resolved simulations lies in enforcing boundary conditions at the liquid-vapor interface.
These conditions include the balance of normal and tangential stresses, continuity of mass flux, and the Stefan condition, which ensures energy conservation under the assumption of local thermodynamic equilibrium at the interface \cite{ishii2010thermo}.
Mathematically, these interfacial constraints give rise to Dirichlet, Neumann, and/or Robin boundary conditions, applied on moving and deforming interfaces.
In the following, we recall the jump conditions that govern mass, momentum, and energy transfer and discuss the closure models used to compute the interfacial mass flux for boiling and evaporating flows.
While this review focuses on boiling, we also consider evaporation, as the two share much of the underlying physics and the numerical algorithms developed for one can often be readily adapted to the other.

\subsection{Jump conditions}
\begin{figure}
\setlength{\unitlength}{0.0025\columnwidth}
\begin{picture}(400,90)
\put(40,-15){\includegraphics[width=0.8\columnwidth, keepaspectratio]{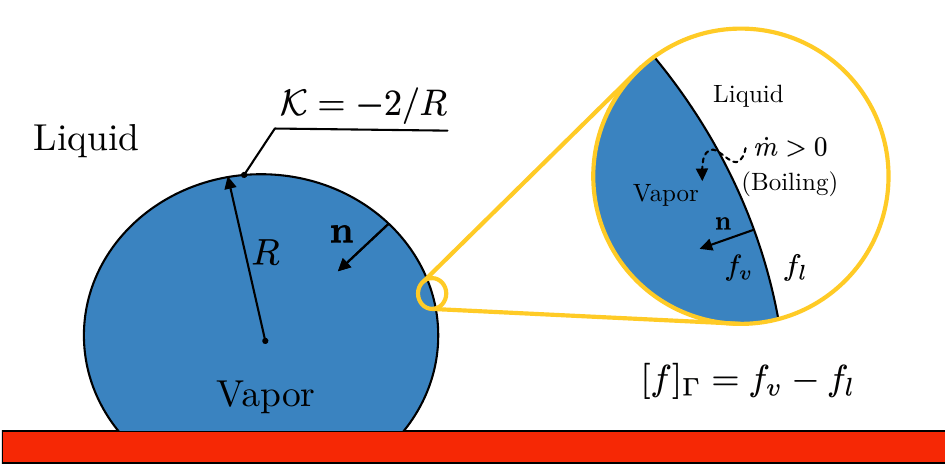}}
\end{picture}
\caption{Vapor bubble growing in a liquid (e.g.\ during pool boiling).
The sketch shows the sign convention used to define the jump conditions.
The interface normal $\bf n$ points from the liquid to the vapor phase. 
During evaporation and boiling, $\dot m$ is positive as liquid is converted into vapor (the mass flow rate is in the same direction of the interface normal). }
\label{fig:jump} 
\end{figure}

The jump conditions at the liquid-vapor interface directly follow from the conservation laws (mass, momentum and energy) 
applied to a control volume enclosing a portion of the interface.
For a generic quantity $f$, we define the jump condition across the interface as  \cite{tanguy2014benchmarks,shao2018computational}:
\begin{equation}
\left[ f \right]_\Gamma = f_v - f_l\, ,
\end{equation}
where the subscripts $v$ and $l$ denote the quantity evaluated on the vapor and liquid sides of the interface, i.e., the side pointed to by the tip of the interface normal vector (vapor) minus the side at the base of the arrow (liquid), see sketch in Fig.~\ref{fig:jump}.

The first jump condition is obtained by imposing the conservation of mass at the interface.
This relation dictates that any discontinuity in the normal velocity is only due to the volumetric expansion or contraction induced by phase-change:
\begin{equation}
    \left[ \mathbf{u} \right]_\Gamma   = \dot{m} \left[ \frac{1}{\rho} \right]_\Gamma  \mathbf{n} \, ,
\label{eq:jump_mass}
\end{equation}
where $\dot{m}$ is the interfacial mass flow rate or vaporization rate (expressed in $\mathrm{ kg/m^2 s}$ and positive for boiling/evaporation) and $\mathbf{n}$  is the unit normal vector that points from the liquid towards the vapor phase.

The conservation of momentum across the interface expresses the balance among pressure forces, viscous stresses, capillary effects, and the momentum flux induced by phase change:
\begin{equation}
    \left[ -p \mathbf{I} + 2\mu \mathbf{D} \right]_\Gamma \cdot \mathbf{n}
    = \sigma \mathcal{K} \mathbf{n}
    - \nabla_s \sigma
    + \dot{m}^2 \left[ \frac{1}{\rho} \right]_\Gamma \mathbf{n}\, ,
    \label{eq:jump_momentum}
\end{equation}
where $\sigma$ is the surface tension,  $\mathcal{K} = \nabla \cdot \mathbf{n}$ the mean curvature of the interface\footnote{Note that the sign of the surface tension term depends on the conventions adopted for the normal vector and the curvature; here we follow the conventions used in Ref.~\cite{Roccon2023}.},  and $\nabla_s = (\mathbf{I} - \mathbf{n} \otimes \mathbf{n}) \cdot \nabla$  denotes the surface gradient operator.
The latter term -- pressure recoil -- is quadratic in $\dot{m}$ and is typically negligible at moderate evaporation rates but becomes significant in strongly evaporating regimes, e.g.\ microlayer evaporation \cite{nikolayev1999boiling,nikolayev2006experimental}.
Projecting Eq.~\eqref{eq:jump_momentum} along the normal direction  recovers the classical Laplace pressure jump due to capillarity, plus the recoil pressure generated by the momentum carried across the interface during phase change:
\begin{equation}
\left[ p \right]_\Gamma = -\sigma \mathcal{K}
+ 2 \left[ \mu \frac{\partial u_n}{\partial n} \right]_\Gamma
- \dot{m}^2 \left[ \frac{1}{\rho} \right]_\Gamma\, ,
\end{equation}
where $\partial u_n / \partial n$ is the normal derivative of the normal velocity component $u_n = {\bf u} \cdot {\bf n}$. 
In contrast, the tangential projection of Eq.~\eqref{eq:jump_momentum} 
results in:
\begin{equation}
    \mathbf{t} \cdot \left[ 2\mu \mathbf{D} \right]_\Gamma \cdot \mathbf{n}
    = \mathbf{t} \cdot \nabla_s \sigma,
    \label{eq:jump_tangential}
\end{equation}
with $\mathbf{t}$ the tangent vector; this relation states that any tangential gradient of surface tension must be balanced by a jump in the tangential viscous stress. 
These tangential stresses (also known as Marangoni stresses) arise whenever the surface tension varies along the interface, either because of temperature gradients or surfactant concentration gradients~\cite{scriven1960marangoni,joseph1976stability}.

The interfacial energy balance requires that the net conductive heat flux arriving at the interface from both phases be entirely consumed as -- or released as -- latent heat:
\begin{equation}
    \left[ \lambda \nabla T \cdot \mathbf{n} \right]_\Gamma = \dot{m}\, L_{\mathrm{vap}},
    \label{eq:jump_energy}
\end{equation}
where $L_{\mathrm{vap}}$ is the latent heat of vaporization and $\lambda$ the thermal conductivity.
This condition couples the local thermal field to the vaporization rate. 
Eq.~\eqref{eq:jump_energy} assumes local thermodynamic equilibrium at the interface:
\begin{equation}
    T_\Gamma = T_{\mathrm{sat}}\, ,
    \label{eq:jump_temp}
\end{equation}
i.e., phase change occurs at the saturation temperature corresponding to the local pressure. 
While this assumption is standard in continuum-scale boiling simulations, it can break down under conditions of intense vaporization~\cite{giustini2016evaporative}.

\subsection{Closure models for the interfacial mass flux}

In interface-resolved simulations of heat and mass transfer, a closure model must be introduced to determine the local phase-change rate. 
Such a closure is necessary because the jump conditions merely enforce conservation across the interface without prescribing the phase-change rate, denoted $\dot m$ above. 
Several closure strategies have been developed for this purpose \cite{kharangate2017review}. 
A first family of models is based directly on the interfacial energy balance \cite{juric1998computations,gibou2007level}, linking the evaporative mass flux to the jump in conductive heat flux across the interface. 
Alternative approaches include the Schrage model \cite{schrage1953theoretical}, derived from kinetic theory, and the Lee model \cite{lee1980pressure}, which drives phase change through  a relaxation toward the saturation temperature. 
A further class of closures  relies on chemical-potential relaxation \cite{pelanti2022arbitrary,sirianni2025mixture} and provides a thermodynamically consistent formulation within compressible multiphase frameworks, e.g. Baer-Nunziato (BN) models \cite{baer1986two}.

\subsubsection{Energy-balance closure}
In many applications interfacial heat transfer is fast enough that local thermodynamic equilibrium can be assumed.
Hence, the most rigorous approach consists of directly enforcing the energy jump condition~\eqref{eq:jump_energy}, with the interface temperature set to the local saturation value.
The mass flux is computed as \cite{juric1998computations,gibou2007level,tsui2014phase}:
\begin{equation}
   \dot{m} = \frac{  ( \lambda_v   \,\nabla T|_v  -\lambda_\ell \,\nabla T|_\ell )  \cdot {\bf n} }{L_{vap}} \, ,
   \label{eq:energy_balance}
\end{equation}
where $v$ and $\ell$ denote the two sides of the interface.
For nucleate boiling at moderate or high pressures, the vapor phase remains close to $T_{\mathrm{sat}}$, and the vapor-side conductive flux is
negligible relative to the liquid-side flux~\cite{tryggvason2011direct,welch2000volume,sun2012development,wang2021phase};
the equation above can thus be simplified as:
\begin{equation}
   \dot{m} \approx \frac{-\lambda_\ell \,\nabla T|_\ell \cdot {\bf n}}{L_{vap}} \, .
   \label{eq:energy_balance_simplified}
\end{equation}
The energy-balance model is parameter-free and thermodynamically consistent, making it the approach of choice for high-fidelity DNS whenever
the computational mesh is fine enough to resolve the thermal boundary layers adjacent to the interface~\cite{welch2000volume,juric1998computations,gibou2007level}.
In practice, evaluating $\nabla T|_\ell \cdot {\bf n}$ at the moving interface is one of the main difficulties of this class of closures.
A first possible strategy is the probe method, in which the temperature is sampled at points located a chosen distance (of the order of the mesh size) from the interface and combined with the saturation condition to evaluate the normal gradient
\cite{udaykumar1996elafint,esmaeeli2004computations,shin2016numerical,irfan2017front,roccon2024boiling,zhong2025front}. 
Alternatively, the gradient can be computed only at the interfacial cell and then extrapolated to the entire domain, or to a band around the interface~\cite{aslam2004partial}.
A third possible option, adopted in sharp-interface VOF-type methods \cite{sato2013sharp}, is to evaluate the one-sided temperature gradients
direction-by-direction in the Cartesian frame, using stencils that embed the saturation temperature at the interface location.
Since the gradient of the phase not containing the cell center cannot be evaluated directly, it is reconstructed by extrapolating the gradient
components from the opposite side of the interface along the normal direction, solving a pseudo-time advection equation \cite{aslam2004partial,osher2002extrapolation}.
A fourth option avoids the explicit evaluation of the one-sided gradient: the phase-change rate is computed from the cell-centered temperature gradient, using the gradient of the liquid volume fraction $\alpha_\ell$ as interfacial area density: $\dot{m}_V = \lambda_\ell \nabla T \cdot \nabla\alpha_\ell / L_{vap}$, where $\dot{m}_V$ denotes a volumetric mass source ($\mathrm{kg/m^3 s}$) rather than an interfacial flux~\cite{wang2021phase,safari2013extended}.

\subsubsection{Hertz–Knudsen–Schrage kinetic model}
An alternative description of phase change is provided by the Hertz-Knudsen-Schrage model~\cite{schrage1953theoretical,knudsen1950kinetic}, which builds on the kinetic theory of gases.  
The net evaporative mass flux is related to the difference between the rates at which molecules leave and arrive at the interface:
\begin{equation}
\dot{m} = \frac{2\hat{\sigma}}{2-\hat{\sigma}}  \sqrt{\frac{M}{2\pi R}}  \left( \frac{p_v}{\sqrt {T_{v,sat}}}  -\frac{p_l}{\sqrt{T_{l,sat}}}  \right) \, ,
\end{equation}
where $\hat{\sigma} \in [0,1]$ is the accommodation coefficient, $M$ is the molar mass of the fluid, $R$ is the universal gas constant, $p_v$ and $T_{v,sat}$ are the vapor pressure and saturation temperature at the interface while $p_l$ and $T_{l,sat}$ are the liquid pressure
and saturation temperature (at the interface).
The accommodation coefficient represents the fraction of vapor molecules impinging on the interface that actually condense, and conversely the fraction of liquid molecules that successfully evaporate.
A major limitation of this model is a proper estimate of $\hat{\sigma}$.
Values reported in the literature span several orders of magnitude, ranging from $\hat{\sigma} \approx 10^{-3}$ to $\hat{\sigma} = 1$~\cite{wang2007characteristics,hardt2008evaporation,marek2001analysis,magnini2013numerical,kharangate2015computational,kharangate2017review}.  
For small deviations from thermodynamic equilibrium, a Taylor expansion about the saturation state $(T_{\mathrm{sat}}, p_{\mathrm{sat}})$ gives the linearized form~\cite{tanasawa1991advances}:
\begin{equation}
   \dot{m} = \frac{2\hat{\sigma}}{2-\hat{\sigma}}  \sqrt{\frac{M}{2\pi R T_{\mathrm{sat}}}}  \frac{\rho_v L_{vap}}{T_{\mathrm{sat}}}  \left( T - T_{\mathrm{sat}} \right) ,
   \label{eq:HKS_linear}
\end{equation}
which explicitly connects $\dot{m}$ to the interfacial superheat.

\subsubsection{Lee phenomenological model}
The phenomenological Lee model~\cite{lee1980pressure} is among the most widely used closures in interface-resolved simulations of boiling.
The main assumption of this model is that phase change is driven by the deviation of the interfacial temperature from the saturation one, with the phase change rate being directly proportional to this temperature departure.
The volumetric mass source\footnote{Note that $\dot{m}_V$ is a volumetric mass source ($\mathrm{kg/m^3s}$) rather than an interfacial mass flux: it enters the governing equations directly, without the interfacial delta function required by the flux-based closures discussed above.} is thus computed as:
\begin{equation}
\dot{m}_V = r \, \alpha_\ell \, \rho_\ell \, \frac{T - T_{\mathrm{sat}}}{T_{\mathrm{sat}}}\, ,
\label{eq:lee_evap}
\end{equation}
where $\alpha_\ell$ is the liquid volume fraction and $r$ is an empirical relaxation parameter ($\mathrm{s}^{-1}$).
The model owes its popularity to its simplicity: no explicit reconstruction of the interface normal nor computation of sharp temperature gradients is required.
However, the relaxation parameter $r$ has no rigorous physical derivation: reported values span several orders of magnitude  ($r \sim 10^3$--$10^7$\,s$^{-1}$)~\cite{kharangate2017review}. 
Accordingly, the model can also be interpreted as a simplified version of the HKS model \cite{bahreini2015numerical,kharangate2017review}.
Recent variants of this approach address the problem-dependent nature  of $r$ by tying the relaxation rate directly to the numerical time step, thereby enforcing interfacial saturation at each step~\cite{rattner2014simple,pan2016saturated,weber2026consistent}.

\subsubsection{Chemical-potential relaxation model}
\label{sec:chem_pot_relaxation}

A different relaxation strategy is adopted within the class of BN models, where mass transfer is driven by the difference in chemical potential between the two phases, see Section~\ref{subsec:bn}.
The physical basis is the thermodynamic equilibrium condition: at a liquid-vapor interface in equilibrium the two phases share the same temperature, pressure and chemical potential, so any local imbalance acts as the thermodynamic force driving phase change.
The mass-transfer source term is written as a relaxation process towards chemical equilibrium \cite{pelanti2022arbitrary,sirianni2025efficient}:
\begin{equation}
  M = \nu\,(g_2 - g_1),
  \label{eq:chem_pot_relaxation}
\end{equation}
where $g_1$ and $g_2$ denote the chemical potentials of the two phases and the relaxation parameter $\nu$ controls the rate at which the interface is driven towards equilibrium, i.e. $g_1 = g_2$\footnote{We use here the subscripts $1$ and $2$ instead of $v$ and $l$ as this model is usually employed in connection with the BN model, a general two-fluid compressible framework.}.
Compared with the Lee model, this approach is thermodynamically consistent by
construction: the relaxation drives the system towards an equilibrium state defined by the second law of thermodynamics.
As mentioned above, this model integrates naturally into the BN framework.
This Gibbs free energy relaxation procedure has been employed to simulate nucleate boiling within a pressure-based, low-Mach four-equation framework~\cite{demou2022pressure}.

\subsubsection{Models for multicomponent evaporation}

Evaporation is a phase-change process in which liquid transforms into vapor at a liquid-vapor interface.
Unlike boiling, which involves the nucleation and growth of vapor bubbles within the bulk liquid or at a heated surface, evaporation occurs exclusively at a pre-existing interface where temperature and pressure drive the mass transfer process~\cite{dhir1998boiling}.
This difference has important consequences for numerical modeling: evaporation does not require additional models for nucleation and bubble dynamics.
An additional layer of complexity arises, however, when the volatile species evaporates into an inert, non-condensable gas, as is typical for a liquid droplet evaporating in air or a  drying liquid film.
In this multicomponent setting the gas phase is itself a mixture, and the phase-change rate is no longer governed by the interfacial energy balance alone but by the transport of the vaporizing species in the gas phase.
Indeed, in simulations of evaporation, one needs to solve an additional advection-diffusion equation for the vapor mass/volume fraction in the gas phase.
The mass flux is then obtained from a species balance at the interface:
\begin{equation}
  \dot{m}\,(Y_{\Gamma} - 1) = \rho_g \mathcal{D}\,
  \left. \frac{\partial Y}{\partial n} \right|_{\Gamma},
  \label{eq:species_balance}
\end{equation}
where $Y$ is the mass fraction of the vaporizing species in the gas, $Y_\Gamma$ its value at the interface, $\mathcal{D}$ the diffusion coefficient and $\rho_g$ the gas density; the left-hand side accounts for the convective (Stefan) flux generated by the mass transfer across the interface, while the right-hand side represents the diffusive flux. 
The interface mass fraction $Y_\Gamma$ is not known a priori but is fixed by the assumption of local thermodynamic equilibrium, which relates the partial pressure of the vapor to the local interface temperature through the Clausius-Clapeyron relation~\cite{tanguy2007level,palmore2019volume,zhao2022}.
Because Eq.~\eqref{eq:species_balance} assumes thermodynamic equilibrium, it can be seen as the counterpart of Eq.~\eqref{eq:jump_energy}, used for boiling problems. 
In the context of evaporation, the conservation of energy at the interface is then used to determine how the latent heat of evaporation/condensation changes the interface temperature, in turn modifying the saturation condition for species mass fraction. 
From a numerical viewpoint, this equilibrium constraint couples the species and energy fields at the interface and takes the form of a Robin (mixed) boundary condition for the vapor mass fraction: Eq.~\eqref{eq:species_balance} links the interface value $Y_\Gamma$ to its normal derivative, with coefficients that depend on the local interface temperature.
Enforcing this condition on a moving and deforming interface is one of the main challenges for multicomponent solvers, especially when based on a sharp-interface representation, and dedicated discretizations have been developed to impose it accurately~\cite{tanguy2007level,chai2020,mialhe2023extended,salimi2024robin,salimi2025low}.

\section{Numerical methods for interface representation}
\label{sec:interface}

Simulating a moving, continuously deforming interface that may undergo topological changes is a challenging task that requires specialized algorithms.
The available methods can be broadly classified into two families: interface tracking and interface capturing methods.
The fundamental difference resides in the definition of the interface: interface tracking approaches explicitly follow the position of the interface, defined with Lagrangian markers, while interface capturing methods represent it implicitly through a continuous field variable.
The interface description has direct consequences on the numerical treatment of topological changes, namely breakage and coalescence: interface tracking methods require closures to manage the connectivity of Lagrangian markers, while topological modifications of the interface are implicitly handled in interface capturing methods.
In the following, the key concepts of the different methods are briefly recalled; readers are referred to previous reviews on the different methods for comprehensive presentations~\cite{mirjalili2017interface,Elghobashi2018,soligo2021turbulent}.

\subsection{Front-tracking method}
Front tracking methods rely on a set of connected marker points to represent the moving interface~\cite{Unverdi1992,Tryggvason2001,das2023vapor}. 
The markers are connected by line elements and form a moving front in 2D problems, while a surface mesh is used for 3D problems.
Using the velocity field from the underlying fixed grid, each marker point located at $\mathbf{x}_f$ is advected according to:
\begin{equation}
\frac{\mathrm{d}\mathbf{x}_f}{\mathrm{d}t} = \mathbf{u}_f\, ,
\label{eq:fronttracking}
\end{equation}
where $\mathbf{u}_f$ is interpolated from the fixed grid to the marker location.
After each time step, the new front information is communicated to the fixed cells, and fluid properties such as density and viscosity are updated in every cell.
Front tracking was among the first interface-resolved methods applied to boiling heat transfer simulations, thanks to the seminal contributions of \cite{juric1998computations} and subsequent works \cite{esmaeeli2004computations,esmaeeli2004computationsp2,tryggvason2005direct}.
Perhaps the most critical challenge of the FT method concerns the spacing and connectivity of the interfacial markers: as the interface stretches and deforms, this spacing may decrease or increase, and this is made worse by topological changes and phase change.
Marker points must therefore be added or removed to ensure the discrete front accurately captures the evolving interface \cite{Tryggvason2001,singh2007three,gennari2025marching,zhong2025front}.
An approach that alleviates these issues is the recently proposed edge-based interface tracking (EBIT) method, which constrains the markers to the edges of the Eulerian grid, thus avoiding the explicit re-meshing procedure and allowing for automatic parallelization~\cite{pan2024edge,long2024edge}.

\subsection{Volume-of-fluid method}
The volume-of-fluid (VOF) method is the most widely used interface-capturing technique for two-phase flow problems \cite{soligo2021turbulent,mirjalili2017interface,kharangate2017review,cmmf2009}. In this approach, the interface is implicitly represented by a color function, $C$, which corresponds to the local volume fraction of a reference phase within each computational cell. Specifically, $C=0$ and $C=1$ denote cells fully occupied by the respective single phases, while intermediate values ($0 < C < 1$) identify cells containing the interface.
In the absence of phase-change, the color function is passively advected by the flow,
\begin{equation}
\frac{\partial C}{\partial t} + \nabla\cdot(\mathbf{u}\,C) = 0\, ,
\label{eq:vof}
\end{equation}
and the local fluid properties are obtained as volume-fraction-weighted averages of the two bulk values. 
The main numerical challenge lies in keeping the advected interface sharp: because $C$ is, by construction, discontinuous across
the interface, its discrete advection tends to smear the transition region \cite{cmmf2009}.
Two families of schemes have been developed to address this challenge.
Geometric schemes explicitly reconstruct the interface within each cell prior to advection: early methods such as the simple line interface calculation (SLIC)~\cite{noh1976slic} use segments aligned with the grid axes, whereas the piecewise-linear interface calculation (PLIC)~\cite{youngs1982time,scardovelli2003interface,pilliod2004second} employs segments of arbitrary orientation, improving the reconstruction accuracy.
Algebraic (or compressive) schemes, by contrast, bypass the need for explicit geometric reconstruction; they advect the color function directly, relying on specialized numerical fluxes to keep the interface confined to a few grid cells~\cite{hirt1981volume,ubbink1999method,xiao2011revisit,ii2012interface,pirozzoli2019algebraic}.
One of the main advantages of the VOF method is the discrete conservation of the volume fraction, which makes it naturally mass-conserving. 
A known drawback, on the other hand, is that the curvature estimation from a discontinuous function is inherently noisy and requires dedicated algorithms, such as height-function techniques, to recover accurate interface normals and curvatures \cite{popinet2018}.
The extension of VOF to boiling flows can be traced back to the seminal work of Welch \& Wilson~\cite{welch2000volume}, who applied the method to film boiling.
Their work inspired following studies where the VOF method was used to investigate different boiling flow configurations~\cite{kunkelmann2009cfd,kharangate2017review,malan2021geometric,sun2014modeling,gennari2022phase,giustini2023modelling,gu2024three,tsui2014phase,lee2019experimental}. 
Closely related to VOF methods are color-function approaches such as the interface-tracking method (ITM)\footnote{We stress that, despite the name, the ITM belongs to the interface-capturing family discussed here: the interface is represented by a continuous volume-fraction field and no Lagrangian markers are employed.} that transport a volume-fraction field but maintain interface sharpness through a dedicated sharpening step rather than geometric reconstruction, and which have also been successfully applied to boiling flow simulations~\cite{sato2013sharp,giustini2017computational,sato2026interface}.

\subsection{Level-set method}

The level-set (LS) method represents the interface implicitly as the zero iso-contour of a continuous scalar field, $\phi$, which is typically defined as a signed distance function that takes positive values in one phase and negative values in the other~\cite{Osher1988,sussman1994level,gibou2018review}.
The interface is advected by solving a transport equation for $\phi$,
\begin{equation}
\frac{\partial \phi}{\partial t} + \mathbf{u}\cdot\nabla\phi = 0\, ,
\label{eq:levelset}
\end{equation}
and geometric quantities -- interface normal and curvature -- are obtained directly from the scalar field $\phi$.
This is the principal advantage of the method: because $\phi$ is continuous and differentiable, normals and curvatures can be evaluated accurately, which makes the level-set method particularly well suited to flows in which surface tension and interface geometry play a dominant role.
The main drawback of the method is that the LS function is not a conserved quantity, so that mass conservation is not guaranteed at the discrete level, an issue that can be further amplified by the reinitialization steps required to maintain the signed-distance property of the LS function~\cite{sussman1994level,luo2019level}.
A widely-used workaround is the introduction of mass-correction steps, in which the spuriously lost (or gained) mass is redistributed in the vicinity of the interface, weighted by factors such as the local interface curvature~\cite{luo2015mass,luo2019level,ge2018efficient}. 
A more advanced alternative is the conservative LS method~\cite{Olsson2005,olsson2007conservative}, which replaces the signed-distance function with a hyperbolic tangent profile, introducing a conservative reinitialization step to ensure discrete mass conservation. 
Alternatively, the coupled level-set/volume-of-fluid (CLSVOF) approach combines the sharp geometric fidelity of the LS representation with the intrinsic mass conservation of the VOF method~\cite{sussman2000coupled,menard2007coupling}.
Among the VOF/LS coupled approaches, it is worth mentioning also the VOSET method~\cite{sun2010coupled}, which follows a different philosophy: only the VOF field is advected, while the LS function is reconstructed geometrically from the VOF data to ensure improved accuracy in calculating geometric quantities (e.g. curvature).
The LS method was extended to boiling flows by Son \& Dhir~\cite{son1997numerical,son1998numerical} and has since been widely used -- together with its CLSVOF and VOSET variants -- for nucleate and film boiling simulations~\cite{li2007numerical,gibou2007level,tanguy2014benchmarks,ningegowda2020mass,dhruv2021investigation,li2021manipulating,chen20223d,chen2023numerical}.

\subsection{Phase-field method}
Phase-field (PF) methods describe the interface as a thin transition layer across which the relevant thermophysical properties vary smoothly from one bulk phase to the other~\cite{Anderson1998,Jacqmin1999,Badalassi2003}.
The interface location is encoded in an order parameter, $\phi$, 
which takes constant values in the bulk phases and varies continuously across the transition region.
This diffuse description is shared with the BN models (see~Section~\ref{subsec:bn}), but the two differ in what the interfacial layer represents.
In PF methods the finite thickness originates from the gradient energy of a free-energy functional, although in practice it is artificially enlarged to render it computationally tractable.
In BN models, by contrast, the diffuse interface represents a mixing region between the two coexisting phases, arising from a compressible non-equilibrium description.
The order parameter (or phase-field) is transported by the following governing equation:
\begin{equation}
\frac{\partial \phi}{\partial t} + \nabla\cdot(\mathbf{u}       \phi) = \mathcal{R}(\phi)\, ,
\label{eq:phasefield}
\end{equation}
where the expression on the right-hand side $\mathcal{R}(\phi)$ depends on the specific formulation adopted \cite{Roccon2023,mirjalili2023assessment}.
We can distinguish two main classes of PF methods depending on the governing equation they rely on.
The first class of methods relies on the Cahn-Hilliard equation~\cite{Cahn1958,Cahn1959a,Cahn1959b}, a fourth-order conservative equation derived from a double-well free-energy functional. 
The second class is based on the Allen-Cahn equation, a second-order non-conservative equation that is computationally less demanding and can be suitably modified to enforce conservation~\cite{sun2004diffuse,chiu2011conservative,kim2014conservative,mirjalili2020conservative,jain2022accurate,hwang2026review}. 
Extension of PF methods to boiling simulations have been proposed in recent years, based either on the Cahn--Hilliard equation~\cite{jafari2015phase,wang2021phase,jafari2016numerical,minozzi2025influence} or on the Allen-Cahn equation~\cite{badillo2012quantitative,tamura2022development,roccon2024boiling,weber2026consistent}.
Nevertheless, the application of PF methods to phase-change phenomena remains at a relatively early stage. 
A key challenge is the dependence of the numerical solution on the interface thickness, which must be small enough to recover the sharp-interface limit, yet large enough to be adequately resolved by the computational grid (i.e.\ three to five points must be present across the interfacial layer~\cite{Soligo2019a,roccon2025flow36}).

\subsection{Lattice-Boltzmann method}
\label{sec:lb}
Lattice-Boltzmann (LB) methods represent an alternative approach to solve the Navier-Stokes equations and to describe the multiphase flow behavior.
Although categorized here with the interface-capturing family for ease of reference -- as the interface is represented through a continuous field -- LB methods are algorithmically distinct as they do not solve the Navier-Stokes equations directly, but rely on a mesoscopic kinetic description in which the evolution of discrete particle distribution functions is tracked on a regular lattice through a streaming-and-collision algorithm \cite{chen1998lattice,succi2001lattice,aidun2010lattice,timm2016lattice}.
For the commonly used Bhatnagar-Gross-Krook collision operator, the distribution functions $f_i$ evolve as:
\begin{equation}
f_i(\mathbf{x}+\mathbf{c}_i\delta t,\, t+\delta t) - f_i(\mathbf{x},t) = -\frac{1}{\tau}\left[f_i(\mathbf{x},t) - f_i^{\mathrm{eq}}(\mathbf{x},t)\right]\, ,
\label{eq:lbm}
\end{equation}
where $\mathbf{c}_i$ are the discrete lattice velocities, $\tau$ is the relaxation time, and $f_i^{\mathrm{eq}}$ is the local equilibrium distribution.
The macroscopic flow quantities (density, velocity, pressure) are recovered as moments of the distribution functions, and the Navier-Stokes equations are obtained in the hydrodynamic limit via a Chapman-Enskog expansion \cite{qian1992lattice}.
Despite their mesoscopic foundation, LB methods are employed in practice as macroscopic solvers.
The interface between phases is described either through pseudopotential models \cite{shan1993lattice}, in which non-ideal interactions between neighboring lattice sites mimic the cohesive forces responsible for phase separation, or by augmenting the hydrodynamic framework with a phase-field equation, wherein a discrete form of the Cahn-Hilliard or Allen-Cahn equation is solved concurrently with the hydrodynamic distributions
\cite{zheng2005lattice,zheng2006lattice,geier2015conservative,ren2016improved,wang2016comparative}.
The main appeal of LB methods lies in the locality of the streaming-collision operator, which makes them particularly well suited for massively parallel architectures~\cite{yang2023implementation,lauricella2025acclb}.
The past decade has seen LB methods progressively extended to interface-resolved simulations of boiling heat transfer~\cite{dong2010numerical,gong2012lattice,biferale2013simulations,safari2013extended,begmohammadi2015simulation,mohammadi2017phase,verdier2020performance,he2022lattice,wang2025mesoscopic}.

\section{Numerical treatment of interfacial conditions}
\label{sec:fluid}

\begin{figure}[!t]
\setlength{\unitlength}{0.0025\columnwidth}
\begin{picture}(400,140)
\put(0,0){\includegraphics[width=1.00\columnwidth, keepaspectratio]{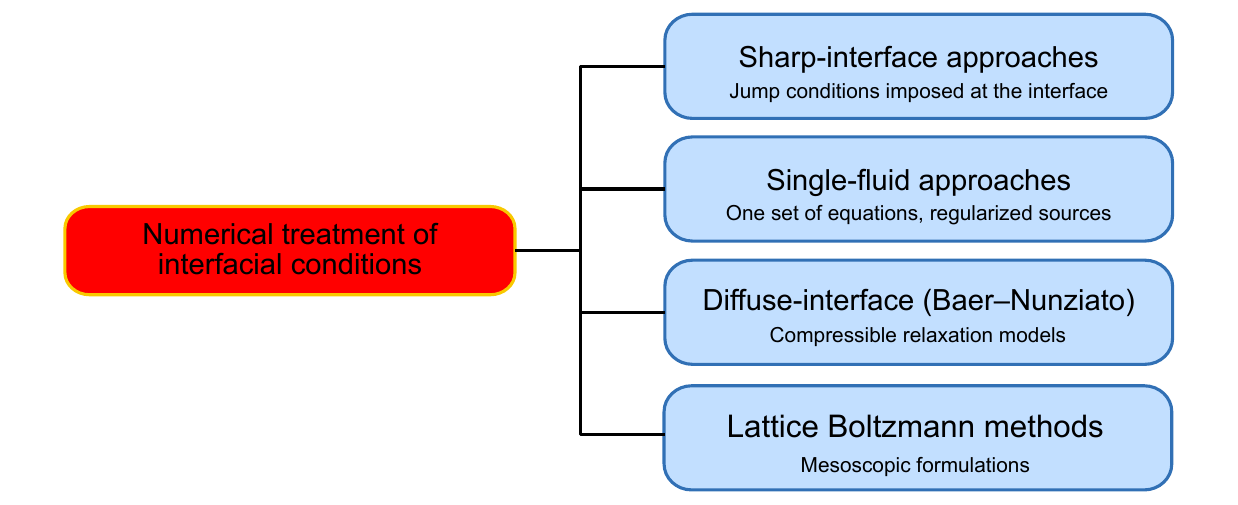}}
\end{picture}
\caption{Families of methods for interface-resolved simulations of boiling flows. 
Sharp-interface approaches treat the liquid–vapor boundary as a mathematical discontinuity and enforce the interfacial jump conditions through one-sided extrapolations. 
Single-fluid approaches solve a single set of conservation equations and account for phase change through interfacial source terms. 
Diffuse-interface approaches represent the interface as a finite-thickness mixture region and model heat and mass transfer through relaxation processes. 
Finally, lattice–Boltzmann methods recover the hydrodynamic behavior from a mesoscopic kinetic description and incorporate phase change through pseudopotential or phase-field formulations.}
\label{fig:summary4} 
\end{figure}

The jump conditions and closures just introduced prescribe what must hold at the interface, but they do not specify how these conditions are enforced numerically. 
That enforcement depends on the numerical representation of the interface and on how the discontinuities in pressure, velocity, and temperature are coupled with the flow field.
Four families of methods have emerged in the literature, see Fig.~\ref{fig:summary4}.

The first three classes -- sharp-interface, single-fluid and diffuse-interface approaches -- share a common starting point: the discretization of the macroscopic continuum equations. 
However, they differ in how jump conditions are imposed at the interface from a mathematical point of view.
Sharp-interface approaches treat the phase boundary as a mathematical discontinuity and impose the jump conditions sharply; the ghost-fluid method (GFM) is the most widely used representative of this class \cite{fedkiw1999non,kang2000boundary,Sussman2007,menard2007coupling,hong2007boundary,Desjardins2008a}.
The single-fluid and diffuse-interface approaches provide a regularized treatment of the interface instead of a sharp one.

The single-fluid approach extends incompressible and compressible multiphase formulations to phase-changing flows in a bottom-up fashion: a single set of conservation equations is written for the mixture, and interfacial mass and energy transfer are accounted for via additional, regularized source terms \cite{cmmf2009,kharangate2017review,darshan2024numerical}.
The diffuse-interface approach (here in the BN sense, distinct from the PF methods) follows instead a top-down logic: starting from a compressible non-equilibrium framework in which the two phases possess distinct velocities, pressures, temperatures, and free energies, successive relaxation steps yield a hierarchy of reduced models in which selected equilibrium conditions are progressively imposed~\cite{flaatten2011relaxation,lund2012hierarchy,linga2019hierarchy,bryngelson2021mfc,demou2022pressure,pelanti2022arbitrary,adebayo2025review}.

As regards boiling flows, the single-fluid and four-equation diffuse-interface formulations exhibit a close correspondence: under low-Mach or incompressible conditions they share the same number of governing equations, equilibrium hypotheses, and physical variables~\cite{Eikelder2026}.
However, they differ in the role attributed to the interface: a numerical regularization of a sharp interface in the single-fluid approach, versus a thermodynamic mixing zone in which both phases coexist at equilibrium in the diffuse-interface models -- and, more fundamentally, in that the four-equation model derives from a compressible non-equilibrium description and retains a thermodynamic pressure--density coupling that the incompressible single-fluid formulation does not.
The two therefore coincide only in the regime where compressibility is negligible.
Notably, in the diffuse interface models the jump conditions are not directly imposed: the interfacial equilibrium conditions are recovered through the relaxation processes while conservation across the interfacial region holds by construction.

The fourth family, the lattice-Boltzmann (LB) method, originates differently: rather than discretizing the macroscopic conservation laws directly, it recovers the hydrodynamic behavior from a mesoscopic kinetic description in which the interface emerges from the particle-distribution dynamics.
Phase-change is incorporated either intrinsically, through the non-ideal interaction force that governs phase coexistence (pseudopotential formulations)~\cite{gong2012lattice,biferale2013simulations}, or through prescribed mass flux source terms, in analogy with the single-fluid strategy (phase-field formulations)~\cite{safari2013extended,mohammadi2017phase}.

\subsection{Sharp-interface method}
\label{subsec:sharp}

\begin{figure}
\setlength{\unitlength}{0.0025\columnwidth}
\begin{picture}(400,180)
\put(0,0){\includegraphics[width=1.00\columnwidth, keepaspectratio]{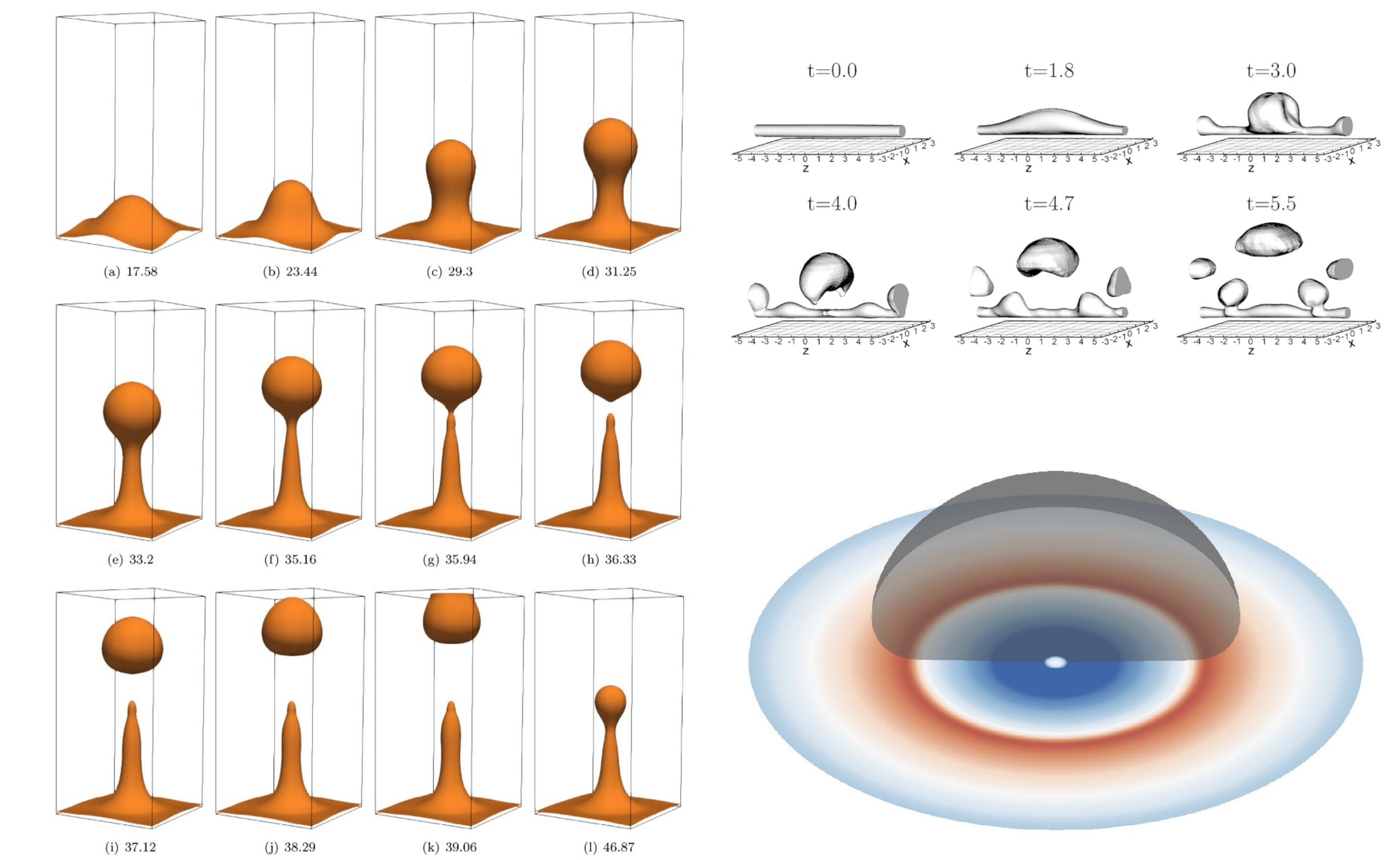}}
\put(-2,225){$(a)$}
\put(200,225){$(b)$}
\put(200,115){$(c)$}
\end{picture}
\caption{Overview of simulation results obtained using sharp-interface approaches. 
Panel~$a$ shows the time evolution of the liquid-vapor interface in a three-dimensional film boiling simulation of water near the critical point ($p_\mathrm{sat}=219~\mathrm{bar}$, $T_\mathrm{sat}=646~\mathrm{K}$), performed with the level-set method~\cite{ningegowda2020mass}.
Panel~$b$ shows the time evolution of the interface during film boiling \cite{son2007level}.
Panel~$c$ shows a volume-of-fluid simulation of nucleate boiling with a resolved microlayer~\cite{long2025direct}.}
\label{fig:sharp} 
\end{figure}
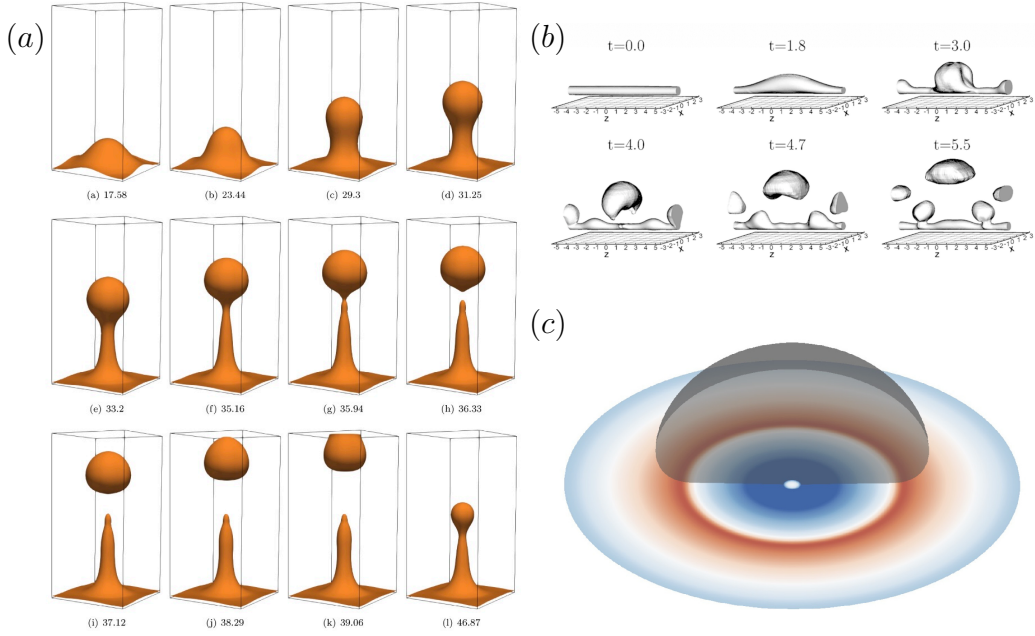

In sharp-interface methods, the liquid--vapor boundary is treated as a mathematical discontinuity of zero thickness, and the interfacial conditions are imposed directly through an ad-hoc numerical treatment of the cells adjacent to the interface. 
The most widely used technique of this class is the GFM~\cite{fedkiw1999non,Sussman2007,menard2007coupling,hong2007boundary,Desjardins2008a}. 
The core idea is to populate a band of ghost cells on each side of the interface with fictitious states, constructed so that the relevant jump conditions are enforced while standard single-phase discretization can be employed in every cell. 
The discontinuous fields are extended across the interface through one-sided extrapolations, so that each phase is advanced as if it occupied the whole domain, with the coupling between phases entering solely through the ghost states.
The GFM is most naturally combined with LS methods: the signed-distance function provides the interface normal and location required to build the ghost states and to perform the extrapolation. 
Because the jump conditions are imposed without smearing the discontinuity over the interfacial cells, the LS/GFM avoids the numerical diffusion of single-fluid approaches and yields accurate interfacial fluxes.

The work of Gibou {\em et al.}~\cite{gibou2007level} established the LS/GFM framework that paved the way for subsequent sharp-interface boiling simulations.
This approach has since been refined and extended by improving the accuracy and robustness of the interfacial flux computation~\cite{tanguy2014benchmarks,villegas2016ghost}.
More recently, the approach has been extended to block-structured Cartesian grids with adaptive mesh refinement (AMR)~\cite{dhruv2019formulation} and implemented on Cartesian collocated grids~\cite{chiramell2024sharp} as well as unstructured grids~\cite{sahut2020evaluation,sahut2021numerical}.
Importantly, when phase change occurs, specific care must be given to the advection of the level-set function: the velocity field exhibits a jump across the interface and thus one of the one-sided extensions must be selected to correctly advect the LS front ($\phi=0$)~\cite{son2007level,tanguy2014benchmarks}.
When applied to boiling, the GFM is employed to enforce the jump conditions derived from mass, momentum, and energy conservation, including the latent-heat effects.
In most LS/GFM formulations for boiling flows, a single pressure Poisson equation is solved over the whole domain, with the velocity and pressure jumps induced by phase change imposed directly at the interface rather than regularized over a finite band~\cite{kang2000boundary,nguyen2001boundary}.
This is possible because the velocity discontinuity induced by the volumetric expansion, and the associated pressure jump, are known quantities, fixed by the mass and momentum balances across the interface; the ghost construction imposes these prescribed jumps so that a standard single-domain solver reproduces the correct discontinuity~\cite{kang2000boundary}.

The temperature, in contrast, is not governed by a jump relating the two sides but by the saturation condition $T_\Gamma = T_\mathrm{sat}$, which fixes only its value at the interface.
The liquid- and vapor-side temperatures are therefore independent unknowns, each obtained by solving the energy equation on its own phase with $T_\mathrm{sat}$ imposed as a Dirichlet condition; the one-sided gradients of the two temperature fields determine the latent-heat flux and hence the phase-change rate~\cite{tanguy2014benchmarks,ningegowda2020mass}.
Overall, the GFM is a powerful method capable of handling the abrupt discontinuities across the interface commonly found in simulations involving liquid and vapor phases and phase-change phenomena, as can be appreciated from Fig.~\ref{fig:sharp}, which shows examples of simulations performed using the GFM. 
It is worth mentioning that, although the GFM is most often coupled with the
LS method, the ghost-fluid concept has also been recently combined with the VOF and AMR~\cite{long2025direct}, see Fig.~\ref{fig:sharp}$c$, and, more recently, with a front-tracking formulation in which the markers are bound to the grid edges~\cite{long2024edge}.

A limitation of sharp-interface methods arises in the treatment of tangential stresses (or tangential jump conditions), for instance, when surface tension varies along the interface because of temperature or composition gradients. 
The resulting Marangoni stress must be balanced by a jump in the tangential viscous stress. 
In an LS/GFM framework, this requires reconstructing the surface gradient of the surface tension and the tangential derivatives of the velocity field from one-sided extrapolations across a discontinuous interface, which is considerably more challenging than imposing a jump in the normal direction.

\subsection{Single-fluid approach}
\label{subsec:single_fluid}

Single-fluid (or one-fluid/single-field) approaches build on the well established framework for incompressible two-phase flows and extend it to phase-changing flows by introducing additional source terms that account for interfacial heat/mass transfer~\cite{cmmf2009,tryggvason2011direct}.
The underlying philosophy is bottom-up: one writes a single set of conservation equations for the mixture and treats the interface as a region where properties rapidly vary and in which mass and energy are exchanged.
This philosophy is opposite to that of diffuse interface models (discussed in Section~\ref{subsec:bn}), which start from a fully non-equilibrium two-phase mixture and progressively impose equilibrium conditions to derive reduced models.

In single-fluid approaches, the interface can be represented using either interface tracking or capturing techniques and a closure model is used to compute the interfacial fluxes.
By assuming that the bulk of the two phases is incompressible, the single-fluid system consists of a continuity equation, a momentum equation, an energy equation, and an equation describing the interface evolution.
Phase-change enters through volumetric source terms in the continuity, momentum, and energy equations, which in its simplest and most general form, read as follow:
\begin{equation}
\nabla \cdot \mathbf{u} = \dot{m}  \left( \frac{1}{\rho_v} - \frac{1}{\rho_l}  \right) \delta_\Gamma \, , 
\label{eq:sf_continuity}
\end{equation}
\begin{equation}
\frac{\partial(\rho \mathbf{u})}{\partial t} + \nabla \cdot (\rho \mathbf{u} \otimes \mathbf{u}) = -\nabla p + \nabla \cdot (2 \mu \mathbf{D}) + \sigma \mathcal{K} \mathbf{n} \delta_\Gamma  + \rho \mathbf{g}\ ,
\label{eq:sf_momentum}
\end{equation}
\begin{equation}
\rho C_p \left( \frac{\partial T}{\partial t} + \mathbf{u}\cdot\nabla T \right) = \nabla\cdot\left(\lambda\nabla T\right) - \dot{m} L_{vap}\, \delta_\Gamma \, ,
\label{eq:sf_energy}
\end{equation}
where $\delta_\Gamma$ is a surface delta function localized at the interface, and the mixture properties $\rho$, $\mu$, $C_p$, $\lambda$ are defined as volume-fraction-weighted averages of the per-phase values.
The continuity equation~\eqref{eq:sf_continuity} reflects the volumetric expansion (for boiling) associated with phase change: away from the interface the flow is divergence-free, while at the interface the source term enforces the kinematic jump condition expressed by Eq.~\eqref{eq:jump_mass}.
The capillary force in Eq.~\eqref{eq:sf_momentum} appears as a singular momentum source localized in a narrow band across the interface, following the continuum surface force formulation~\cite{brackbill1992continuum,popinet2018} or alternative formulations~\cite{gueyffier1999volume,francois2006,mirjalili2023assessment}.
By contrast, the recoil contribution to the normal-stress jump requires no ad hoc source term, as it is already accounted for by the advective term written in conservative form. 
The latent heat sink in Eq.~\eqref{eq:sf_energy} accounts for the energy absorbed (or released) by phase change and couples the temperature field to the interfacial mass flux, consistent with the energy jump condition, cf.\ Eq.~\eqref{eq:jump_energy}.

From an implementation viewpoint, a central question, common to all single-fluid formulations, is how to evaluate the delta function $\delta_\Gamma$ and distribute the associated source terms across the discrete numerical grid.
Sharp evaluations are possible when the interface representation provides explicit geometric information -- for example, in VOF methods with interface reconstruction or in FT methods with explicit Lagrangian markers -- but most implementations regularize $\delta_\Gamma$ over a few grid cells around the interface, replacing the singular distribution with a smoothed approximation.
The specific form of $\delta_\Gamma$ is generally dictated by how the interface is represented.
A recognized drawback of the delta-function formulation is that it smears the density, viscosity, and pressure profiles across the interface~\cite{nguyen2001boundary}.
A related difficulty is that the strongly localized source terms -- the volumetric expansion in the continuity equation and the singular capillary and recoil contributions in the momentum equation -- act as localized contributions when solving the pressure Poisson equation, whose stiffness grows with the density ratio and local interface curvature, to the point of compromising the robustness of the solver itself.
This has motivated dedicated treatments that enforce discrete momentum--mass consistency and ad-hoc numerical schemes for large density ratios and phase-change problems~\cite{DING2007,dodd2014fast,mirjalili2021consistent,dodd2025coupled,poblador2025momentum,weber2026consistent}.
In the following, we discuss in detail the coupling of the single-fluid framework with FT, VOF, LS, and PF methods.

\begin{figure}[!t]
\setlength{\unitlength}{0.0025\columnwidth}
\begin{picture}(400,170)
\put(0,0){\includegraphics[width=1.00\columnwidth, keepaspectratio]{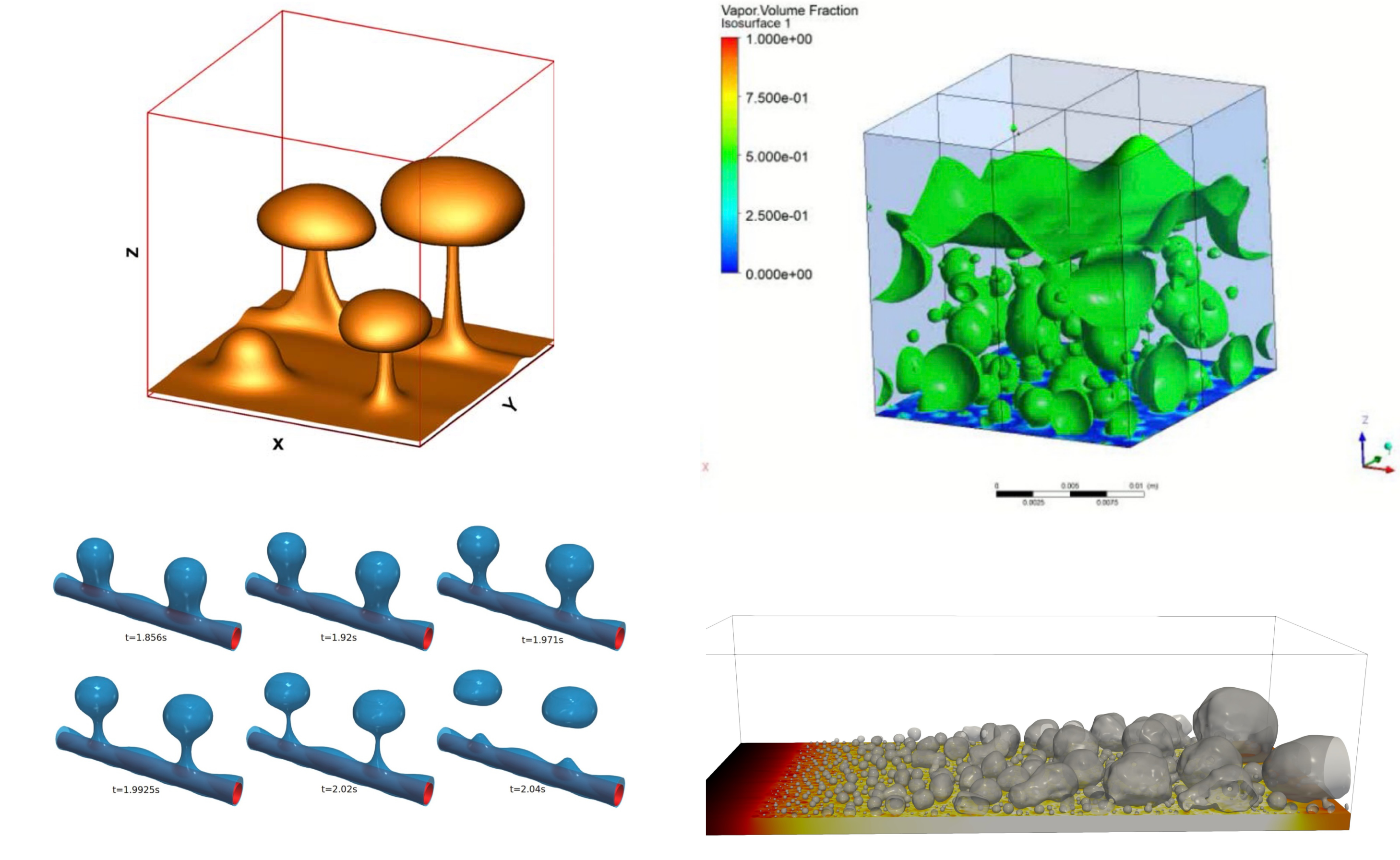}}
\put(-2,230){$(a)$}
\put(190,230){$(b)$}
\put(-2,90){$(c)$}
\put(190,90){$(d)$}
\end{picture}
\caption{Overview of simulation results obtained using the single-fluid approach.
Panel~$a$ shows a film boiling simulation performed with the front-tracking method, starting from an initial bimodal interface
perturbation, reproduced from Esmaeeli \& Tryggvason~\cite{esmaeeli2004computationsp2}.
Panel~$b$ shows a pool boiling simulation performed with the volume-of-fluid method where mixed-wettability surface is used, reproduced from Li {\em et al.}~\cite{li2021manipulating}.
Panel~$c$ shows the vapor bubble formation and release pattern for 3D film boiling over a horizontal cylinder for $\Delta T=10~K$, adapted from Kumar \& Premachandran \cite{kumar2022coupled}.
Panel~$d$ shows a flow boiling simulation performed using the phase-field method proposed by Weber {\em et al.}~\cite{weber2026consistent}.}
\label{fig:single} 
\end{figure}
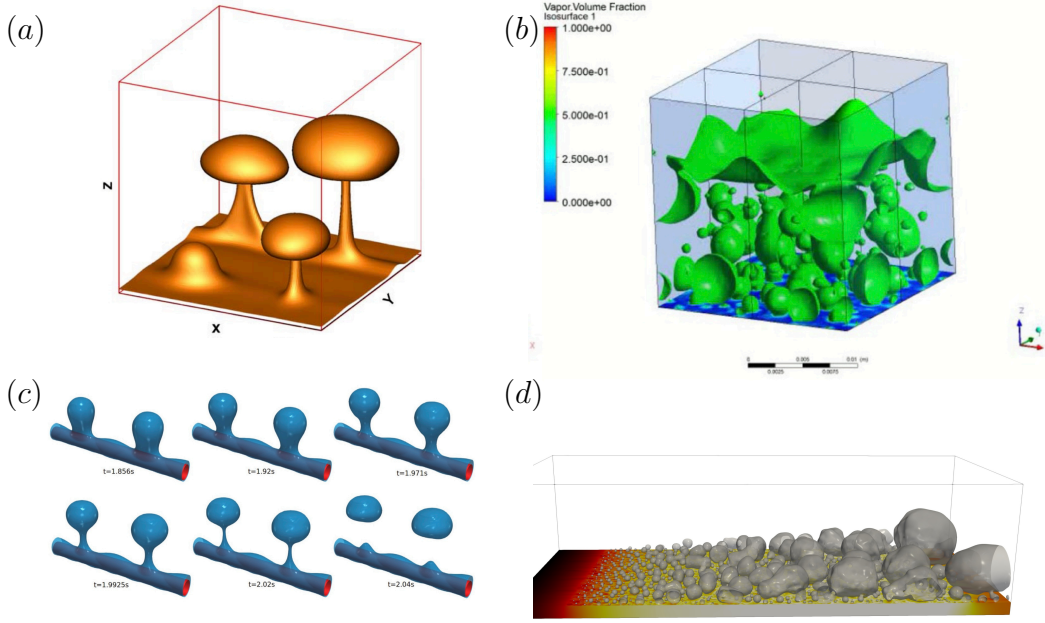

\subsubsection{Coupling with front-tracking}
The FT method was first used in combination with a single-fluid approach to simulate boiling flows by Juric \& Tryggvason~\cite{juric1998computations}, and further improved in subsequent contributions~\cite{esmaeeli2004computations,esmaeeli2004computationsp2}, see Fig.~\ref{fig:single}$a$.
Because the markers carry an explicit geometric description of the interface, the mass flux $\dot{m}$ can be evaluated directly on the front from the local temperature gradients and then spread to the Eulerian grid through a regularized delta function, in the same way as the surface tension force is distributed in simulations of isothermal flows~\cite{Tryggvason2001}.
Four aspects require care when phase change is simulated with a front-tracking method.
First, the saturation temperature must be enforced at the front: in the original formulation this requires an iterative adjustment of the interfacial source term~\cite{juric1998computations}, a procedure later eliminated by Esmaeeli \& Tryggvason~\cite{esmaeeli2004computations}.
Second, the markers must be advected with a velocity that accounts for the phase-change contribution: the front velocity is the interpolated bulk velocity plus the displacement induced by the local mass flux~\cite{esmaeeli2004computations,rouzbahani2023numerical}.
Third, the discrete mass source and the interfacial density entering this advection velocity must be defined consistently, since different combinations yield markedly different phase-change rates~\cite{shin2016numerical}.
Finally, phase change continuously stretches and compresses the front, requiring dynamic re-meshing of the markers~\cite{zhong2025front}.

\subsubsection{Coupling with volume-of-fluid}
The coupling of the single-fluid framework with the VOF method is one of the most widely employed strategies for boiling simulations, as demonstrated by the large body of work built on it~\cite{kharangate2017review,chen2024review} and the example shown in Fig.~\ref{fig:single}$b$.
In the context of VOF-based simulations, we can identify three main challenges.
A first challenge -- which is not specific to boiling simulations -- is that the geometric information needed to localize the mass/heat source terms -- the interface normal and curvature -- must be reconstructed from a discontinuous volume-fraction field, which makes both the curvature, defining the capillary stress, and the gradients, entering $\dot{m}$, susceptible to numerical noise and prone to the generation of spurious currents~\cite{jamet2002theory,popinet2018}.
This has motivated the development of improved geometric evaluation techniques within the VOF framework itself, such as height functions~\cite{popinet2018}, as well as the hybrid CLSVOF and VOSET approaches~\cite{sussman2000coupled,sun2010coupled}.

\begin{figure}[!t]
\setlength{\unitlength}{0.0025\columnwidth}
\begin{picture}(400,80)
\put(0,-10){\includegraphics[width=1.00\columnwidth, keepaspectratio]{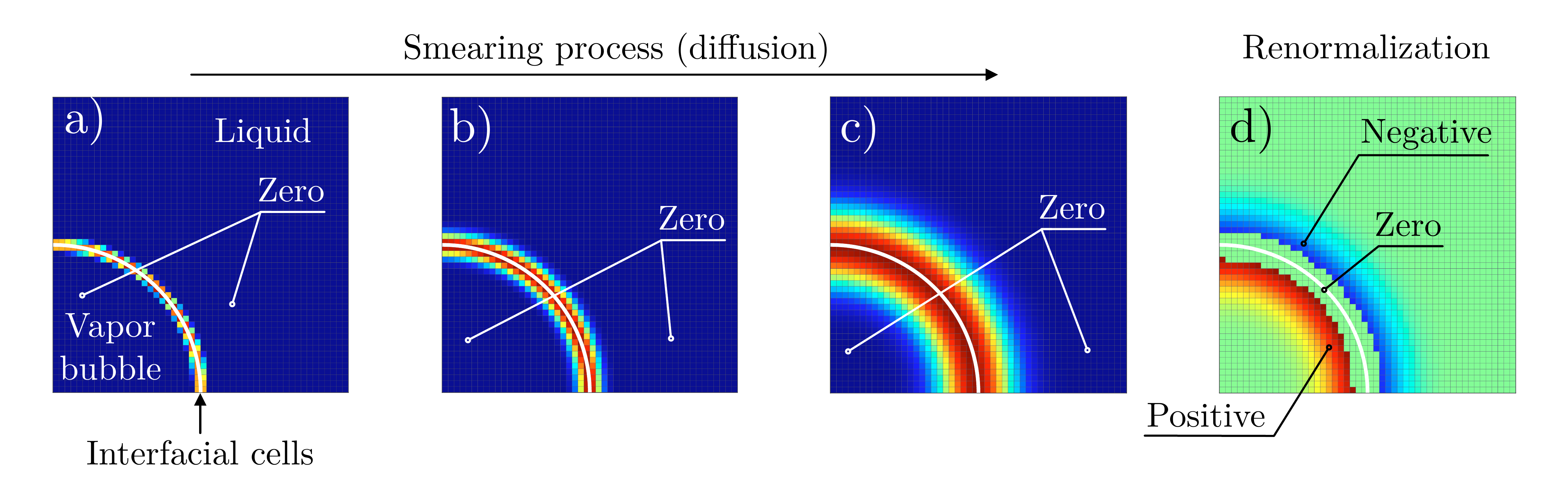}}
\end{picture}
\caption{Distribution of the mass-source term using the method proposed by Hardt \& Wondra~\cite{hardt2008evaporation}. 
The mass flow rate is first evaluated in the interfacial cells (panel~$a$). 
It is then distributed over the neighboring cells by solving a diffusion equation in pseudo-time (panels~$b,c$). 
Finally, the resulting flux is combined with the VOF field to assign its sign and perform the re-normalization (panel~$d$).}
\label{fig:hardt} 
\end{figure}

The second and third challenges are tightly coupled and involve how the mass source terms are distributed across the interface and the resulting velocity field used to advect the VOF function.
The most widely adopted approach is the smeared source-term formulation of Hardt \& Wondra~\cite{hardt2008evaporation}, in which the interfacial mass flux is first computed at the interface and then distributed across a finite band using a smoothed delta function constructed from the volume-fraction field.
The different steps of this process are shown in Fig.~\ref{fig:hardt}.
This approach has been widely used in VOF-based boiling simulations and has been extended in several directions, including the consistent treatment of the velocity jump~\cite{kunkelmann2009cfd}, the coupling with microlayer models at the contact-line~\cite{kunkelmann2010numerical}, and the combination of a regularized mass source with a sharp, ghost-cell imposition of the saturation temperature~\cite{zhang2018direct}.
The continuum-field representation of the source terms adopted by Hardt \& Wondra~\cite{hardt2008evaporation} may however violate local mass conservation in the vicinity of a highly curved interface, an issue that can be mitigated by reducing the diffusion constant governing the smearing~\cite{zanutto2026modelling}.

Alternative approaches -- which do not employ the Hardt \& Wondra method~\cite{hardt2008evaporation} -- have also been developed, and rely on the use of a divergence-free extension of the velocity field for the advection of the VOF function~\cite{palmore2019volume,scapin2020,malan2021geometric,gennari2022phase}.
For instance, Scapin {\em et al.}~\cite{scapin2020} advect the volume fraction with an interface velocity built as the sum of a divergence-free extension of the velocity field and an irrotational contribution due to phase-change, the latter obtained from a velocity potential satisfying a constant-coefficient Poisson equation with the phase-change source term.

\subsubsection{Coupling with level-set}
The single-fluid approach has also been successfully coupled with the LS method, as in the seminal works of Son \& Dhir~\cite{son1997numerical,son1998numerical}, see also Fig.~\ref{fig:single}$c$.
The main advantage of the LS representation within the single-fluid framework is geometric: the smooth signed-distance function provides accurate interface normal, curvature, and the temperature gradients required by the mass-flux closure -- precisely the quantities on which the single-fluid source-term distribution depends. 
Coupling the standard LS method with the single-fluid approach, however, poses some challenges from the mass-conservation point of view. 
The phase-change source enters continuity as a distributed volume source/sink, spread across the same interfacial band where the smoothed properties are defined~\cite{son1998numerical}, which is precisely the region where the standard LS method already loses mass through reinitialization. 
The two errors therefore overlap: the source term does not by itself corrupt the mass balance, but it places a production term exactly where the LS transport is least reliable, and the errors compound. 
Unlike the LS/GFM previously discussed, where the mass-transfer flux is imposed as a sharp jump condition at the interface, here the flux is smeared into the transport region. 
This can be mitigated through CLSVOF approaches~\cite{mudawar2023coupled}, which restore conservation through the volume-fraction field, or through conservative LS formulations which have also been employed in the context of boiling simulations~\cite{balcazar2023unstructured}, still within the single-fluid framework.

\subsubsection{Coupling with phase-field}
When the single-fluid approach is coupled with a PF method, the source terms appearing in the governing equations can be naturally distributed over the finite-thickness interface by exploiting the property of the phase-field~\cite{Roccon2023}, without requiring an explicitly constructed $\delta_\Gamma$.
In paractice, in the PF governing equation, the phase-change contribution is usually treated as a source term in the transport equation for the order parameter, which is distributed using the gradient of the phase-field or an approximation of it, e.g. using a polynomial expression~\cite{Roccon2023}.

Despite the PF method offers a natural framework for distributing the source terms thus allowing for an accurate computation of the surface tension forces and interfacial quantities, two aspects deserve particular care.
First, as in other interface-capturing methods coupled with single-fluid approaches, the velocity jump across the interface requires the source terms and the advecting velocity to be constructed consistently, so that the equilibrium interfacial profile is maintained during the computation~\cite{roccon2024boiling,roccon2025phase}.
This is generally done by distributing the source term using phase-field function~\cite{wang2021phase}.
In this context, Weber {\em et al.}~\cite{weber2026consistent} employed the conservative diffuse-interface model~\cite{mirjalili2020conservative} -- a modified version of the Allen-Cahn equation -- to simulate boiling heat transfer, improving the treatment of the velocity jump at the interface and thereby allowing the use of direct solvers for the pressure Poisson equation.
Second, the interface thickness acts as a numerical parameter, and its effect must be assessed to ensure accuracy and convergence to the sharp-interface limit.

\subsection{Baer–Nunziato diffuse-interface model}
\label{subsec:bn}

The most comprehensive diffuse-interface formulations for multiphase systems are the non-equilibrium BN and Romenski models \cite{baer1986two,romenski2015multiphase}. 
Of the two, the BN model \cite{baer1986two} is the more widely used, providing a general continuum framework for compressible multiphase flows under mechanical and thermal non-equilibrium.
In the following, the hierarchy of BN models is derived starting from the full non-equilibrium seven-equation formulation and progressively imposing equilibrium conditions: kinetic equilibrium reduces the seven-equation model to the six-equation model, mechanical equilibrium further reduces it to the five-equation model, and thermal equilibrium finally yields the four-equation model.
All the BN models require an equation of state (EOS) for each phase, see Section~\ref{sec:eos}.

\subsubsection{Seven equations model}
The seven-equation BN model allows the two phases to possess distinct velocities (kinetic disequilibrium), pressures (mechanical disequilibrium), and temperatures (thermal disequilibrium).
The corresponding system of governing equations is composed of one equation for the volume fraction (complemented by the condition $\alpha_1+\alpha_2=1$), two phasic mass conservation equations, two phasic momentum conservation equations and two phasic energy conservation equations.
The system of governing equations, without irreversible processes such as heat transfer and viscous stresses, read as follows \cite{saurel1999multiphase}:
\begin{equation}
\frac{\partial \alpha_1}{\partial t} + {\bf u}_I \cdot \nabla \alpha_1 =  \mathcal{P}\, ,
\end{equation}
\begin{equation}
\frac{\partial (\alpha_1 \rho_1)}{\partial t} +  \nabla \cdot (\alpha_1 \rho_1 {\bf u}_1) =  \mathcal{M}\, , 
\end{equation}
\begin{equation}
\frac{\partial (\alpha_2 \rho_2)}{\partial t} +  \nabla \cdot (\alpha_2 \rho_2 {\bf u}_2) = - \mathcal{M}\, , 
\end{equation}
\begin{equation}
\frac{\partial (\alpha_1 \rho_1 {\bf u}_1)}{\partial t} +  \nabla \cdot (\alpha_1 \rho_1 {\bf u}_1 \otimes {\bf u}_1) + \nabla ( \alpha_1 p_1)  =  p_I \nabla \alpha_1 + \mathcal{M}{\bf u}_I +  \mathcal{F}  \, ,
\end{equation}
\begin{equation}
\frac{\partial (\alpha_2 \rho_2 {\bf u}_2)}{\partial t} +  \nabla \cdot (\alpha_2 \rho_2 {\bf u}_2 \otimes {\bf u}_2) + \nabla ( \alpha_2 p_2)  =  p_I \nabla \alpha_2  - \mathcal{M}{\bf u}_I -  \mathcal{F}\, ,
\end{equation}
\begin{equation}
\frac{\partial (\alpha_1 E_1)}{\partial t} + \nabla \cdot [\alpha_1  (E_1 + p_1){\bf u}_1] =    - p_I \mathcal{P} + \mathcal{Q} + \mathcal{F} {\bf u}_I + \left(g_{I} + \frac{|{\bf u}_1|^2}{2} \right) \mathcal{M}\, ,
\end{equation}
\begin{equation}
\frac{\partial (\alpha_2  E_2)}{\partial t} + \nabla \cdot [\alpha_2  (E_2 + p_2){\bf u}_2] =     p_I \mathcal{P} - \mathcal{Q} - \mathcal{F} {\bf u}_I - \left(g_{I} + \frac{|{\bf u}_2|^2}{2} \right) \mathcal{M}\, ,
\end{equation}
where the subscripts $1,2$ have been used to identify quantities that refer to the two phases, ${\bf u}_{i}$ and $E_i$ represent the velocity vector and the total energy of the $i$-th phase.
The terms at right-hand side represent the interactions between the two phases.
Specifically, $\mathcal{M}$ is the mass transfer, $\mathcal{F}$ the drag force (exchange of momentum), $\mathcal{Q}$ the heat transfer between phases and $\mathcal{P}$ the pressure transfer.
These terms are defined as follows:
\begin{equation}
\mathcal{P}=\mu (p_1 - p_2);
\quad \mathcal{F}=  \lambda({\bf u}_2 - {\bf u}_1);
\quad \mathcal{Q}=\theta (T_2 - T_1);
\quad \mathcal{M}=\nu (g_2 - g_1)\, ,
\end{equation}
where the relaxation parameters $\mu$, $\lambda$, $\theta$ and $\nu$, denote the rate at which the different quantities relax to equilibrium.
Finally, the quantities $p_I$, ${\bf u}_I$ and $g_I$ denote the interface pressure, interface velocity and interface chemical potential, respectively. 
The definitions of these quantities must be consistent with the second law of thermodynamics, namely the entropy production for the mixture must be positive \cite{flaatten2011relaxation,pelanti2022arbitrary}.
Available definitions for these interface terms can be found in the literature \cite{saurel1999multiphase,saurel2003multiphase,saurel2009simple,pelanti2014mixture,perrier2021derivation}.
The BN model in its seven-equations formulation has been only rarely used for simulations of heat and mass transfer processes, as its numerical solution is rather challenging. 
This is primarily due to the character of the governing equations, which contain non-conservative terms and closure relations that must be carefully formulated to avoid the generation of spurious oscillations \cite{saurel1999multiphase,tokareva2010hllc,dumbser2016new,sirianni2025mixture,bhoriya2026physical}. 

\subsubsection{Six-equation model}
A first reduction of the seven equation BN model can be obtained by considering the asymptotic limit of stiff velocity relaxation ($\lambda \rightarrow \infty $).
In this formulation, pressure non-equilibrium effects are retained while a single velocity field is considered: ${\bf u}_2={\bf u}_1={\bf u}$.
As a consequence, the two phasic momentum conservation equations reduce to a single mixture momentum equation, the drag force contribution $\mathcal{F}$ disappears, and the interface velocity vanishes from the volume-fraction and energy equations.
The resulting system retains one equation for the volume fraction, two phasic mass conservation equations, one mixture momentum equation, and two phasic energy conservation equations -- six equations in total. 
The corresponding system is reported in~\ref{app:56bn}, to which we refer the reader for details.

One of the main advantages of this model is the elimination of the drag force term $\mathcal{F}$, which can introduce small spatial and fast temporal relaxation scales, making the system very stiff \cite{kapila2001two}. 
By assuming velocity relaxation, the six equation model enforces a single velocity field.
This simplifies the hyperbolic structure of the system, which now involves only three characteristic wave speeds, leading to simpler and more robust Riemann solvers and flux constructions \cite{saurel2009simple}.
The six-equation single-velocity model is not used as a standalone model but is typically used with instantaneous pressure relaxation as a numerical strategy to approximate solutions to the five-equation model.
Indeed, with respect to the latter, the six-equation formulation includes better preservation of volume-fraction positivity when dealing with shocks and rarefaction waves, and a speed of sound that is monotonic with respect to the volume fraction~\cite{saurel2009simple,pelanti2014mixture,saurel2018diffuse}.

\subsubsection{Five-equation model}
From the six equation model, full mechanical equilibrium can be imposed by assuming also pressure equilibrium (in addition to velocity) while allowing thermal and chemical non-equilibrium. 
This corresponds to the expansion of the system in the limit of stiff mechanical relaxation ($\mu \rightarrow \infty$ and $\lambda \rightarrow \infty $) \cite{stewart1984two,kapila2001two}, also known as the Kapila model \cite{kapila2001two}. 
The system of governing equations consists of a volume fraction transport equation, two phasic mass conservation equations, one momentum equation, and one energy equation, whose explicit form is reported in \ref{app:56bn}.

A distinctive feature of this formulation is the appearance of a velocity-divergence term $K \nabla \cdot {\bf u}$ in the volume fraction transport equation, together with two additional source contributions $K_h \mathcal{Q}$ and $K_m \mathcal{M}$ that couple the volume fraction evolution to heat and mass transfer.
The three coefficients $K$, $K_h$ and $K_m$ represent the expansion and compression coefficient in the mixture regions, the interface temperature coefficient and the interface density coefficient \cite{saurel2008modelling,pelanti2022arbitrary}.
The primary advantage of the five-equation model lies in its computational efficiency and robustness, achieved by reducing the number of governing equations. 
However, a disadvantage is the limitation in physical fidelity for the phase compressibility \cite{saurel2007shock,lemartelot2013liquid}. 
While the Kapila formulation restores this physics by including a velocity divergence term, it also introduces numerical stiffness, which can make the solution unstable under violent expansions or compressions \cite{schmidmayer2020assessment,bryngelson2021mfc}. 

Additional challenges include the difficulty of preserving volume fraction positivity in the presence of shock waves \cite{saurel2009simple,ten2017acoustic} and the non-monotonic behavior of the mixture speed of sound, $c_{eq}$, determined by the Wood formula \cite{wood1930textbook,wallis2020one}, which make difficult the use of Riemann solvers due to the loss of convexity. 
To mitigate the spurious pressure and velocity oscillations that arise at material interfaces from the non-conservative volume-fraction transport, high-order shock- and interface-capturing schemes have been also developed \cite{johnsen2006implementation}.

\subsubsection{Four-equation model}

\begin{figure}
\setlength{\unitlength}{0.0025\columnwidth}
\begin{picture}(400,150)
\put(8,0){\includegraphics[width=1.00\columnwidth, keepaspectratio]{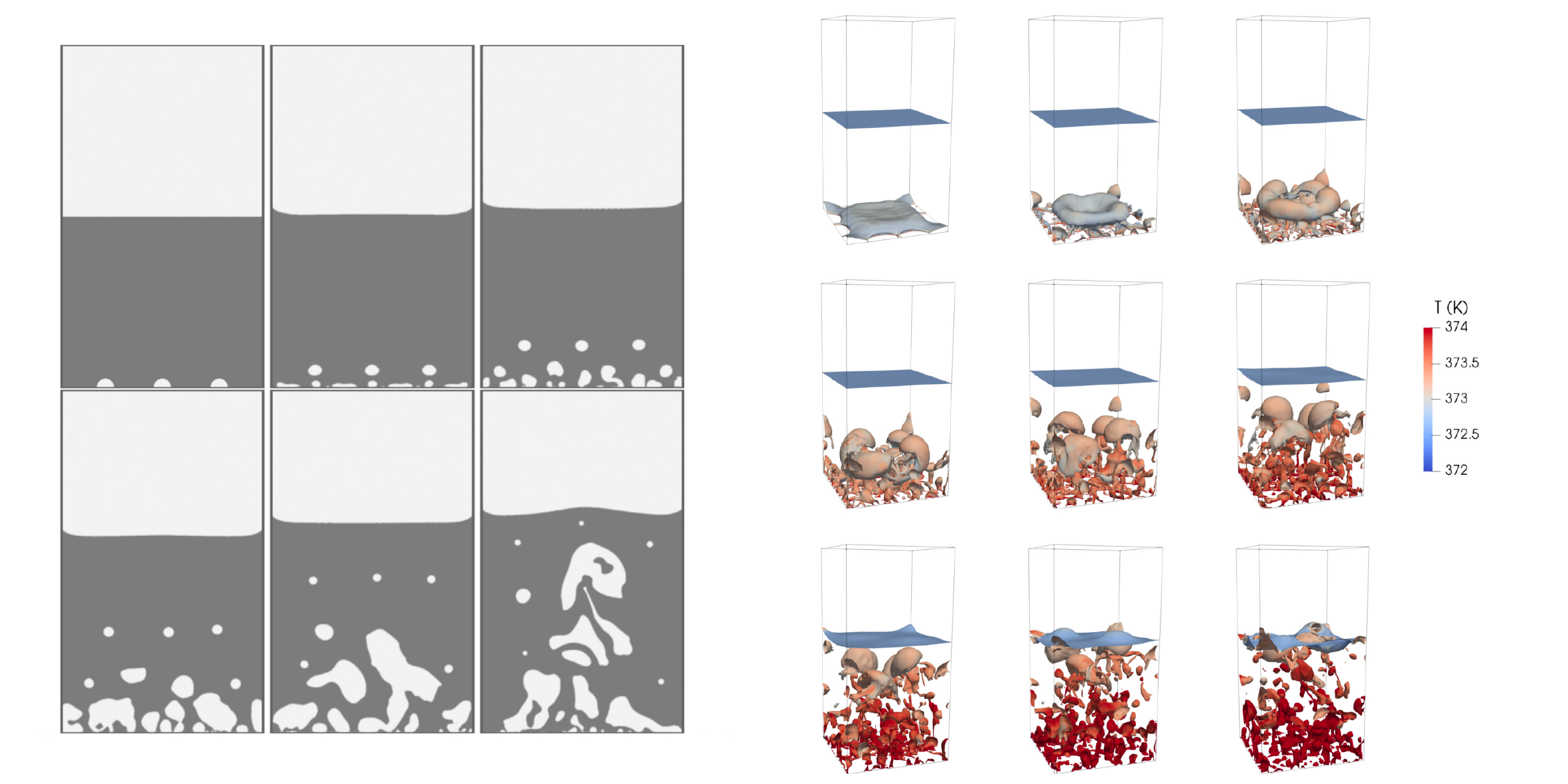}}
\put(0,180){$(a)$}
\put(195,180){$(b)$}
\end{picture}
\caption{Graphical overview of computations performed using the four-equation BN model.
Panel $a$ shows the volume fraction at six different times ($t=0$, $t=50~ms$, $t=100~ms$, $t=200~ms$,  $t=300~ms$ and  $t=400~ms$). 
The computational domain has dimensions $7 \times 12~cm$ and the bottom wall is heated, reproduced from Le Martelot {\em et al.} \cite{le2014towards}.
Panel $b$ shows the iso-contour of the volume fraction colored by the temperature field for the three-dimensional nucleate boiling in water in a computational domain with dimensions $7 \times 7 \times 12~cm$.
First snapshot refers to $t=0.262 s$, and each subsequent snapshot at intervals $\Delta t \simeq 0.046~s$, reproduced from Demou {\em et al.} \cite{demou2022pressure}.}
\label{fig:bn} 
\end{figure}

The hierarchy of BN models presented above has been widely employed to investigate compressible two-phase flow problems, including bubble cavitation and collapse
\cite{rasthofer2017large,tiwari2013diffuse,johnsen2009numerical}, shock--bubble interactions \cite{coralic2014finite,schmidmayer2020assessment,beig2015maintaining,hejazialhosseini2013vortex},
and droplet atomization \cite{meng2018numerical}.
For boiling flows, a further simplification follows from assuming thermal equilibrium \cite{le2014towards,saurel2018diffuse,demou2022pressure}.
Since modeling phase change requires accounting for conductive heat transfer, the temperature field becomes continuous across the interface, allowing a single energy equation to be used.
The five-equation model can thus be reduced by imposing instantaneous thermal equilibrium ($\theta \rightarrow \infty$), yielding a four-equation model with two phasic mass conservation equations, one momentum equation, and one energy equation~\cite{pelanti2022arbitrary}:
\begin{equation}
\frac{\partial (\alpha_1 \rho_1)}{\partial t} +  \nabla \cdot (\alpha_1 \rho_1 {\bf u}) =  \mathcal{M}\, , 
\end{equation}
\label{foureq1}
\begin{equation}
\frac{\partial (\alpha_2 \rho_2)}{\partial t} +  \nabla \cdot (\alpha_2 \rho_2 {\bf u}) = - \mathcal{M}\, , 
\end{equation}
\label{foureq2}
\begin{equation}
\frac{\partial (\rho {\bf u})}{\partial t} +  \nabla \cdot ( \rho  {\bf u} \otimes {\bf u} + p \mathcal{I})  =  0\, ,
\end{equation}
\begin{equation}
\frac{\partial (\rho E)}{\partial t} + \nabla \cdot \left[ (\rho E + p){\bf u} \right] = 0\, .
\end{equation}

The formulation above (obtained from the parent BN model) is here presented in its most basic formulation.
To describe boiling flows, modeling of additional physical effects is however usually required.
These effects, namely viscous stresses, surface tension, heat conduction, and buoyancy can be directly incorporated on the right hand side of the momentum and energy equations~\cite{saurel2018diffuse}.

However, the four-equation BN model has been scarcely used in its original formulation due to its inherent numerical challenges; instead, a reformulation with a different set of primary variables is preferred~\cite{demou2022pressure,saurel2016general}.
To remove the time-step limitations associated to acoustic waves in the low-Mach configurations typical of pool and flow boiling, it is possible to exploit mass conservation and the mixture EOS to derive a system of four transport equations: volume fraction, temperature, velocity, and pressure~\cite{demou2022pressure,scapin2022mass}.
Compared to the standard density-based formulation, the pressure-based formulation is computationally more efficient for low-speed flows as it avoids costly preconditioning and spurious pressure oscillations by solving directly for the thermodynamic pressure~\cite{demou2022pressure}.
Further, Scapin~{\em et al.}~\cite{scapin2022mass} augmented the system to five equations by solving two equations for the mass density of each phase instead of the volume fraction, which has been shown to more accurately ensure mass conservation.
Two examples of simulations of boiling flows performed using the four-equation BN model are shown in Fig.~\ref{fig:bn}.

\subsection{Lattice-Boltzmann method}
\label{subsec:lb}

\begin{figure}
\setlength{\unitlength}{0.0025\columnwidth}
\begin{picture}(400,100)
\put(0,0){\includegraphics[width=1.00\columnwidth, keepaspectratio]{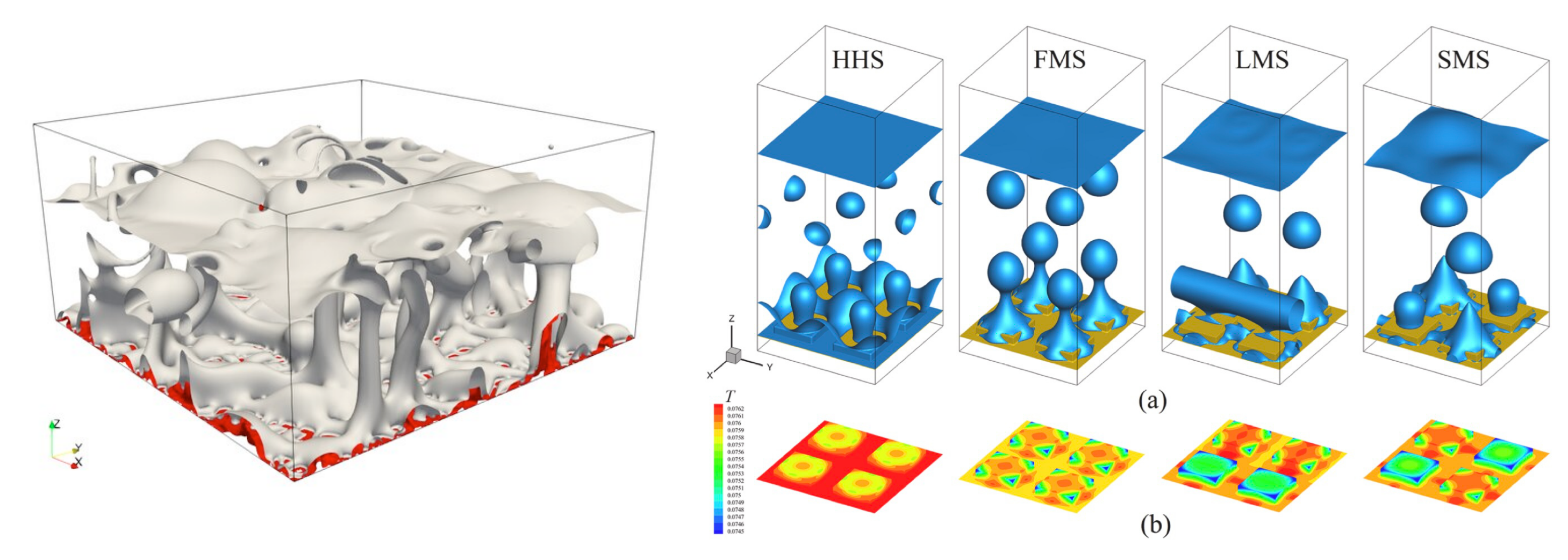}}
\put(0,125){$(a)$}
\put(175,125){$(b)$}
\end{picture}
\caption{Overview of simulation results obtained using the LB method.
Panel~$a$ shows simulation of pool boiling in the transition regime \cite{fei2020mesoscopic} for $t=63.09$.
Panel~$b$ shows the results obtained from simulation of boiling heat transfer on structured surfaces: homogeneous hydrophilic
surface (HHS), fully mixed surface (FMS), linear mixed surface (LMS), and staggered mixed surface (SMS) \cite{luo2024enhanced}.}
\label{fig:lbm} 
\end{figure}

Unlike the aforementioned approaches, LB methods do not impose the jump conditions as explicit constraints.
Surface tension and the associated normal-stress jump emerge from the mesoscopic dynamics, through the non-ideal interaction force in pseudopotential models~\cite{shan1993lattice} or the free-energy functional in phase-field formulations~\cite{zheng2005lattice,zheng2006lattice,geier2015conservative,ren2016improved}.
The two routes differ in how the interfacial mass flux enters the formulation~\cite{li2016lattice,yang2025lattice}.
In pseudopotential-based models, phase change is driven directly by the EOS, the mass flux being an output of the computation.
In phase-field-based models, the mass flux is computed from the local temperature field and introduced through regularized source terms, similarly to the single-fluid approach.
The Stefan condition and the latent heat are correspondingly either embedded implicitly in the thermodynamics of the model or imposed explicitly as interfacial energy sinks.

In the pseudopotential approach, the non-ideal EOS that sustains phase coexistence also controls phase change: when the local temperature departs from saturation, the resulting mechanical disequilibrium drives evaporation without any interfacial source term~\cite{li2016lattice}.
In the original Shan--Chen formulation, the effective EOS is implicitly dictated by the choice of the pseudopotential function; the model has since been extended to incorporate realistic non-ideal EOSs, such as the cubic Peng--Robinson or the Carnahan--Starling equations, improving the representation of liquid--vapor coexistence and saturation properties~\cite{YuanSchaefer2006}.
The first boiling simulations within this framework date back to Zhang \& Chen~\cite{zhang2003lattice}.
This formulation was later improved by H\'azi \& M\'arkus~\cite{hazi2009bubble,markus2011simulation}, who derived the target temperature equation from the local entropy balance for a non-ideal fluid:
\begin{equation}
\frac{\partial T}{\partial t} + {\mathbf u}\cdot\nabla T = \frac{1}{\rho c_v}\nabla\cdot\left(\lambda \nabla T\right) - \frac{T}{\rho c_v}\left(\frac{\partial p_{EOS}}{\partial T}\right)_{\rho} \nabla \cdot {\mathbf u} \, ,
\label{eq:lbm_energy}
\end{equation}
where the last term is the one responsible for latent heat absorption and release at the interface, across which the flow divergence is non-zero.
Latent heat is therefore not an input parameter: it is set implicitly by the EOS, and its correct reproduction requires the pseudopotential to be thermodynamically consistent, i.e.\ the mechanical stability condition of the lattice interaction force must recover the Maxwell equal-area construction~\cite{biferale2013simulations}.
Since the two conditions do not coincide in general, the coexistence densities and the effective latent heat deviate from the target EOS unless the pseudopotential form or the forcing scheme is suitably constructed~\cite{biferale2013simulations,li2016lattice}.

The energy equation~\eqref{eq:lbm_energy} can be solved using two main approaches.
The first relies on a second distribution function that evolves alongside the hydrodynamic one, with source terms constructed to recover equation~\eqref{eq:lbm_energy}~\cite{markus2011simulation,biferale2012convection}.
While this strategy retains a fully local LB discretization, the Chapman--Enskog limit of the temperature equation is contaminated by spurious terms that, while negligible in single-phase incompressible flows, become significant across the interface~\cite{li2014effect}.
For this reason, several authors~~\cite{li2015lattice,fei2020mesoscopic}~adopt a hybrid strategy in which the flow field is solved with the LB algorithm while Eq.n~\eqref{eq:lbm_energy} is discretized with standard finite-difference schemes.
\sloppy{This hybrid approach circumvents the error terms of the double-distribution-function formulation at the cost of partially sacrificing locality, and has become the prevailing choice in recent boiling studies~\cite{li2015lattice,zhang2021lattice}.}

Both strategies have since been applied broadly to simulate wall-superheat, heat-flux, and wettability effects on bubble dynamics and the boiling curve~\cite{lee2016critical,zhang2016mesoscale,lee2017numerical,lee2017conjugate,zhang2017mesoscale,zhang2018study,ma20193d,wang2022effects}, heterogeneous nucleation from cavities~\cite{mu2017nucleate,zhou2019lbm}, surface-microstructure and combined wettability--microstructure enhancement~\cite{chang2019lattice,zhou2019lattice,mondal2021numerical,wang2023boiling,li2018enhancement,ma2018simulations,yu2018boiling,xu2021experimental,wang2023pool}, and channel flow boiling~\cite{gong2014numerical,chen2022micro,zhang2023lattice,song2023channel,luo2024exploring}.
Despite this versatility, most pseudopotential boiling studies were long limited to moderate liquid--vapor density ratios by numerical instability and spurious interfacial currents.
A series of improvements to the collision operator, the forcing scheme, and the near-wall thermal treatment has progressively relaxed this limitation, from density ratios of $\mathcal{O}(10^2)$~\cite{Fang2017,Mukherjee2021} to $\mathcal{O}(10^3)$ in recent nucleate- and film-boiling simulations~\cite{fei2020mesoscopic,le2023,xu2025}.

In the phase-field LB approach, the interface is transported by a Cahn--Hilliard or Allen--Cahn equation (or modified versions of these equations) solved through an auxiliary distribution function, and phase-change terms are introduced exactly as in the single-fluid framework.
The first phase-change extension within this framework was proposed by Dong {\em et al.}~\cite{dong2009}, who coupled the model of Zheng {\em et al.}~\cite{zheng2006lattice} with the thermal scheme of Inamuro {\em et al.}~\cite{inamuro2002}, adding source terms constructed under the assumption that the heat conducted to the interface is entirely converted into latent heat; the approach was subsequently applied to nucleate boiling from single and multiple nucleation sites~\cite{dong2010numerical} and to bubble growth and departure in quiescent and flowing liquids~\cite{sun2013a,sun2013b,sun2016,sun2017}.

However, Safari {\em et al.}~\cite{safari2013extended} pointed out that this construction lacks consistency at high density ratios and leaves the interface velocity ambiguous, since the source term enters the phase-field equation while the momentum equation is left unchanged.
In consistent formulations, a mass-transfer source term is added to the phase-field governing equation, and the corresponding volume expansion is imposed through the velocity-divergence constraint localized at the interface~\cite{safari2013extended,haghani2021phase,abadi2018hybrid}.
The mass flux is computed from the one-sided temperature gradient, and the latent heat appears as an explicit sink in the temperature equation.
Following this strategy, the model of Safari {\em et al.}~\cite{safari2013extended} was extended to nucleate pool boiling~\cite{begmohammadi2015simulation} and to three-dimensional configurations with multiple nucleation sites and film boiling~\cite{sadeghi2016three,sadeghi2017three}, while Mohammadi-Shad \& Lee~\cite{mohammadi2017phase} computed the mass-transfer source term directly from the Rankine-Hugoniot jump condition in combination with a sharp-interface energy solver.
This construction inherits the properties of its macroscopic counterpart: the Stefan condition is imposed in regularized form, the saturation properties and latent heat are free inputs decoupled from the EOS, but phase change requires a pre-existing interface, since a uniform order parameter yields no mass transfer and bubble growth simulations must be initialized with seeded nuclei~\cite{li2016lattice}.
To overcome this limitation, source terms based on the sensible heat of the superheated phase have recently been proposed, enabling phase change initiated from a single phase~\cite{xiong2025phase}.

Recently, the conservative Allen–Cahn equation 
has become increasingly preferred over
its Cahn–Hilliard counterpart, since it conserves mass and involves only second-order spatial derivatives.
Verdier {\em et al.}~\cite{verdier2020performance} observed that the earlier phase-change source terms~\cite{dong2009,safari2013extended} assume the liquid at saturation and neglect its thermal conductivity, and proposed instead a source term computed from the departure of the interface temperature from saturation.
Haghani-Hassan-Abadi {\em et al.}~\cite{haghani2021phase} coupled the conservative Allen-Cahn equation with an LB momentum solver and a finite-difference energy solver.
Two examples of boiling flows simulations performed using the LB method are shown in Fig.~\ref{fig:lbm}.

\section{Equations of state and compressibility effects}
\label{sec:eos}

An EOS -- a relation between temperature, pressure, and density -- is typically required in three distinct regards within boiling flow simulations: i) to close the hyperbolic system of equations when the compressible diffuse-interface framework is employed; ii) when one of the two phases is treated as compressible in single-fluid or sharp-interface approaches; iii) in pseudopotential LB methods, where the EOS is embedded in the mesoscopic interaction force and drives phase change.
The latter  has been discussed in Section~\ref{subsec:lb} and is not further considered here.

A variety of EOS formulations are available in the literature~\cite{adebayo2025review}, ranging from simple models suitable for idealized conditions to more sophisticated expressions capable of capturing complex thermodynamic behaviors.
Commonly adopted EOSs include the ideal-gas, stiffened-gas, Noble-Abel stiffened-gas, and cubic models.
In the following, we discuss the available EOSs in the context of BN models and low-Mach/fully compressible formulations.

\subsection{Equation of state for Baer-Nunziato models}
In BN models, the thermodynamic closure requires an EOS for each phase.
The mixture EOS is then not prescribed independently but emerges from the pressure and temperature equilibrium conditions applied simultaneously with the mixture density and energy definitions, yielding a nonlinear algebraic system that must be solved at each point in the domain.
Widely adopted EOSs are the stiffened-gas \cite{harlow1968numerical} and the Noble--Abel stiffened-gas \cite{le2016noble,radulescu2019noble}, where the latter extends the classical stiffened gas formulation by accounting for short-range repulsive effects.
The stiffened gas EOS reads as follows:
\begin{equation}
p_k = (\gamma_k -1) \rho_k e_k - \gamma_k p_{\infty,k}
\end{equation}
where $\gamma_k$ is the adiabatic specific heat ratio of the $k$-th phase and $p_{\infty,k}$ the corresponding reference pressure.
The Noble--Abel stiffened gas EOS is:
\begin{equation}
p_k = \frac{\rho_k (\gamma_k -1) (e_k -q_k)}{(1-\rho_k b_k)} - \gamma_k p_{\infty,k}
\end{equation}
where the numerator accounts for thermal agitation, the denominator for short-range distance repulsion and $q_k$ represents the heat bond of the corresponding phase.
The parameters $q_k$ and $b_k$ are constant coefficients characteristic of the thermodynamic properties of the fluid \cite{le2016noble}.
Because an EOS is required for each phase in BN-derived models (i.e. for $k=1,2$), 
there is no need for complex formulations covering both liquid and vapor thermodynamics; each phasic EOS needs only be accurate within its respective domain of applicability.

The wide use of stiffened-gas EOSs is also motivated by two numerical reasons.
First, the pressure-, temperature-, and chemical-potential-relaxation procedures through which heat and mass transfer phenomena are modeled require the repeated search of equilibrium conditions.
Using stiffened-gas EOSs, these reduce to the solution of explicit or low-order algebraic relations~\cite{pelanti2014mixture,pelanti2022arbitrary}, whereas a more complex EOS would turn each relaxation step into a nested iterative procedure.
Second, both EOSs are convex~\cite{le2016noble}: the speed of sound is real and well defined at every admissible state, which guarantees the hyperbolicity of the system, and the isentropes are convex, which ensures the classical shock--rarefaction wave structure assumed by approximate Riemann solvers~\cite{menikoff1989riemann,toro2013riemann}.
A new formulation able to circumvent the restriction to simple analytic EOSs was recently proposed, in which arbitrary EOSs can be employed~\cite{sirianni2025efficient}.

\subsection{Low-Mach and fully compressible formulations}

A common assumption in interface-resolved simulations of boiling heat transfer is that both phases are incompressible, so that the density is piecewise constant and changes only across the interface.
This assumption is justified for boiling at moderate superheat and away from the critical point, where the vapor density varies slowly enough that acoustic effects are negligible and the resulting Mach number is small.
Owing to its simplicity and to the fact that no EOS is required, the incompressible approach was among the first used to extend no-phase-change interface-capturing and -tracking methods to boiling heat transfer~\cite{welch2000volume,juric1998computations,son1998numerical}, and it remains the most widely used approach.

However, this assumption no longer holds in two situations.
First, vapor compressibility can play a significant role even at low Mach numbers in confined geometries or under rapid bubble growth~\cite{prosperetti2017vapor,urbano2022semi,bibal2024compressible}.
Second, at elevated pressures, approaching the critical point, the density ratio between the phases decreases and the assumption of an incompressible liquid becomes less accurate.
To explore the regime where compressibility plays a non-negligible role, two approaches are available: low-Mach-number variable-density solvers and fully compressible solvers.
The first class retains the assumption of a slowly varying thermodynamic pressure and neglects acoustic propagation, while introducing a temperature-dependent density field.
From a mathematical point of view, the pressure is decomposed into a thermodynamic component and a hydrodynamic one:
\begin{equation}
  p(\mathbf{x},t) = p_0(t) + p_1(\mathbf{x},t)\, , \qquad p_1/p_0 = \mathcal{O}(M^2)\, ,
  \label{eq:pressure_splitting}
\end{equation}
where $M$ is the Mach number.
The spatially uniform $p_0(t)$ enters the EOS and determines the density and other properties, while the hydrodynamic correction $p_1$ plays a purely mechanical role, enforcing the velocity-divergence constraint through an elliptic equation as in incompressible flows.
This circumvents the need to resolve acoustic waves, whose propagation would otherwise impose a stiff time-step restriction.
It also preserves the elliptic character of the pressure equation while allowing the vapor density to respond to variations of the local temperature and of the background thermodynamic pressure, e.g. in closed -- or strictly periodic -- domains.
Pressure-based solvers using this approach have been developed and applied to multi-species evaporation~\cite{palmore2019volume,demou2022pressure,mialhe2023extended,salimi2025low,salimi2026evaporation}, and recently also to film boiling \cite{bourdon2025direct}.

When the phase change is directly driven by pressure variations -- as in cavitation and/or depressurization in closed environments -- acoustic effects can no longer be neglected and the fully compressible equations must be solved.
Work in this direction is ongoing: semi-implicit pressure-based solvers, which avoid the acoustic time-step restriction and recover the incompressible limit at low Mach number, have recently been extended from single-phase real-fluid flows~\cite{urbano2022semi} to liquid-vapor phase change~\cite{bibal2024compressible}.
These solvers rely on cubic EOSs, which describe the liquid, the vapor, and the evolving saturation conditions at the interface using a single thermodynamically consistent formulation.
An alternative route instead retains a density-based, Riemann-solver formulation: Wu \& Grenier~\cite{wu2026low} combine a Lagrange-Projection scheme with an LS/GFM treatment of the interface and a Mach-dependent correction that recovers a centered discretization in the low-Mach limit while reverting to the standard relaxation solver at higher Mach numbers.

\section{Modeling of small-scale physics}
\label{sec:sss}


Interface-resolved simulations offer the advantage of explicitly capturing the interactions of the liquid-vapor with the different flow scales without resorting to any subgrid models for turbulence mixing and scalar/energy transport, see e.g.\ the subgrid model for scalar transport at a bubble interface \cite{bothe2013volume}.
Despite the high fidelity offered by the framework, this remains a continuum-scale framework, still limited by the available computational resources.
As many phenomena occur at length scales well below the achievable grid resolution or beyond the continuum limit, they cannot be described from first principles.
Among these, three are particularly important: i) nucleation dynamics; ii) microlayer formation and contact line evaporation; iii) disjoining pressure and thin-film intermolecular forces. 

All these phenomena originate at nanometer to sub-micron scales that lie beyond the direct resolution of continuum simulations.
Their influence, however, extends far beyond their microscopic origin. 
These processes directly affect bubble growth rates, departure diameter, heat transfer partitioning between latent and sensible contributions, and the initiation of surface dryout. 
In the following sections, each of these phenomena is examined in detail, with emphasis on their physical origin and the modeling strategies currently employed to represent them in interface-resolved simulations.

\subsection{Nucleation dynamics}


Nucleation is the process by which a vapor embryo first emerges within a metastable liquid. The primary theoretical baseline for understanding this phenomenon is provided by the classical nucleation theory (CNT)~\cite{volmer1926keimbildung,becker1935kinetische,frenkel1939general,blander1975bubble}, which estimates the energy required to form a nucleus under homogeneous conditions from simple energetic arguments. 
Specifically, it balances the bulk free-energy gain associated with the new phase against the interfacial cost of forming the nucleus.
While elegant, CNT relies on simplifying assumptions that make a thorough validation against experiments difficult~\cite{karthika2016review,garbin2025bubbles}.
Two aspects, in particular, drive real systems away from CNT predictions: 
i) practical nucleation is heterogeneous, seeded at wall defects where gas entrapment lowers the energetic barrier~\cite{atchley1989crevice,giacomello2013geometry}; ii) the initial embryo is nanometric, a scale at which the macroscopic definition of a sharp surface tension breaks down.
Accounting for this curvature dependence, for instance through a Tolman-length correction, has been shown to improve CNT predictions~\cite{gallo2023nanoscale}.

A series of numerical tools is available to shed light on tne nucleation dynamics.
MD simulations represent the most fundamental route, but their cost confines them to nanoscopic domains and timescales far shorter than those of a boiling cycle~\cite{shahmardi2021effects}.
Between the atomistic limit and the continuum descriptions, coarse-grained and enhanced-sampling molecular models, together with density-functional and phase-field descriptions, have been developed with the specific goal of reaching larger scales than pure MD~\cite{zipoli2013improved,salvalaglio2016overcoming,lutsko2008density,marengo2021surface}.
Among these, fluctuating hydrodynamics has emerged as a promising approach for nucleation simulations, augmenting the continuum equations with a stochastic stress that accounts for thermal fluctuations~\cite{gallo2020nucleation,gallo2023nanoscale}.
While progress in understanding the origin of boiling is ongoing, a fully predictive description of nucleation under realistic conditions remains unavailable, owing to its phenomenological complexity and its non-equilibrium, multiscale nature.

\begin{figure}[!t]
\setlength{\unitlength}{0.0025\columnwidth}
\begin{picture}(400,80)
\put(0,-15){\includegraphics[width=1.00\columnwidth, keepaspectratio]{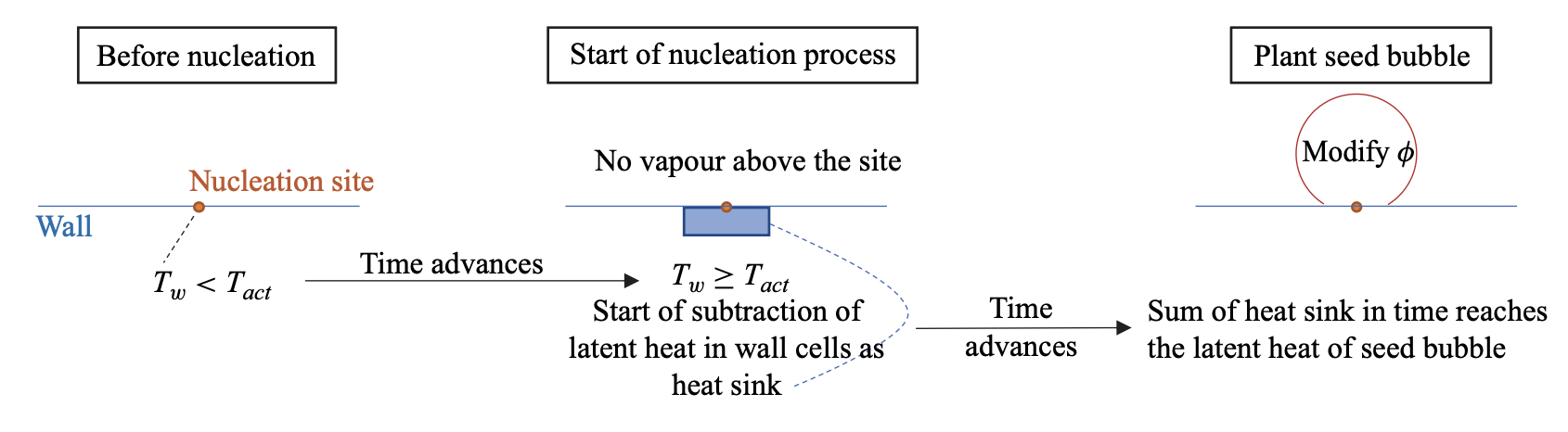}}
\end{picture}
\caption{Example of the algorithm used to prescribe the activation (or not) of a nucleation site, reproduced from Sato {\em et al.}~\cite{sato2026interface}.}
\label{fig:nuc} 
\end{figure}

Since nucleation originates at the atomistic scale, it is absent from the governing equations obtained at the continuum scale, and interface-resolved simulations must therefore rely on closures to prescribe it~\cite{kharangate2017review,chen2024review,garbin2025bubbles,magaletti2020unraveling}.
The main exception is that of pseudopotential LB methods, in which the nonideal EOS that sustains phase coexistence also governs phase change: vapor embryos emerge from the mechanical disequilibrium induced by the local superheat~\cite{li2020enhancement}.
The simplest approach is to directly seed the simulations with small bubbles, imposing vapor nuclei of prescribed size at selected locations.
This bypasses the nucleation problem entirely and is often used when the focus is on bubble growth, departure, or bubble-flow interactions.
A second class of approaches prescribes random nucleation embryos via a nucleation site density and an activation temperature, relating the activation of a site to its local superheat~\cite{sato2017nucleate,sato2018pool,qiu2022numerical,kaiser2024subcooled,sato2026interface}, as briefly sketched in Fig.~\ref{fig:nuc}.
The site density can be obtained either through empirical correlations~\cite{kocamustafaogullari1983interfacial,hibiki2003active} or by fitting experimental data~\cite{kossolapov2021experimental,sato2026interface}.
Within this second class, Li {\em et al.}~\cite{li2018development} proposed a method tailored for water as the working fluid, which was later employed for three-dimensional simulations of subcooled flow boiling~\cite{chen20223d}.
Similarly, Youssoufi {\em et al.}~\cite{youssoufi2025direct} adopted a nucleation model based on a Halton sequence and calibrated against experimental data~\cite{kim2002pool}.
Alternative approaches rely on coupling an atomistic region to a continuum solver to predict nucleation~\cite{gennari2024coupled}; more recently, data-driven methods have also demonstrated their potential to act as surrogates for MD simulations~\cite{noe2020machine}.

\subsection{Microlayer and contact-line evaporation}
\label{subsec:microlayer_contactline}

\begin{figure}[!t]
\setlength{\unitlength}{0.0025\columnwidth}
\begin{picture}(400,160)
\put(0,-15){\includegraphics[width=1.00\columnwidth, keepaspectratio]{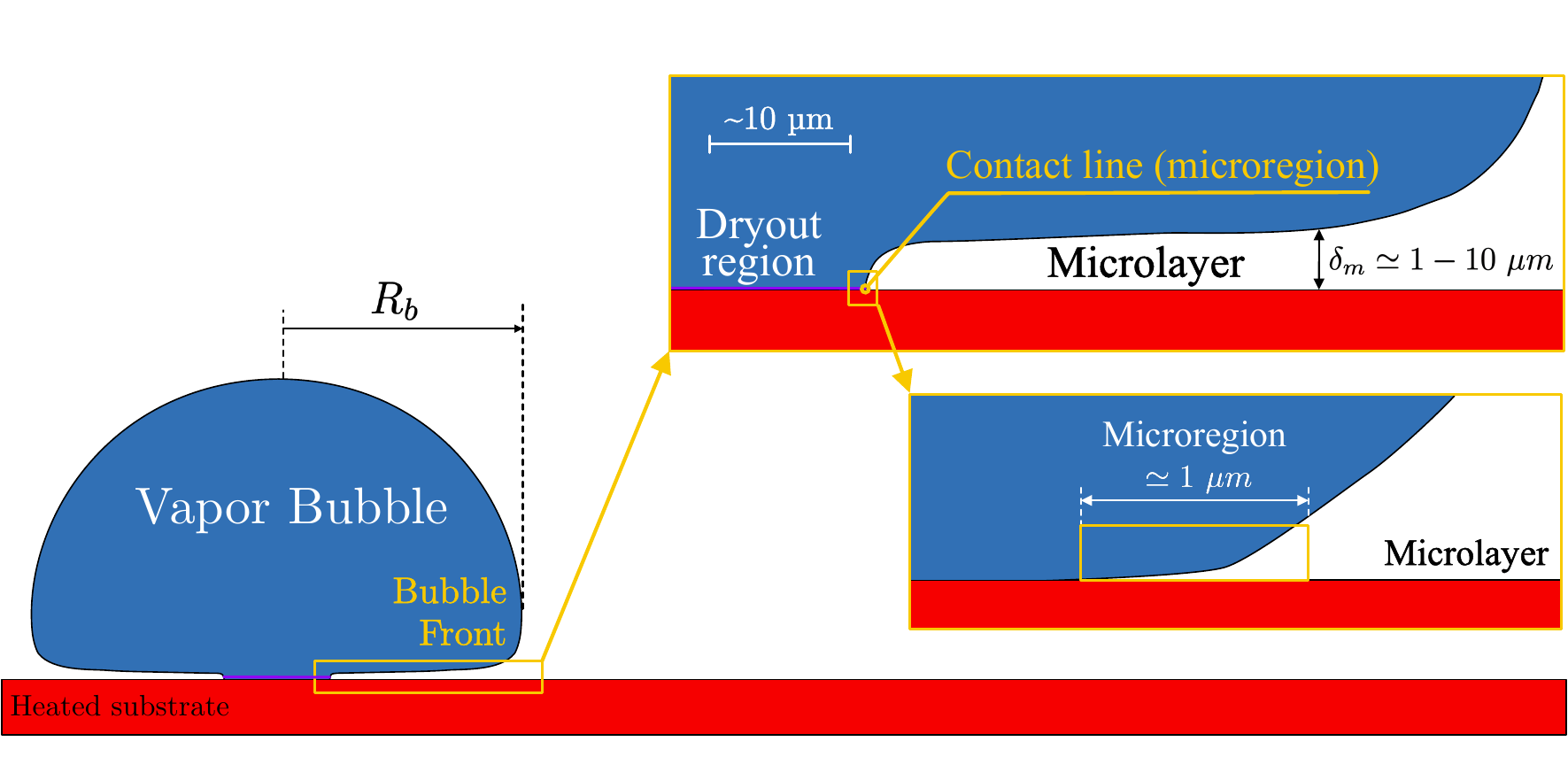}}
\put(-7,110){$(a)$}
\put(150,145){$(b)$}
\put(215,65){$(c)$}
\end{picture}
\caption{Panel~$a$ shows the growth of a vapor bubble of radius $R_b$ on a superheated substrate.
The right panels show successive close-ups of the region beneath the bubble front: panel~$b$ refers to the microlayer region extending outward from the dry spot while panel~$c$ zooms into the contact-line microregion.
We stress that the microlayer and the microregion are distinct: the microlayer is the deposited thin film behind the receding contact line, whereas the microregion is the contact-line closure region.
We also remark that the shape of the microlayer is regime dependent, with a ridge near the contact line and a bump-shaped microlayer both possible~\cite{guion2018simulations,tecchio2024jfm,zhang2024predicting,magnini2026evaporation}.}
\label{fig:microa} 
\end{figure}
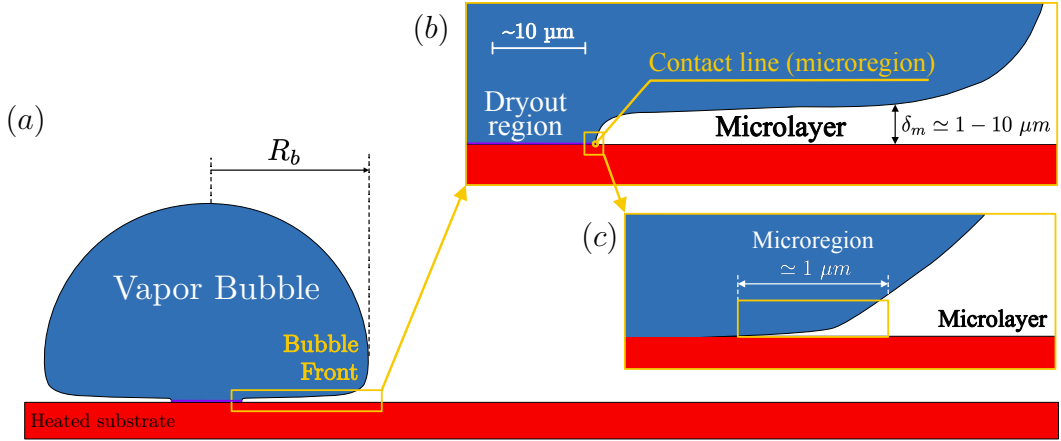

\begin{figure}[!t]
\setlength{\unitlength}{0.0025\columnwidth}
\begin{picture}(400,155)
\put(0,-15){\includegraphics[width=1.00\columnwidth, keepaspectratio]{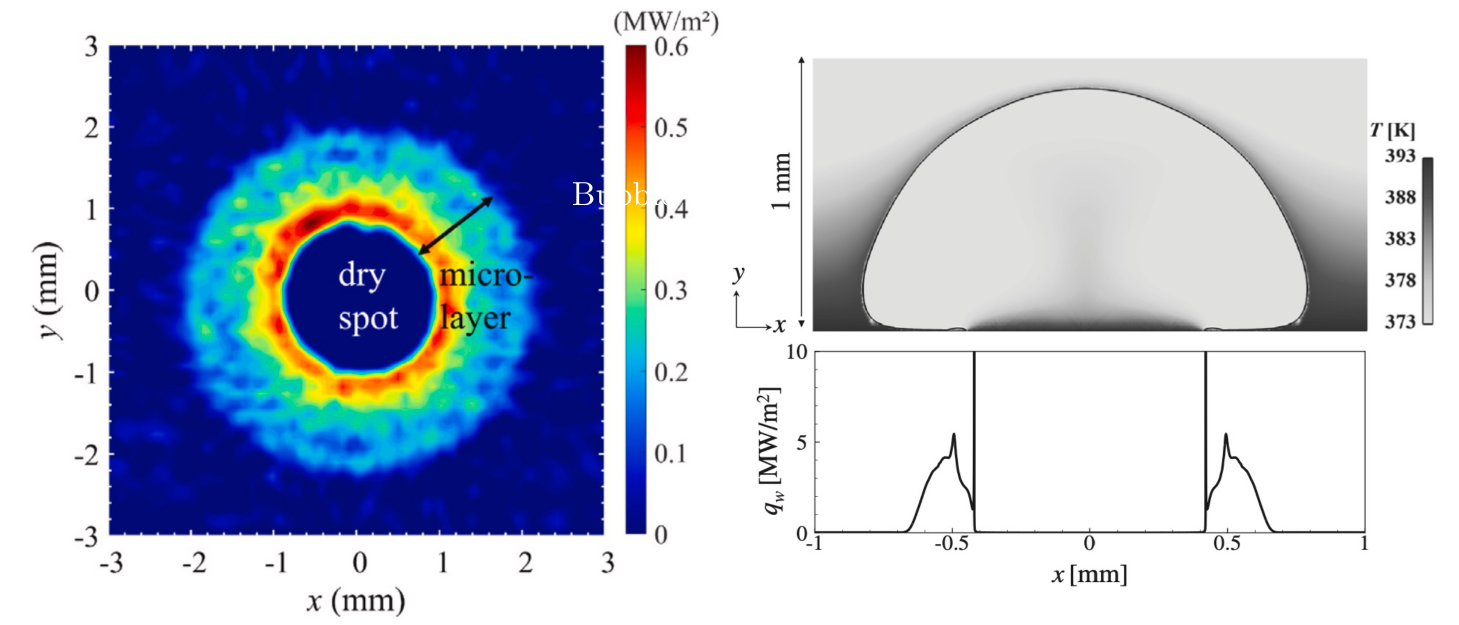}}
\put(-7,140){$(a)$}
\put(200,140){$(b)$}
\end{picture}
\caption{Panel~$a$ shows results obtained from infrared thermography beneath a growing bubble~\cite{tecchio2024ijhmt}.
Panel~$b$ shows results obtained from an interface-resolved DNS of a growing bubble in which the microlayer is fully resolved~\cite{urbano2018direct}.}
\label{fig:microb} 
\end{figure}

In the nucleate boiling regime, a substantial fraction of the heat transfer typically occurs in a small region at the base of a growing bubble~\cite{jung2015experimental,yabuki2016microscale,utaka2018measurement,magnini2026evaporation}.
Two mechanisms contribute to the heat transfer here: i) the microlayer, a liquid film a few microns thick that forms between the wall and the bubble interface and that extends laterally beneath much of the bubble foot (Fig.~\ref{fig:microa}$b$); ii) contact-line evaporation (or microregion), the intense but confined evaporation occurring in close proximity of the three-phase contact line (Fig.~\ref{fig:microa}$c$).
The microlayer contribution stems from its geometry: only a few microns thick yet extending up to several millimeters in the radial direction~\cite{cooper1969microlayer,moore1961measurement}, it combines a large surface area with a low conductive thermal resistance.
Heat is therefore readily transferred from the substrate to the liquid-vapor interface, making the microlayer an efficient pathway that sustains high wall heat fluxes.
By contrast, the contact-line region significantly impacts the global heat flux due to extreme confinement. As the film thins toward the contact line, its conductive resistance drops sharply; although the local flux is ultimately capped by the interfacial thermal resistance and disjoining pressure, its peak values are still orders of magnitude above the wall average~\cite{stephan1994new,wayner1976interline}.
The microlayer contribution is demonstrated by the experiments and numerical results in Fig.~\ref{fig:microb}: 
it spreads over a larger area and reaches its peak close to the contact-line region.
Notably, the contact-line contribution cannot be directly measured, as it lies below the available resolution.

Depending on the relative importance of these two mechanisms, two regimes can be identified~\cite{burevs2021modelling}: i) a microlayer regime, in which a film is entrained and, thanks to its large area and low thermal resistance, contributes to the overall heat flux alongside contact-line evaporation; ii) a contact-line regime, in which no microlayer forms and the evaporation is confined to the contact-line microregion~\cite{sodtke2006high,burevs2021modelling,graffiedi2025unraveling}.
The transition between these two regimes can be interpreted as a dewetting transition, with the microlayer forming only once the bubble expansion velocity exceeds a certain critical value~\cite{urbano2018direct,schweikert2019transition,burevs2021modelling}.
The relative contribution of the microlayer and contact-line mechanisms to the heat flux remains an open issue.
In the microlayer regime, reported contributions differ widely in the literature.
Some studies~\cite{utaka2014heat,chen2015heat,utaka2018measurement,baglietto2019boiling} report that the microlayer contribution varies with the wall superheat, ranging from 14\% up to 70\%, while others estimate it to be either roughly 50\%~\cite{yabuki2014heat} or 20\% to 25\%~\cite{kim2009review,myers2005time,jung2014experimental}.
In the contact-line regime, by contrast, evaporation at the triple line has been estimated to contribute between 5\% and 25\% of the total heat flux~\cite{graffiedi2025unraveling}.

Regardless of the relative importance, the accurate simulation and modeling of these heat transfer mechanisms is a key problem for numerical simulations.
The two regimes differ fundamentally in how they can be treated numerically as they occur at different scales, see Fig.~\ref{fig:microa}$b,c$.
The microlayer, with a typical thickness $\delta_m \simeq \mathcal{O}(\mathrm{\mu m})$, can be resolved from first principles because, from a fluid-dynamics perspective, it behaves as a thin liquid film~\cite{tecchio2024jfm}.
The contact-line regime, by contrast, originates below the continuum scale~\cite{fischer2015development,huber2017direct}, and has to be modeled via sub-grid closures~\cite{stephan1992analysis,lay1995shape}.

\subsubsection{Microlayer modeling}
A first natural question is whether a microlayer forms at all.
This is not straightforward, as it requires identifying the transition between the microlayer regime and the contact-line regime, in which no microlayer is present.
The precise onset of this transition remains an open question: although several criteria have been proposed~\cite{urbano2018direct,burevs2021modelling,schweikert2019transition}, no universal agreement has been achieved.
When a microlayer is present, resolving it from first principles is feasible but computationally expensive, since the microlayer thickness, $\mathcal{O}(\mu\mathrm{m})$, and the bubble radius, $\mathcal{O}(\mathrm{mm})$, must be captured within the same simulation. 
AMR alleviates this cost, as the finest resolution is required only beneath the bubble~\cite{long2025direct}; nevertheless, these studies remain confined to single-bubble configurations~\cite{urbano2018direct,guion2018simulations,burevs2022comprehensive,zhang2023direct,saini2024direct,giustini2024hydrodynamic}.
Hence, larger-scale or multi-bubble simulations must resort to models to account for the microlayer contribution.

These microlayer models usually rely on two ingredients: an initial shape for the microlayer and an evolution equation for the film thickness and its heat-transfer contribution.
For the initial microlayer thickness, a commonly used approach relies on the theoretical model proposed by Cooper \& Lloyd~\cite{cooper1969microlayer}:
\begin{equation}
    \delta_0 (R_b,t)= C_0 \sqrt{\nu t}\, ,
\end{equation}
where $\delta_0$ is the maximum initial microlayer thickness at time $t$ and radius $R_b$, $\nu$ is the kinematic viscosity of the working fluid and $C_0$ is a model coefficient\footnote{Reported values for the constant $C_0$ can be found in Tab.~2 of Sinha {\em et al.}~\cite{sinha2022microlayer}.}.
Once the bubble-front motion, $R_b(t)$, is known, the microlayer thickness can be expressed as a function of the distance from the nucleation site.
Another possible choice is to employ correlations obtained from experimental studies~\cite{koffman1982analysis,utaka2013microlayer,yabuki2014heat}, given in the form:
\begin{equation}
\delta_0=C_1 r^{\gamma}\, ,
\end{equation}
where $C_1$ is a constant and $r$ is the distance from the nucleation site\footnote{Reported values for $C_1$ and $\gamma$ can be found in Tab.~1 of Sinha {\em et al.}~\cite{sinha2022microlayer}.
Note that the constants are dimensional and are often tabulated in non-standard units.}.
More sophisticated estimates of the initial microlayer thickness are also available~\cite{olander1969analytical,smirnov1975calculation}.
A common feature of these expressions is that they do not retain an explicit dependence on surface tension.
In addition, they cannot reproduce the non-monotonic, bumped microlayer profiles reported in experiments~\cite{chen2020measurement,tecchio2024ijhmt} and simulations~\cite{magnini2026evaporation}, in which the thickness first increases with radial distance and then decreases.
In this context, recent works~\cite{tecchio2024jfm,zhang2024predicting,le2026structure,magnini2026evaporation} have shown that the initial microlayer thickness is well described by the Landau-Levich theory~\cite{landau1988dragging}, originally derived for adiabatic film deposition.
In this framework, the initial thickness is given by~\cite{zhang2024predicting}
\begin{equation}
\delta_0 \simeq R_m\, \mathrm{Ca}^{2/3}\, ,
\end{equation}
where $\mathrm{Ca}=\mu_l u_m/\sigma$ is the capillary number, expressing the ratio between viscous and surface tension forces, and $R_m$ is the radius of curvature of the outer meniscus.
Other models that incorporate surface-tension effects can be found in Jung \& Kim~\cite{jung2018hydrodynamic} and Sun {\em et al.}~\cite{sun2018transient}.

Once the initial microlayer thickness is defined, its time evolution can be obtained by assuming that the film changes size only through evaporation, hence neglecting transport -- though this can also be incorporated~\cite{sun2018transient,chen2023modeling} -- which gives~\cite{cai2024new}:
\begin{equation}
    \frac{d \delta }{d t} = - \frac{\dot{m}_{ml}}{\rho_l} =  - \frac{\dot{q}_{ml}}{\rho_l L_{vap}} \, ,
    \label{eq:mlevo}
\end{equation}
where $\dot{m}_{ml}$ is the evaporation rate of the microlayer, $\dot{q}_{ml}$ the corresponding heat flux and $\rho_l$ the liquid density.
Given the small microlayer thickness, 
$\dot{q}_{ml}$ is commonly estimated by a one-dimensional conduction model,  $\dot{q}_{ml} = \lambda_l (T_w - T_{lv})/\delta$, where the liquid-vapor interface temperature $T_{lv}$ is usually set equal to the saturation temperature~\cite{sato2015depletable,ling2021interface}.
When this model is combined with kinetic models, however, it leads to an overestimation of the microlayer evaporation rate~\cite{giustini2016evaporative,giustini2019microlayer}: assuming thermal equilibrium yields heat fluxes roughly an order of magnitude larger than the values observed experimentally~\cite{burevs2022comprehensive,kossolapov2021can,tecchio2024ijhmt}.
A possible workaround is to adopt a different value of the accommodation coefficient within the microlayer~\cite{giustini2019microlayer} or to introduce an interfacial heat-transfer resistance (IHTR)~\cite{burevs2022comprehensive,tecchio2024ijhmt,cai2024new,long2025direct,magnini2026evaporation}.
Placed in series with the conductive resistance of the film, the IHTR reduces the resulting evaporative microlayer heat flux, thereby improving the agreement with experimental data.
These two corrections are in fact closely related, since the IHTR is inversely proportional to the accommodation coefficient~\cite{burevs2022comprehensive}.
The physical origin of this discrepancy remains debated and may be attributed either to the accuracy of the Schrage model or to non-equilibrium effects present in the microlayer.
As a result, both the accommodation coefficient and the IHTR are typically tuned against experimental data~\cite{burevs2022comprehensive,long2025direct}.

As regards their numerical implementation, the microlayer models discussed above are generally coupled to the solver through a layer of wall-adjacent computational cells or nodes \cite{sato2015depletable,li2021numerical}.
Eq~\eqref{eq:mlevo} is used to compute the time evolution of the microlayer thickness and the corresponding heat flux, which acts in the energy equation as an additional source/sink term.
The local thickness allows for the distinction between regions where the microlayer has been completely depleted (dryout region) from those where it is still present.

\subsubsection{Contact-line modeling}
The heat-transfer contribution due to contact-line evaporation, also referred to as the micro-region or nano-region, is difficult to quantify experimentally because it occurs on a submicron region~\cite{sodtke2006high}, see Fig.~\ref{fig:microa}$c$.
It also cannot be simulated within continuum approaches as the relevant physics originates below the continuum scale.
As a result, there is no general consensus on how the micro-region should be incorporated into interface-resolved simulations and its modeling is an active area of research. 

Existing models can be broadly divided into two classes~\cite{torres2024coupling}, distinguished by the prescribed wetting condition and, as a consequence, by the presence or absence of an extremely thin adsorbed layer ($\delta_{a} \simeq 100~\text{\AA}$~\cite{chung2011review}) covering the wall: i) perfectly wetting micro-region models~\cite{wayner1976interline,stephan1994new}, in which the equilibrium contact angle is set to zero, an adsorbed film persists on the wall, and the apparent contact angle emerges from the disjoining-pressure/evaporation balance; ii) partially wetting micro-region models~\cite{janevcek2013apparent,mathieu2003etudes,nikolayev2010dynamics}, in which a finite contact angle is prescribed and the film terminates at a true contact line.
The theoretical basis for contact-line evaporation was established by Wayner and co workers~\cite{wayner1976interline}, who showed that the evaporative flux in the thin-film region is controlled by the balance between disjoining pressure and interfacial thermal resistance.
Building on this framework, Stephan \& Busse~\cite{stephan1992analysis} developed a micro-region model for grooved evaporator walls that was subsequently coupled to a VOF framework~\cite{fuchs2006transient,kunkelmann2009cfd,kunkelmann2010numerical}.
A closely related formulation, based on a set of fourth-order ordinary differential equations, was developed within the Dhir group~\cite{lay1995shape,dhir2005mechanistic} and applied to nucleate boiling heat transfer prediction.
Recent works have adopted more sophisticated models: Huber {\em et al.}~\cite{huber2017direct} and Torres {\em et al.}~\cite{torres2024coupling} both take the five-equation micro-region model of Mathieu~\cite{mathieu2003etudes} as their reference, the former work assessing its contribution and the latter coupling it to the DNS, whereas Wei {\em et al.}~\cite{wei2025effect} employ the model of Jane\v{c}ek \& Nikolayev~\cite{janevcek2013apparent}.

A key difficulty is the multiscale coupling of these models with interface-resolved DNS: the micro-region heat flux depends on the local wall temperature, which is itself part of the solution.
In addition, a strong scale separation is present between the contact-line microregion and the resolved flow scales.
Hence, dedicated coupling strategies are required~\cite{kunkelmann2009cfd,torres2024coupling,wei2025effect,kind2025development}.
An alternative strategy bypasses an explicit contact-line description and instead models the thin liquid film as a depletable sub-grid layer~\cite{sato2015depletable}, absorbing the near-wall evaporation and the contact-line singularity into a single resolved-scale closure.

\subsection{Disjoining pressure, thin liquid film and numerical coalescence}
\label{subsec:disjoining}
The third limitation of continuum simulations emerges when two interfaces are separated by a thin liquid film.
As the film thickness approaches molecular scales, the intermolecular forces governing the film stability and rupture can no longer be represented in the continuum equations, and the outcome of the interactions is determined by the numerical method rather than the physics.
In interface-resolved simulations this issue is referred to as numerical coalescence~\cite{Dodd2016,Soligo2019,soligo2021turbulent}: once two interfaces are separated by less than roughly one grid cell,  interface-capturing methods lead to coalescence.
Interface-tracking methods avoid implicit coalescence but still require a prescribed criterion to decide whether the two interfaces merge, e.g.\ a critical distance or contact time~\cite{tryggvason2013multiscale}.

Unlike the phenomena just discussed, this modeling  issue is shared with interface-resolved simulations of drop- and bubble-laden flows \cite{Dodd2016,soligo2021turbulent} and is not specific to boiling.
Numerical coalescence is commonly observed at high void fraction, where it can bias bubble-bubble and bubble-departure statistics, and is most pronounced near the CHF condition and in transition boiling, where frequent bubble-bubble interactions occur.
This has a direct consequence for boiling predictions.
Since the coalescence of neighboring bubbles into wall-blanketing vapor patches is one of the proposed causes of the boiling crisis, and since this process is what interface-capturing methods represent in a grid-sensitive manner, any CHF prediction resting on this mechanism inherits a grid dependence.

In addition to CHF, the effect of numerical coalescence must be analyzed also in other aspects: some studies deliberately introduce surfactants to modify surface tension and promote nucleation~\cite{hetsroni2004boiling,yin2020experimental}, thereby requiring modeling disjoining pressure explicitly.
A cost-effective solution is offered by coarse-grained approaches in which the unresolved film dynamics are accounted for via a repulsive force between approaching interfaces~\cite{Devita2019,montessori2019mesoscale,jin2024direct,liu2025phase}.

\section{Insights from interface-resolved boiling simulations}
\label{sec:insights}

\begin{figure}[!t]
\setlength{\unitlength}{0.0025\columnwidth}
\begin{picture}(400,170)
\put(0,-10){\includegraphics[width=1.00\columnwidth, keepaspectratio]{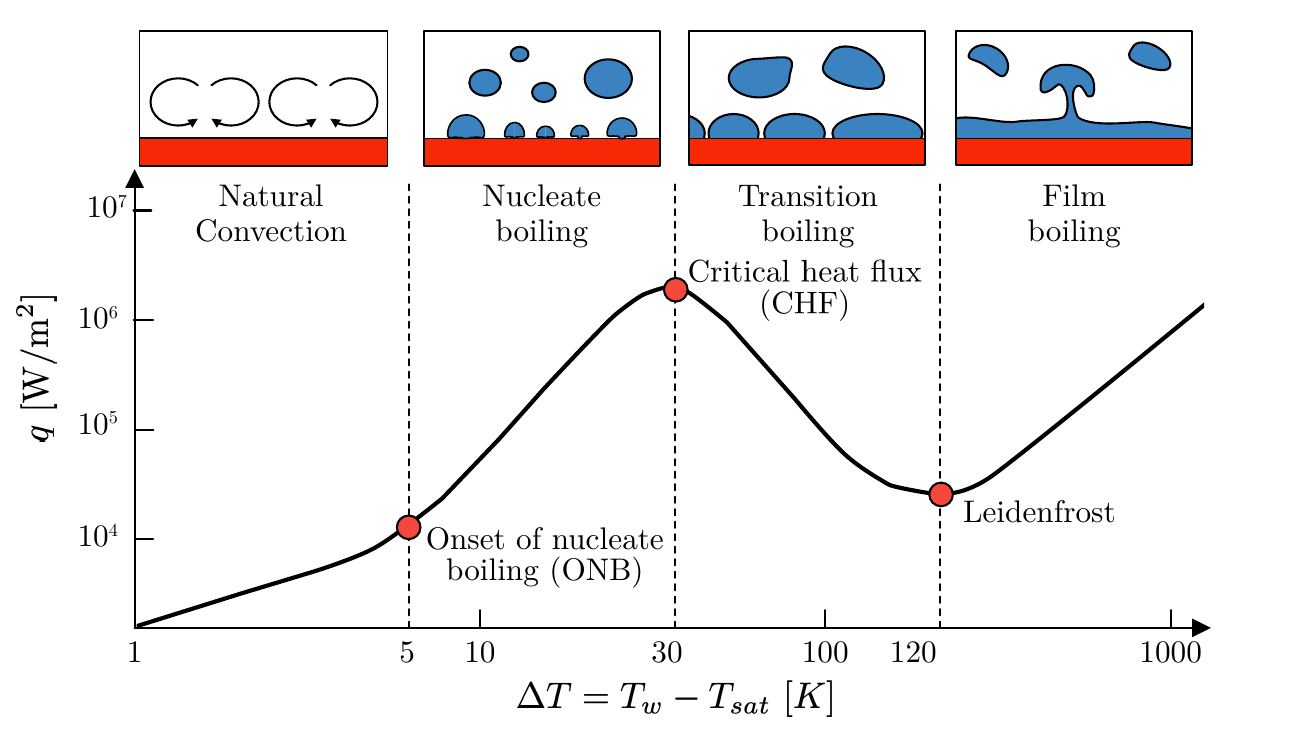}}
\end{picture}
\caption{Qualitative behavior of the boiling curve: wall-heat flux as a function of the wall superheat.
For each regime, the top panel shows a graphical summary of the system behavior.
Red circles are used to highlight the critical points: onset of nucleate boiling (ONB); critical heat flux (CHF) and Leidenfrost point.
The reported values are only indicative and do not refer to a specific configuration (e.g.\ pool/flow boiling).}
\label{fig:curve}
\end{figure}

In the previous sections, we described how the interface is represented, how the interfacial conditions are imposed, and how the unresolved small-scale physics can be modeled.
Here, we analyze what can be learned from interface-resolved simulations of boiling flows, organizing the discussion around the boiling curve, which provides a natural framework to link the simulation results to the boiling physics.
The boiling curve~\cite{nukiyama1966maximum}, sketched in Fig.~\ref{fig:curve}, relates the wall heat flux to the wall superheat $\Delta T = T_w - T_\mathrm{sat}$, and is qualitatively similar for pool and flow boiling, though clear differences exist~\cite{dhir1991nucleate}.

Below the onset of nucleate boiling (ONB) heat is removed by single-phase natural or forced convection, whereas above it the curve enters the nucleate boiling regime, where the heat flux increases monotonically as bubbles repeatedly grow, depart, and induce agitation in the near-wall liquid.
This is the regime of greatest interest, owing to its high heat-transfer rates.
As nucleation becomes more vigorous, neighboring bubbles coalesce into vapor patches that hinder the supply of liquid to the wall -- one of several proposed explanations for the boiling crisis~\cite{theofanous2002boilingp1,theofanous2002boilingp2} -- and the heat flux reaches the CHF condition that identifies the upper limit of safe nucleate-boiling operation~\cite{dhir1991nucleate,bongarala2024boiling}.
Beyond CHF lies the transition boiling regime, where an unstable, intermittent vapor film alternately forms and collapses on the substrate and the heat flux decreases as the superheat is increased, until -- past the Leidenfrost point -- the substrate becomes entirely covered by a stable vapor film.
In this film boiling regime the wall is thermally insulated by the vapor layer and the heat flux is low, rising only gradually as the wall temperature is increased substantially.

\begin{sidewaystable}
\centering
\small
\begin{tabular}{lllllll}
\hline
Reference & Config. & Fluid  & Cond. & Method & Models & Regime \\
\hline
Esmaeeli \& Tryggvason~\cite{esmaeeli2004computationsp2} & Pool & Water         & SAT      & FT     & --      & FB \\
Son \& Dhir~\cite{son2008three}                          & Pool & Water         & SAT      & LS     & --      & FB \\
Son \& Dhir~\cite{son2008numerical}                      & Pool & Water         & SAT      & LS     & NUC, ML & NB \\
Yazdani et al.~\cite{yazdani2016high}                    & Pool & Water, R134a  & SAT      & VOF    & NUC, ML & NB \\
Sato \& Ni\v{c}eno~\cite{sato2017nucleate}               & Pool & Water         & SAT      & ITM    & NUC, ML & NB \\
Sato \& Ni\v{c}eno~\cite{sato2018pool}                   & Pool & Water         & SAT      & ITM    & NUC, ML & NB, TB, FB \\
Dhruv et al.~\cite{dhruv2019formulation}                 & Pool & Water         & SUB      & LS     & NUC     & NB \\
Li et al.~\cite{li2021manipulating}                      & Pool & Water         & SAT      & VOF    & NUC     & NB \\
Dhruv et al.~\cite{dhruv2021investigation}               & Pool & FC-72         & SUB      & LS     & NUC     & NB \\
Ma \& Cheng~\cite{ma20193d}                              & Pool & --            & SAT      & LB     & --      & NB, TB, FB \\
Ma \& Cheng~\cite{ma2019dry}                             & Pool & --            & SAT      & LB     & --      & NB, TB, FB \\
Li et al.~\cite{li2020enhancement}                       & Pool & --            & SAT      & LB     & --      & NB \\
Fei et al.~\cite{fei2020mesoscopic}                      & Pool & Water         & SAT      & LB     & --      & NB, TB, FB \\
Chen et al.~\cite{chen20223d}                            & Flow & Water         & SUB      & VOSET  & NUC     & NB \\
Demou et al.~\cite{demou2022pressure}                    & Pool & Water         & SAT      & BN     & --      & NB \\
Qiu et al.~\cite{qiu2022numerical}                       & Pool & Water         & SAT      & VOF    & NUC, ML & NB \\
Guo et al.~\cite{guo2023numerical}                       & Flow & Water         & SAT      & VOF    & --      & NB \\
Chen et al.~\cite{chen2023modeling}                      & Flow & Water         & SAT      & VOSET  & NUC, ML & NB \\
Mudawar et al.~\cite{mudawar2023coupled}                 & Flow & nPFH          & SUB      & CLSVOF & --      & NB \\
Kaiser et al.~\cite{kaiser2024subcooled}                 & Flow & Water         & SUB      & ITM    & NUC, ML & NB \\
Luo \& Tagawa~\cite{luo2024enhanced}                     & Pool & --            & SAT      & LB     & --      & NB, TB, FB \\
Gu et al.~\cite{gu2024three}                             & Pool & --            & SAT      & VOF    & --      & FB \\
Wu et al.~\cite{wu2025numerical}                         & Flow & Water         & SUB      & VOF    & --      & NB \\
Wang et al.~\cite{wang2025mesoscopic}                    & Pool & --            & SAT      & LB     & --      & TB \\
Vachhani et al.~\cite{vachhani2025numerical}             & Pool & FC-72         & SUB      & LS     & NUC     & NB \\
Youssoufi et al.~\cite{youssoufi2025direct}              & Pool & FC-72         & SUB, SAT & LS     & NUC     & NB \\
Sato et al.~\cite{sato2026interface}                     & Flow & Water         & SUB      & ITM    & NUC     & NB \\
\hline
\end{tabular}
\caption{Representative three-dimensional interface-resolved simulations of pool and flow boiling (not exhaustive).
Abbreviations: SAT, saturated; SUB, subcooled; NUC, nucleation model; ML, microlayer model; NB, nucleate boiling; TB, transition boiling; FB, film boiling; --, not applicable or not reported.
The employed fluids are also reported when specified.}
\label{tab:landscape}
\end{sidewaystable}

First, we mention that the treatment of the heated substrate is a further modeling choice that bears on the interpretation of interface-resolved results across all regimes.
Three main alternatives are commonly adopted: i) imposing a fixed wall temperature; ii) imposing a fixed wall heat flux; iii) solving the conjugate heat-transfer problem in the solid.
The first two are numerically convenient but neglect the local, transient cooling of the wall beneath a growing bubble; a conjugate treatment instead captures the quench-and-rewet cycle and the resulting non-uniform surface temperature~\cite{aktinol2012numerical,giustini2017computational,kim2015enhanced}.
The effect of this choice depends on the solid-to-fluid effusivity ratio -- a measure of the material ability to exchange heat -- and the substrate thickness: a thin, low-effusivity heater deviates considerably from an isothermal wall, whereas a thick, high-effusivity substrate behaves as one. 
The resulting wall temperature also has a direct effect on nucleation when temperature-dependent models are used: the activation of a site depends on the local wall temperature, which cannot be prescribed as a boundary condition and must instead be obtained from the conjugate solution in the substrate~\cite{sato2017nucleate,sato2026interface}.

\begin{figure}[!t]
\setlength{\unitlength}{0.0025\columnwidth}
\begin{picture}(400,130)
\put(0,-10){\includegraphics[width=1.00\columnwidth, keepaspectratio]{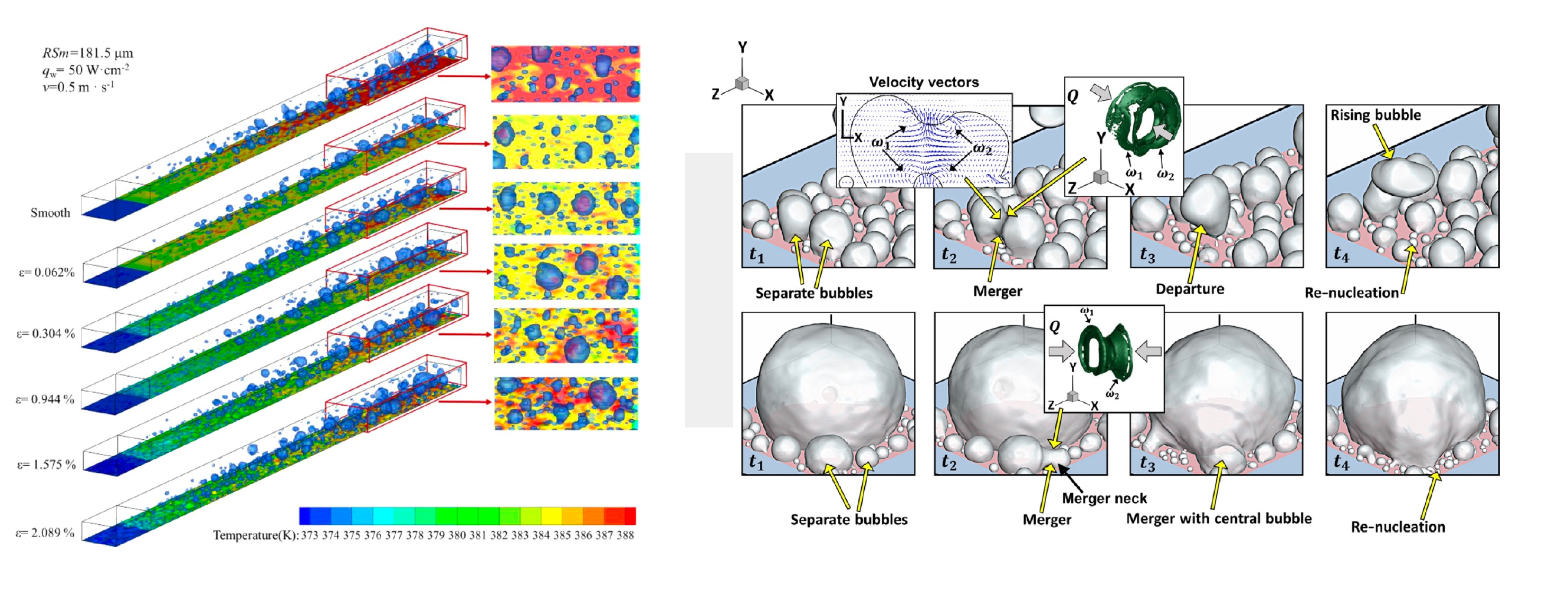}}
\put(-7,140){$(a)$}
\put(200,140){$(b)$}
\end{picture}
\caption{Rendering of three-dimensional simulation of boiling heat transfer performed in the nucleate boiling regime: flow boiling (panel~$a$) and pool boiling (panel~$b$).
Panel~$a$ shows the vapor distribution and bottom wall temperature contour in a micro-channel during boiling heat transfer for different relative roughness~\cite{guo2023numerical}.
Panel~$b$ shows the bubble dynamics in pool boiling simulations in Earth gravity conditions (top row) and micro-gravity conditions (bottom row)~\cite{dhruv2021investigation}.}
\label{fig:3d}
\end{figure}

An overview of three-dimensional interface-resolved simulations of boiling heat transfer is reported in Tab.~\ref{tab:landscape}, while Fig.~\ref{fig:3d} shows some renderings obtained from flow and pool boiling simulations~\cite{guo2023numerical,dhruv2021investigation}.
Across these contributions, interface-resolved simulations span nucleate, transition, and film boiling, but the coverage is uneven, with most studies concentrated in the nucleate and film boiling regimes.
Contributions are also scarce in the transition between the two -- from near-CHF conditions through the Leidenfrost point to the onset of film boiling -- with only a few studies covering all three regimes~\cite{sato2018pool,ma20193d,ma2019dry,fei2020mesoscopic,luo2024enhanced}.
This is not surprising, since the transition is the most challenging regime to capture in simulations, for several reasons.
First, the high heat flux observed in the region close to the CHF condition leads to a strong coupling between the thermal and flow field, which can cause stability problems~\cite{kaiser2024subcooled}.
Second, the physics becomes highly non-linear with the appearance of hysteresis phenomena~\cite{gabbana2025flow} and accurate multi-regime models for microlayer, contact-line and nucleation dynamics are needed~\cite{sato2018pool,kaiser2024subcooled}.
Together, these factors make the transition regime the least explored of the three.

Focusing on the nucleate boiling regime, a significant number of three-dimensional resolved simulations now access quantities that experiments cannot measure directly, including the spatial distribution of wall heat flux, the partitioning between liquid and vapor pathways, and the liquid-film thickness beneath growing and sliding bubbles~\cite{sato2026interface}.
However, state-of-the-art nucleate-boiling simulations remain limited to single realizations, already requiring order $10^6$ core-hours, so that conclusions inevitably rest on a handful of conditions rather than parametric sweeps~\cite{sato2026interface,chen20223d}. 
Recent simulations~\cite{yazdani2016high,dhruv2019formulation,dhruv2021investigation,sato2026interface} show that both the computed bubble shapes and the resulting heat-transfer coefficients closely match the experimental observations.
Numerical data have also been compared against correlations providing the basis for a physical explanation of the discrepancies -- developing thermal boundary layers and bubble-induced agitation rather than fully developed single-phase convection~\cite{sato2026interface} -- a mechanism consistent with experimental observations~\cite{aguiar2024infrared}.
This marks a shift from qualitative agreement in bubble shape towards quantitative validations of the governing mechanisms.
The time- and space-resolved data provided by simulations pave therefore the way for unveiling the role of hydrodynamics in triggering the boiling crisis.
These simulations further reveal how the growth, detachment, and motion of bubbles inject velocity fluctuations into the surrounding liquid that substantially exceed those of the equivalent single-phase flow, enhancing the wall-normal transport of subcooled liquid towards the heated substrate~\cite{sato2026interface}.

Moving to film boiling, the physics of this regime allows for some simplifications.
In particular, since a stable vapor film blankets the wall, the liquid never wets the substrate and the microregion, contact-line, and microlayer models that are needed in the nucleate regime are no longer required.
Nucleation is likewise absent: the heat transfer proceeds by conduction and radiation across the film and by the growth and release of vapor at the interface, removing the need for the empirical site-density and nucleation-criterion closures discussed in Section~\ref{sec:sss}.
Conjugate heat transfer also plays a less important role, since the near-wall thermal resistance is set by the film rather than by the transient response of the solid substrate.
The dynamics are therefore primarily hydrodynamic in nature: the vapor--liquid interface is unstable to Rayleigh--Taylor modes, and the characteristic mushroom-shaped columns and the bubble-release frequency they set are governed by the most-amplified wavelength.
This makes film boiling a well-posed benchmark in both two and three dimensions, with established correlations~\cite{berenson1961film,klimenko1981film,klimenko1982film} available for quantitative validation of the predicted wall heat flux.
These features make film boiling configurations far cheaper to simulate than nucleate boiling: with no active nucleation-site model required and no need to resolve a contact line, statistically converged results can be readily found~\cite{welch2000volume,esmaeeli2004computationsp2,ahammad2016application,kumar2022coupled,rouzbahani2023numerical,gu2024three}.
This is why a consistent portion of simulations have been performed in this regime, from seminal two-dimensional studies~\cite{welch2000volume,esmaeeli2004computationsp2} to more recent works addressing film boiling on complex geometries and moving objects~\cite{singh2021two,bourdon2025direct,kumar2024flow}. 

\sloppy
Returning to the nucleate regime, where sub-grid models cannot be avoided, a final question concerns how strongly the predicted quantities depend on the particular closures adopted.
Sato \& Ni\v{c}eno~\cite{sato2018pool} varied the microlayer slope constant over roughly a factor of two, shifting the predicted bubble departure diameter from about $4$ to $3$~mm; all the resulting values, however, fall within the scatter of the reference experimental data.
Likewise, Bure\v{s} \& Sato~\cite{burevs2022comprehensive} report a comparable sensitivity of the bubble growth rate to the IHTR, and an even stronger sensitivity to the portion of the interface over which it is applied: imposing it over the whole interface under-predicts the growth, whereas omitting it entirely overpredicts it. These authors found the thickness of the microlayer profiles to also be extremely sensitive to the assumed wetting conditions.
Similarly, the wall vapor coverage predicted by the subcooled flow boiling simulations of Sato {\em et al.}~\cite{sato2026interface} was found to depend on the prescribed contact angle.
Similar considerations can be made with respect to grid convergence: the integral wall heat flux converges under refinement in~\cite{sato2018pool}, whereas the subcooled flow boiling simulations of Kaiser {\em et al.}~\cite{kaiser2024subcooled} show a spread of about $6$~K in the predicted wall superheat between the two finest grids.
A further difficulty is that, because of the computational cost, convergence is often tested on a configuration simpler than the one of interest.

Taken together, these results provide empirical evidence of the modeling limitations discussed in Section~\ref{sec:sss}: quantitative predictions of boiling vary measurably with the calibrated sub-grid closures, the available experimental data are often unable to discriminate among the resulting values, and the sensitivity to these closures is neither consistently characterized nor systematically explored across studies. 
The accuracy of the hydrodynamic solver alone therefore does not determine the accuracy of the prediction.


\section{Future perspectives and open challenges}
\label{sec:final}

Interface-resolved simulations have progressed remarkably over nearly three decades, driven by advances in algorithms for interface representation and by the growing availability of computing resources.
They represent a powerful tool for understanding the fundamental physics of boiling heat transfer.
Their limitations, however, should be kept in mind: many of the phenomena that control the macroscopic behavior (nucleation rates, microlayer thickness, contact angle, thin-film intermolecular forces) originate below the smallest resolved scale, and some below the continuum description.
Interface-resolved simulations of boiling are therefore hydrodynamically resolved but microscopically modeled, and the quality of their predictions depends directly on the closures supplied for the unresolved small-scale physics.
The open challenges and future directions outlined below stem from this observation.

One future direction concerns choosing the resolved range of scales: simulations of boiling demand grid resolutions of the order of $10^7$--$10^8$ points even for laboratory-scale configurations~\cite{vachhani2025numerical,youssoufi2025direct,sato2026interface}, and remain out of reach in most cases of practical interest.
The range of scales that must be represented simultaneously, from the thermal boundary layer at the wall down to the microlayer and the contact-line region, spans several decades, and current simulations resolve only its upper part.
In this context, AMR is advantageous whenever the finest features are confined to a limited portion of the domain, as in single-bubble configurations~\cite{long2025direct,dhruv2019formulation} or when the resolution is increased locally around the microlayer. %
Its extension to many-bubble configurations is however less straightforward, since every bubble carries its own refined region and the fraction of the domain held at the finest level grows accordingly, limiting the benefit of the approach.
No single remedy therefore enables large-scale simulations of boiling heat transfer at substantially reduced cost, and studies at conditions of practical relevance remain single realizations obtained at considerable computational cost~\cite{vachhani2025numerical,sato2026interface}.

A second direction concerns the accuracy of the closures themselves.
Here, MD can offer physical insight into nucleation, wettability, and interfacial phenomena at the atomistic level~\cite{shahmardi2021effects,lavino2021surface,wu2020molecular,lin2024recent}, while FH has emerged as a particularly promising framework for capturing the spontaneous nucleation of vapor bubbles in metastable liquids~\cite{gallo2018thermally,gallo2020nucleation,magaletti2020unraveling,gallo2025complex}.
Both can be further exploited to improve the closures supplied to interface-resolved simulations, either directly or as training data for data-driven formulations~\cite{koumoutsakos2024roads}.

Another challenge concerns the constitutive assumptions made on the vapor and liquid phases.
Most of the existing work has been carried out under conditions where the incompressible assumption is well justified, whereas high-pressure, near-critical conditions, in which compressibility effects can no longer be neglected, remain only marginally explored by interface-resolved simulations.
Many energy-transition applications, moreover, operate with non-ideal mixtures in regimes where vapor compressibility and real-gas effects can be important.
Under these conditions, incompressible single-fluid formulations break down, and low-Mach formulations, BN-derived four-equation models, and fully compressible diffuse-interface solvers~\cite{demou2022pressure,le2014towards,saurel2016general,pelanti2022arbitrary} constitute the natural framework.
Among these, the four-equation model deserves particular attention: it is thermodynamically consistent, well suited to modern GPU-based implementations, and accommodates compressibility at moderate computational cost, yet its use in the boiling literature remains limited compared to incompressible formulations.
High-pressure conditions are also more demanding computationally: as the pressure increases, both the surface tension and the vapor-to-liquid density ratio decrease, so that bubbles depart at smaller diameters and a finer grid is required~\cite{Kossolapov2024}.

Finally, validation, benchmarking, and direct comparison against experiments represent an important collaborative direction.
Considerable effort has already been made in this direction -- for instance the DEBORA database~\cite{garnier2001local,bois2024benchmark} and the experiments performed on the International Space Station~\cite{oikonomidou2022bubble,sielaff2022multiscale} -- but wider open access to codes and datasets is still needed.
Recent advances in experimental diagnostics offer a timely opportunity in this respect, as they resolve the same micrometer-scale features that interface-resolved simulations aim to compute directly: microlayer thickness, contact-line motion, and individual bubble thermal footprints~\cite{richenderfer2018investigation,voulgaropoulos2022simultaneous}.
This paves the way for a closer synergy between simulations and experiments, allowing new hypotheses to be tested and validated.

\section*{Acknowledgments}
AR gratefully acknowledges financial support from the MIT International Science and Technology Initiatives Global Seed Funds - Friuli Venezia Giulia (MIT-FVG) program whereas LB was funded by the 

The authors thank Elias Balaras, Matteo Bucci, Marica Pelanti, Aritra Mujherjee, Annafederica Urbano and Lorenz Weber for the discussions and comments.
A repository collecting tools for the development and testing of interface-resolved simulation methods is available at \url{https://github.com/aroccon/boiling_codes}.

\appendix

\section{Six- and five-equation BN models}
\label{app:56bn}

For the sake of completeness, we report in the following the full system of equations for the six and five equations model.
The six-equation model is obtained from the seven-equation model by considering the asymptotic limit of stiff velocity relaxation ($\lambda \rightarrow \infty $).
The resulting governing equations read as follows \cite{perigaud2005compressible,saurel2009simple,pelanti2014mixture}:
\begin{equation}
\frac{\partial \alpha_1}{\partial t} + {\bf u} \cdot \nabla \alpha_1 =   \mathcal{P},
\end{equation}
\begin{equation}
\frac{\partial (\alpha_1 \rho_1)}{\partial t} +  \nabla \cdot (\alpha_1 \rho_1 {\bf u}) =  \mathcal{M}\, , 
\end{equation}
\begin{equation}
\frac{\partial (\alpha_2 \rho_2)}{\partial t} +  \nabla \cdot (\alpha_2 \rho_2 {\bf u}) = - \mathcal{M}\, , 
\end{equation}
\begin{equation}
\frac{\partial (\rho {\bf u})}{\partial t} +  \nabla \cdot (\rho  {\bf u} \otimes {\bf u}) + \nabla ( \alpha_1 p_1 + \alpha_2 p_2)  =  0  ,
\end{equation}
\begin{equation}
\frac{\partial (\alpha_1 E_1)}{\partial t} + \nabla \cdot [\alpha_1  ( E_1 + p_1){\bf u}] =  - p_I \mathcal{P}  +  \mathcal{Q}  + \left(g_{I} + \frac{|{\bf u}|^2}{2} \right) \mathcal{M}, 
\end{equation}
\begin{equation}
\frac{\partial (\alpha_2 E_2)}{\partial t} + \nabla \cdot [\alpha_2  ( E_2 + p_2){\bf u}] =     p_I \mathcal{P} - \mathcal{Q} - \left(g_{I} + \frac{|{\bf u}|^2}{2} \right) \mathcal{M}, 
\end{equation}
where it can be observed that thanks to the velocity relaxation the interface velocity vanishes.

A further reduction can be obtained by assuming full mechanical equilibrium (velocity and pressure) while allowing thermal and chemical non-equilibrium. 
This corresponds to the expansion of the system in the limit of stiff mechanical relaxation ($\mu \rightarrow \infty$ and $\lambda \rightarrow \infty $), when the system of equations reduces to the model proposed by Kapila \cite{kapila2001two}. 
Please note that alternative five-equation model formulations can also be derived \cite{massoni2002some,allaire2002five}.
Following the derivation of Kapila \cite{kapila2001two}, the resulting governing equations, accounting also for heat and mass transfer, become \cite{pelanti2022arbitrary}:
\begin{equation}
\frac{\partial \alpha_1}{\partial t} + {\bf u} \cdot \nabla \alpha_1 =  K \nabla \cdot {\bf u}  + K_h \mathcal{Q} + K_m \mathcal{M}\, ,
\end{equation}
\begin{equation}
\frac{\partial (\alpha_1 \rho_1)}{\partial t} +  \nabla \cdot (\alpha_1 \rho_1 {\bf u}) =  \mathcal{M}\, , 
\end{equation}
\begin{equation}
\frac{\partial (\alpha_2 \rho_2)}{\partial t} +  \nabla \cdot (\alpha_2 \rho_2 {\bf u}) = - \mathcal{M}\, , 
\end{equation}
\begin{equation}
\frac{\partial (\rho {\bf u})}{\partial t} +  \nabla \cdot ( \rho  {\bf u} \otimes {\bf u}) + \nabla p  =  0\, ,
\end{equation}
\begin{equation}
\frac{\partial (\rho E)}{\partial t} + \nabla \cdot \left[ (\rho E + p){\bf u} \right] = 0\, ,
\end{equation}
where the coefficients $K, K_h$ and $K_m$ that appear in the volume fraction transport equations represent the expansion and compression coefficient in the mixture regions, the interface temperature coefficient and the interface density coefficient~\cite{saurel2008modelling,pelanti2022arbitrary}.
The first coefficient is defined as follows:
\begin{equation}
K= \frac{\alpha_1 \alpha_2 (\rho_2 c_2^2 - \rho_1 c_1^2)}{\alpha_2 \rho_1 c_1^2 + \alpha_1 \rho_2 c_2^2}\, ,
\end{equation}
where $\alpha_1$ and $\alpha_2=1-\alpha_1$ are the two volume fractions, $\rho_1$ and $\rho_2$ the two densities and $c_1$ and $c_2$ the speeds of sounds in the two phases. 
The interface temperature coefficient is defined as:
\begin{equation}
K_h = \frac{\alpha_1 \alpha_2}{\alpha_2 \rho_1 c_1^2 + \alpha_1 \rho_2 c_2^2} \left(\frac{\Gamma_1}{\alpha_1} + \frac{\Gamma_2}{\alpha_2}\right) \, ,
\end{equation}
being $\Gamma_1$ and $\Gamma_2$ the Gruneisen coefficients of the two phases.
Finally, the interface density coefficient is defined as:
\begin{equation}
K_m =\left( \frac{c_1^2}{\alpha_1} + \frac{c_2^2}{\alpha_2}\right) \left( \frac{\rho_1 c_1^2}{\alpha_1} +  \frac{\rho_2 c_2^2}{\alpha_2} \right)^{-1} \, .
\end{equation}

\bibliographystyle{elsarticle-num} 
\bibliography{totalbib.bib}

@article{kang2000boundary,
  title={A boundary condition capturing method for multiphase incompressible flow},
  author={Kang, M. and Fedkiw, R. P. and Liu, X.-D.},
  journal={J. Sci. Comput.},
  volume={15},
  number={3},
  pages={323--360},
  year={2000},
}

@article{Osher1988,
	Author = {Osher, S. and Sethian, J.},
	Journal = {J. Comput. Phys.},
	Pages = {12--49},
	Title = {Fronts Propagating with Curvature-Dependent Speed: Algorithms Based on Hamilton-Jacobi Formulations},
	Volume = {79},
	Year = {1988}
}

@article{Sussman2007,
	Author = {Sussman, M. and Smith, K.M. and Hussaini, M.Y. and Ohta, M.},
	Journal = {J. Comput. Phys.},
	Pages = {469--505},
	Title = {{A sharp interface method for incompressible two-phase flows}},
	Volume = {221},
	Year = {2007}
}

@article{Tryggvason2001,
Author = {Tryggvason, G. and Bunner, B. and Esmaeeli, A. and Juric, D. and Tauber, W. and Han, J. and Nas, S. and Jan, Y.},
Pages = {708--759},
Journal = {J. Comput. Phys.},
Title = {A Front-Tracking Method for the Computations of Multiphase Flow},
Volume = {169},
Year = {2001}
}

@article{Unverdi1992,
	Author = {Unverdi, S. and Tryggvason, G.},
	Journal = {J. Comput. Phys.},
	Title = {A Front-Tracking Method for Viscous , Incompressible , Multi-fluid Flows},
	Year = {1992},
	volume ={100},
	number ={1}
}

@article{Cahn1958,
Author = {Cahn, J.W. and Hilliard, J.E.},
title = {Free Energy of a Nonuniform System. {I}. Interfacial Free Energy},
Journal = {J. Chem. Phys.},
year = {1958},
volume = {28},
pages = {258-267},
}

@article{Cahn1959a,
Author = {Cahn, J.W.},
Journal = {J. Chem. Phys.},
Number = {5},
Pages = {1121-1124},
Title = {Free Energy of a Nonuniform System. {II}. Thermodynamic Basis},
Volume = {30},
Year = {1959}
}

@article{Cahn1959b,
Author = {Cahn, J.W. and Hilliard, J.E.},
Journal = {J. Chem. Phys.},
Pages = {688},
Title = {Free Energy of a Nonuniform System. {III}. Nucleation in a Two-Component Incompressible Fluid},
Volume = {31},
Year = {1959}
}

@article{Elghobashi2018,
author = {Elghobashi, S.},
title = {Direct numerical simulation of turbulent flows laden with Droplets or Bubbles},
Journal = {Annu. Rev. Fluid Mech.},
number = {1},
volume={51},
pages={217--244},
year={2019},
}

@article{mirjalili2020conservative,
  title={A conservative diffuse interface method for two-phase flows with provable boundedness properties},
  author={Mirjalili, Shahab and Ivey, Christopher B and Mani, Ali},
  Journal = {J. Comput. Phys.},
  volume={401},
  pages={109006},
  year={2020},
}

@article{olsson2007conservative,
  title={A conservative level set method for two phase flow II},
  author={Olsson, E. and Kreiss, G. and Zahedi, S.},
  Journal = {J. Comput. Phys.},
  volume={225},
  number={1},
  pages={785--807},
  year={2007},
}

@article{geier2015conservative,
  title={Conservative phase-field lattice {Boltzmann} model for interface tracking equation},
  author={Geier, M. and Fakhari, A. and Lee, T.},
  journal={Phys. Rev. E},
  volume={91},
  number={6},
  pages={063309},
  year={2015},
  publisher={APS}
}

@article{zheng2005lattice,
  title={Lattice {Boltzmann} interface capturing method for incompressible flows},
  author={Zheng, H.W. and Shu, C and Chew, YT},
  journal={Phys. Rev. E},
  volume={72},
  number={5},
  pages={056705},
  year={2005},
  publisher={APS}
}

@article{zheng2006lattice,
  title={A lattice Boltzmann model for multiphase flows with large density ratio},
  author={Zheng, HW and Shu, Chang and Chew, Yong-Tian},
  journal = {J. Comput. Phys.},
  volume={218},
  number={1},
  pages={353--371},
  year={2006},
  publisher={Elsevier}
}

@article{ren2016improved,
  title={Improved lattice {Boltzmann} modeling of binary flow based on the conservative {Allen-Cahn} equation},
  author={Ren, Feng and Song, Baowei and Sukop, Michael C and Hu, Haibao},
  journal={Phys. Rev. E},
  volume={94},
  number={2},
  pages={023311},
  year={2016},
  publisher={APS}
}

@article{wang2016comparative,
  title={Comparative study of the lattice {Boltzmann} models for {Allen-Cahn} and {Cahn-Hilliard} equations},
  author={Wang, H.L. and Chai, Z.H. and Shi, B.C. and Liang, H},
  journal={Phys. Rev. E},
  volume={94},
  number={3},
  pages={033304},
  year={2016},
  publisher={APS}
}

@article{wang2021phase,
  title={A phase-field method for boiling heat transfer},
  author={Wang, Z. and Zheng, X. and Chryssostomidis, C. and Karniadakis, G. E.},
  journal={J. Comput. Phys.},
  volume={435},
  pages={110239},
  year={2021},
  publisher={Elsevier}
}

@article{badillo2012quantitative,
  title={Quantitative phase-field modeling for boiling phenomena},
  author={Badillo, Arnoldo},
  journal={Phys. Rev. E},
  volume={86},
  number={4},
  pages={041603},
  year={2012},
  publisher={APS}
}

@article{irfan2017front,
  title={A front tracking method for direct numerical simulation of evaporation process in a multiphase system},
  author={Irfan, M. and Muradoglu, M.},
  journal={J. Comput. Phys.},
  volume={337},
  pages={132--153},
  year={2017},
  publisher={Elsevier}
}

@article{juric1998computations,
  title={Computations of boiling flows},
  author={Juric, Damir and Tryggvason, Gr{\'e}tar},
journal = {Int. J. Multiph. Flow},
  volume={24},
  number={3},
  pages={387--410},
  year={1998},
  publisher={Elsevier}
}

@article{verdier2020performance,
  title={Performance portability of lattice Boltzmann methods for two-phase flows with phase change},
  author={Verdier, Werner and Kestener, Pierre and Cartalade, Alain},
  journal={Comput. Methods Appl. Mech. Eng.},
  volume={370},
  pages={113266},
  year={2020},
  publisher={Elsevier}
}

@article{brackbill1992continuum,
  title={A continuum method for modeling surface tension},
  author={Brackbill, Jeremiah U and Kothe, Douglas B and Zemach, Charles},
  journal={J. Comput. Phys.},
  volume={100},
  number={2},
  pages={335--354},
  year={1992},
  publisher={Elsevier}
}

@article{sussman1994level,
  title={A level set approach for computing solutions to incompressible two-phase flow},
  author={Sussman, Mark and Smereka, Peter and Osher, Stanley},
  journal={J. Comput. Phys.},
  volume={114},
  number={1},
  pages={146--159},
  year={1994},
  publisher={Elsevier}
}

@article{lee1980pressure,
  title={A pressure iteration scheme for two-phase flow modeling},
  author={Lee, Wen Ho},
  journal={Multiphase transport fundamentals, reactor safety, applications},
  volume={1},
  pages={407--431},
  year={1980},
  publisher={Washington, DC}
}

@article{Anderson1998,
	Author = {Anderson, D.M. and McFadden, G.B. and Wheeler, A.A.},
	Journal = {Annu. Rev. Fluid Mech.},
	Number = {1},
	Pages = {139-165},
	Title = {Diffuse interface methods in fluid mechanics},
	Volume = {30},
	Year = {1998}}

@article{Badalassi2003,
	Author = {Badalassi, V.E. and Ceniceros, H.D. and Banerjee, S.},
	Journal = {J. Comput. Phys.},
	Number = {2},
	Pages = {371--397},
	Title = {{Computation of multiphase systems with phase field models}},
	Volume = {190},
	Year = {2003}}

@article{Dodd2016,
	Author = {Dodd, M.S. and Ferrante, A.},
	Journal = {J. Fluid Mech.},
	Pages = {356--412},
	Title = {{On the interaction of Taylor length scale size droplets and isotropic turbulence}},
	Volume = {806},
	Year = {2016}}

@article{Olsson2005,
	Author = {Olsson, E. and Kreiss, G.},
	Journal = {J. Comput. Phys.},
	Number = {1},
	Pages = {225--246},
	Title = {A Conservative Level Set Method for Two Phase Flow},
	Volume = {210},
	Year = {2005}}

@article{DING2007,
	Author = {H. Ding and P.D.M. Spelt and C. Shu},
	Journal = {J. Comput. Phys.},
	Number = {2},
	Pages = {2078--2095},
	Title = {Diffuse interface model for incompressible two-phase flows with large density ratios},
	Volume = {226},
	Year = {2007}}

@book{tryggvason2011direct,
	Author = {G. Tryggvason and R. Scardovelli and S. Zaleski},
	Title = {Direct Numerical Simulations of Gas-Liquid Multiphase Flows},
	Publisher = {Cambridge university press},
	Year = {2011}
}

@article{Desjardins2008a,
	Author = {Desjardins, O. and Moureau, V. and Pitsch, H.},
	Journal = {J. Comput. Phys.},
	Pages = {8395--8416},
	Title = {{An accurate conservative level set/ghost fluid method for simulating turbulent atomization}},
	Volume = {227},
	Year = {2008}}

@article{Jacqmin1999,
	Author = {D. Jacqmin},
	Journal = {J. Comput. Phys.},
	Number = {1},
	Pages = {96--127},
	Title = {{Calculation of two-phase Navier--Stokes flows using phase-field modeling}},
	Volume = {155},
	Year = {1999}
}

@book{cmmf2009,
author = {Prosperetti, A. and Tryggvason, G.},
publisher={Cambridge University Press},
title = {Computational Methods for Multiphase Flow},
Year = {2009}
}

@article{popinet2018,
title={Numerical Models of Surface Tension},
author={Popinet, St{\'e}phane},
Journal = {Annu. Rev. Fluid Mech.},
volume={50},
pages={1--28},
year={2018}
}

@article{scriven1960marangoni,
  title={The marangoni effects},
  author={Scriven, L.E. and Sternling, C.V.},
  journal={Nature},
  volume={187},
  number={4733},
  pages={186},
  year={1960},
  publisher={Nature Publishing Group}
}

@article{Soligo2019a,
  title={{Coalescence of surfactant-laden drops by phase field method}},
  author={Soligo, G. and Roccon, A. and Soldati, A.},
  journal={J. Comput. Phys.},
  volume={376},
  number={},
  pages={1292--1311},
  year={2019},
}

@article{shan1993lattice,
  title={{Lattice Boltzmann model for simulating flows with multiple phases and components}},
  author={Shan, X. and Chen, H.},
  journal={Phys. Rev. E},
  volume={47},
  number={3},
  pages={1815},
  year={1993},
}

@article{gibou2018review,
  title={A review of level-set methods and some recent applications},
  author={Gibou, F. and Fedkiw, R. and Osher, S.},
  journal={J. Comput. Phys.},
  volume={353},
  pages={82--109},
  year={2018},
  publisher={Elsevier}
}

@article{Devita2019,
title={On the effect of coalescence on the rheology of emulsions},
volume={880},
journal={J. Fluid Mech.},
author={De Vita, F. and Rosti, M. E. and Caserta, S. and Brandt, L.},
year={2019},
pages={969--991}}

@article{tryggvason2013multiscale,
  title={Multiscale considerations in direct numerical simulations of multiphase flows},
  author={Tryggvason, G. and Dabiri, S. and Aboulhasanzadeh, B. and Lu, J.},
  journal={Phys. Fluids},
  volume={25},
  number={3},
  pages={031302},
  year={2013},
}

@article{mathai2020bubbly,
  title={Bubbly and buoyant particle-laden turbulent flows},
  author={Mathai, V. and Lohse, D. and Sun, C.},
  journal={Annu. Rev. Condens. Matter Phys.},
  volume={11},
  pages={529--559},
  year={2020},
}

@book{joseph1976stability,
  title={Stability of fluid motions. I, II},
  author={Joseph, D. D.},
	publisher={Springer Science \& Business Media},
  year={1976}
}

@article{chiu2011conservative,
  title={A conservative phase field method for solving incompressible two-phase flows},
  author={Chiu, Pao-Hsiung and Lin, Yan-Ting},
  Journal = {J. Comput. Phys.},
  volume={230},
  number={1},
  pages={185--204},
  year={2011},
  publisher={Elsevier}
}

@article{mirjalili2017interface,
  title={Interface-capturing methods for two-phase flows: An overview and recent developments},
  author={Mirjalili, S. and Jain, S. S. and Dodd, M.},
  journal={Center for Turbulence Research Annual Research Briefs},
  volume={2017},
  pages={117-135},
  year={2017}
}

@article{mirjalili2021consistent,
  title={Consistent, energy-conserving momentum transport for simulations of two-phase flows using the phase field equations},
  author={Mirjalili, S. and Mani, A.},
  Journal = {J. Comput. Phys.},
  volume={426},
  pages={109918},
  year={2021},
  publisher={Elsevier}
}

@article{kim2014conservative,
  title={A conservative {Allen--Cahn} equation with a space--time dependent Lagrange multiplier},
  author={Kim, Junseok and Lee, Seunggyu and Choi, Yongho},
  journal={Int. J. Eng. Sci.},
  volume={84},
  pages={11--17},
  year={2014},
  publisher={Elsevier}
}

@article{jain2022accurate,
  title={Accurate conservative phase-field method for simulation of two-phase flows},
  author={Jain, Suhas S},
  journal={J. Comput. Phys.},
  volume={469},
  pages={111529},
  year={2022},
  publisher={Elsevier}
}

@article{bothe2013volume,
  title={A volume-of-fluid-based method for mass transfer processes at fluid particles},
  author={Bothe, Dieter and Fleckenstein, Stefan},
  journal={Chem. Eng. Sci.},
  volume={101},
  pages={283--302},
  year={2013},
  publisher={Elsevier}
}

@book{ishii2010thermo,
  title={Thermo-fluid dynamics of two-phase flow},
  author={Ishii, Mamoru and Hibiki, Takashi},
  year={2010},
  publisher={Springer Science \& Business Media}
}

@article{welch2000volume,
  title={A volume of fluid based method for fluid flows with phase change},
  author={Welch, Samuel WJ and Wilson, John},
  journal={J. Comput. Phys.},
  volume={160},
  number={2},
  pages={662--682},
  year={2000},
}

@article{hardt2008evaporation,
  title={Evaporation model for interfacial flows based on a continuum-field representation of the source terms},
  author={Hardt, S. and Wondra, F.},
  journal={J. Comput. Phys.},
  volume={227},
  number={11},
  pages={5871--5895},
  year={2008},
  publisher={Elsevier}
}

@article{kharangate2017review,
  title={Review of computational studies on boiling and condensation},
  author={Kharangate, Chirag R and Mudawar, Issam},
  journal={Int. J. Heat Mass Transf.},
  volume={108},
  pages={1164--1196},
  year={2017},
  publisher={Elsevier}
}

@article{ningegowda2020mass,
  title={A mass-preserving interface-correction level set/ghost fluid method for modeling of three-dimensional boiling flows},
  author={Ningegowda, Bittagowdanahalli M and Ge, Zhouyang and Lupo, Giandomenico and Brandt, Luca and Duwig, Christophe},
  journal={Int. J. Heat Mass Transf.},
  volume={162},
  pages={120382},
  year={2020},
  publisher={Elsevier}
}

@article{gibou2007level,
  title={A level set based sharp interface method for the multiphase incompressible {Navier--Stokes} equations with phase change},
  author={Gibou, Fr{\'e}d{\'e}ric and Chen, Liguo and Nguyen, Duc and Banerjee, Sanjoy},
  journal={J. Comput. Phys.},
  volume={222},
  number={2},
  pages={536--555},
  year={2007},
  publisher={Elsevier}
}

@article{tanguy2007level,
  title={A level set method for vaporizing two-phase flows},
  author={Tanguy, S{\'e}bastien and M{\'e}nard, Thibaut and Berlemont, Alain},
  journal={J. Comput. Phys.},
  volume={221},
  number={2},
  pages={837--853},
  year={2007},
  publisher={Elsevier}
}

@article{tamura2022development,
  title={Development of a Phase-Field Method for Phase Change Simulations Using a Conservative {Allen--Cahn} Equation},
  author={Tamura, A. and Katono, K.},
  journal={J. Nucl. Eng. Rad. Sci.},
  volume={8},
  number={3},
  pages={031402},
  year={2022},
}

@article{safari2013extended,
  title={Extended lattice {Boltzmann} method for numerical simulation of thermal phase change in two-phase fluid flow},
  author={Safari, Hesameddin and Rahimian, M. H. and Krafczyk, Manfred},
  journal={Phys. Rev. E},
  volume={88},
  number={1},
  pages={013304},
  year={2013},
  publisher={APS}
}

@article{dong2010numerical,
  title={A numerical investigation of bubble growth on and departure from a superheated wall by lattice {Boltzmann} method},
  author={Dong, Zhiqiang and Li, Weizhong and Song, Yongchen},
  journal={Int. J. Heat Mass Transf.},
  volume={53},
  number={21-22},
  pages={4908--4916},
  year={2010},
  publisher={Elsevier}
}

@article{gueyffier1999volume,
  title={Volume-of-fluid interface tracking with smoothed surface stress methods for three-dimensional flows},
  author={Gueyffier, Denis and Li, Jie and Nadim, Ali and Scardovelli, Ruben and Zaleski, St{\'e}phane},
  journal={J. Comput. Phys.},
  volume={152},
  number={2},
  pages={423--456},
  year={1999},
  publisher={Elsevier}
}

@article{mirjalili2023assessment,
  title={Assessment of an energy-based surface tension model for simulation of two-phase flows using second-order phase field methods},
  author={Mirjalili, Shahab and Khanwale, Makrand A and Mani, Ali},
  journal={J. Comput. Phys.},
  volume={474},
  pages={111795},
  year={2023},
  publisher={Elsevier}
}

@article{hong2007boundary,
  title={On boundary condition capturing for multiphase interfaces},
  author={Hong, Jeong-Mo and Shinar, Tamar and Kang, Myungjoo and Fedkiw, Ronald},
  journal={J. Sci. Comput.},
  volume={31},
  pages={99--125},
  year={2007},
  publisher={Springer}
}

@article{he2022lattice,
  title={A lattice Boltzmann model for liquid-vapor-solid flow with thermal phase change},
  author={He, Qiang and Huang, Weifeng and Yin, Yuan and Li, Decai and Wang, Yuming},
  journal={Comput. Math. with Appl.},
  volume={114},
  pages={60--72},
  year={2022},
  publisher={Elsevier}
}

@article{begmohammadi2015simulation,
  title={Simulation of pool boiling and periodic bubble release at high density ratio using lattice {Boltzmann} method},
  author={Begmohammadi, A. and Farhadzadeh, M. and Rahimian, M. H.},
  journal={Int. Commun. Heat Mass Transf.},
  volume={61},
  pages={78--87},
  year={2015},
  publisher={Elsevier}
}

@article{shao2018computational,
  title={A computational framework for interface-resolved {DNS} of simultaneous atomization, evaporation and combustion},
  author={Shao, Changxiao and Luo, Kun and Chai, Min and Wang, Haiou and Fan, Jianren},
  Journal = {J. Comput. Phys.},
  volume={371},
  pages={751--778},
  year={2018},
  publisher={Elsevier}
}

@article{francois2006,
  title={A balanced-force algorithm for continuous and sharp interfacial surface tension models within a volume tracking framework},
  author={M. M. Francois and S. J. Cummins and E. D. Dendy and D. B. Kothe and J. M. Sicilian and M. W. Williams},
  Journal = { J. Comput. Phys.},
  volume={213},
  pages={141--173},
  year={2006},
}

@article{Roccon2023,
  title = {Phase-field modeling of complex interface dynamics in drop-laden turbulence},
  author = {Roccon, Alessio and Zonta, Francesco and Soldati, Alfredo},
  journal = {Phys. Rev. Fluids},
  volume = {8},
  issue = {9},
  pages = {090501},
  numpages = {42},
  year = {2023},
  month = {Sep},
}

@article{Soligo2019, 
title={Breakage, coalescence and size distribution of surfactant-laden droplets in turbulent flow}, 
volume={881}, 
journal={J. Fluid Mech.}, 
author={Soligo, Giovanni and Roccon, Alessio and Soldati, Alfredo}, 
year={2019}, 
pages={244–282}}

@article{sun2004diffuse,
  title={Diffuse interface modeling of two-phase flows based on averaging: mass and momentum equations},
  author={Sun, Y and Beckermann, C},
  journal={Physica D},
  volume={198},
  number={3-4},
  pages={281--308},
  year={2004},
  publisher={Elsevier}
}

@article{son1998numerical,
  title={Numerical simulation of film boiling near critical pressures with a level set method},
  author={Son, G. and Dhir, V. K.},
  journal={J. Heat Transf.},
  volume={120},
  number={1},
  pages={183--192},
  year={1998}
}

@article{sun2014modeling,
  title={Modeling of the evaporation and condensation phase-change problems with {FLUENT}},
  author={Sun, D. and Xu, J. and Chen, Q.},
  journal={Numer. Heat Transf. B.},
  volume={66},
  number={4},
  pages={326--342},
  year={2014},
  publisher={Taylor \& Francis}
}

@article{kunkelmann2009cfd,
  title={{CFD} simulation of boiling flows using the volume-of-fluid method within {OpenFOAM}},
  author={Kunkelmann, C. and Stephan, P.},
  journal={Numer. Heat Transf. A},
  volume={56},
  number={8},
  pages={631--646},
  year={2009},
  publisher={Taylor \& Francis}
}

@article{tanguy2014benchmarks,
  title={Benchmarks and numerical methods for the simulation of boiling flows},
  author={Tanguy, S{\'e}bastien and Sagan, Micha{\"e}l and Lalanne, Benjamin and Couderc, Fr{\'e}d{\'e}ric and Colin, Catherine},
	Journal = {J. Comput. Phys.},
  volume={264},
  pages={1--22},
  year={2014},
  publisher={Elsevier}
}

@article{haghani2021phase,
  title={Phase-change modeling based on a novel conservative phase-field method},
  author={Haghani-Hassan-Abadi, R. and Fakhari, A. and Rahimian, M.-H.},
  Journal = {J. Comput. Phys.},
  volume={432},
  pages={110111},
  year={2021},
  publisher={Elsevier}
}

@article{salimi2025low,
  title={A low {Mach} number diffuse-interface model for multicomponent two-phase flows with phase change},
  author={Salimi, S. Z. and Mukherjee, Aritra and Pelanti, Marica and Brandt, Luca},
  Journal = {J. Comput. Phys.},
  volume={523},
  pages={113683},
  year={2025},
  publisher={Elsevier}
}

@article{mohammadi2017phase,
  title={Phase-field lattice {Boltzmann} modeling of boiling using a sharp-interface energy solver},
  author={Mohammadi-Shad, M. and Lee, T.},
  journal={Phys. Rev. E},
  volume={96},
  number={1},
  pages={013306},
  year={2017},
  publisher={APS}
}

@article{esmaeeli2004computations,
  title={Computations of film boiling. {Part I}: numerical method},
  author={Esmaeeli, A. and Tryggvason, G.},
  journal={Int. J. Heat Mass Transf.},
  volume={47},
  number={25},
  pages={5451--5461},
  year={2004},
  publisher={Elsevier}
}

@article{tanasawa1991advances,
  title={Advances in condensation heat transfer},
  author={Tanasawa, I.},
  journal={Adv. Heat Transf.},
  volume={21},
  pages={55--139},
  year={1991},
}

@article{jafari2015phase,
  title={Phase-field modeling of vapor bubble growth in a microchannel},
  author={Jafari, R. and Okutucu-{\"O}zyurt, T.},
  journal={J. Comput. Multiph. Flows},
  volume={7},
  number={3},
  pages={143--158},
  year={2015},
}

@book{schrage1953theoretical,
  title={A Theoretical Study of Interphase Mass Transfer},
  author={Schrage, R. W.},
  year={1953},
  publisher={Columbia University Press}
}

@article{magnini2013numerical,
  title={Numerical investigation of hydrodynamics and heat transfer of elongated bubbles during flow boiling in a microchannel},
  author={Magnini, Mirco and Pulvirenti, Beatrice and Thome, John Richard},
  journal={Int. J. Heat Mass Transf.},
  volume={59},
  pages={451--471},
  year={2013},
  publisher={Elsevier}
}

@article{malan2021geometric,
  title={A geometric {VOF} method for interface resolved phase change and conservative thermal energy advection},
  author={Malan, L. C. and Malan, A. G. and Zaleski, S. and Rousseau, P. G.},
	Journal = {J. Comput. Phys.},
  volume={426},
  pages={109920},
  year={2021},
  publisher={Elsevier}
}

@article{udaykumar1996elafint,
  title={Elafint: a mixed {Eulerian-Lagrangian} method for fluid flows with complex and moving boundaries},
  author={Udaykumar, H. S. and Shyy, W. and Rao, M. M.},
  journal={Int. J. Numer. Methods Fluids},
  volume={22},
  number={8},
  pages={691--712},
  year={1996},
  publisher={Wiley Online Library}
}

@article{shin2016numerical,
  title={Numerical simulation of a rising bubble with phase change},
  author={Shin, S. and Choi, B.},
  journal = {Appl. Therm. Eng.},
  volume={100},
  pages={256--266},
  year={2016},
  publisher={Elsevier}
}

@article{gallo2020nucleation,
  title={Nucleation and growth dynamics of vapour bubbles},
  author={Gallo, Mirko and Magaletti, Francesco and Cocco, Davide and Casciola, Carlo Massimo},
  journal={J. Fluid Mech.},
  volume={883},
  pages={A14},
  year={2020},
  publisher={Cambridge University Press}
}

@article{gennari2024coupled,
  title={Coupled atomistic--continuum simulations of nucleate boiling},
  author={Gennari, Gabriele and Smith, Edward R and Pringle, Gavin J and Magnini, Mirco},
  journal={Int. J. Therm. Sci.},
  volume={200},
  pages={108954},
  year={2024},
  publisher={Elsevier}
}

@article{bures2024coarse,
  title={A coarse grid approach for single bubble boiling simulations with the volume of fluid method},
  author={Bure{\v{s}}, L. and Bucci, M. and Sato, Y. and Bucci, M.},
  journal={Comput. Fluids},
  volume={271},
  pages={106182},
  year={2024},
  publisher={Elsevier}
}

@article{Kossolapov2024, 
title={Bubble departure and sliding in high-pressure flow boiling of water}, 
volume={987}, 
journal={J. Fluid Mech.}, 
author={Kossolapov, A. and Hughes, M. T. and Phillips, B. and Bucci, M.}, 
year={2024}, 
pages={A35}
}

@article{gallo2018thermally,
  title={Thermally activated vapor bubble nucleation: The {Landau-Lifshitz--Van der Waals} approach},
  author={Gallo, M. and Magaletti, F. and Casciola, C. M.},
  journal={Phys. Rev. Fluids},
  volume={3},
  number={5},
  pages={053604},
  year={2018},
  publisher={APS}
}

@article{lavino2021surface,
  title={Surface topography effects on pool boiling via non-equilibrium molecular dynamics simulations},
  author={Lavino, A. D. and Smith, E. and Magnini, M. and Matar, O. K.},
  journal={Langmuir},
  volume={37},
  number={18},
  pages={5731--5744},
  year={2021},
  publisher={ACS Publications}
}

@article{shahmardi2021effects,
  title={Effects of surface nanostructure and wettability on pool boiling: A molecular dynamics study},
  author={Shahmardi, A. and Tammisola, O. and Chinappi, M. and Brandt, L.},
  journal={Int. J. Therm. Sci.},
  volume={167},
  pages={106980},
  year={2021},
  publisher={Elsevier}
}

@article{demou2022pressure,
  title={A pressure-based diffuse interface method for low-{Mach} multiphase flows with mass transfer},
  author={Demou, Andreas D and Scapin, Nicol{\`o} and Pelanti, Marica and Brandt, Luca},
  journal={J. Comput. Phys.},
  volume={448},
  pages={110730},
  year={2022},
  publisher={Elsevier}
}

@article{silvi2021understanding,
  title={Understanding dry-out mechanism in rod bundles of boiling water reactor},
  author={Silvi, Liril D and Chandraker, Dinesh K and Ghosh, Sumana and Das, Arup K},
  journal={Int. J. Heat Mass Transf.},
  volume={177},
  pages={121534},
  year={2021},
  publisher={Elsevier}
}

@article{chen2018numerical,
  title={Numerical simulation of thermal property effect of heat transfer plate on bubble growth with microlayer evaporation during nucleate pool boiling},
  author={Chen, Zhihao and Wu, Feifei and Utaka, Yoshio},
  journal={Int. J. Heat Mass Transf.},
  volume={118},
  pages={989--996},
  year={2018},
  publisher={Elsevier}
}

@article{LEONG2017,
title = {A critical review of pool and flow boiling heat transfer of dielectric fluids on enhanced surfaces},
journal = {Appl. Therm. Eng.},
volume = {112},
pages = {999-1019},
year = {2017},
issn = {1359-4311},
author = {K.C. Leong and J.Y. Ho and K.K. Wong},
}

@article{chen2023modeling,
  title={Modeling and study of microlayer effects on flow boiling in a mini-channel},
  author={Chen, Yujie and Jin, Shuqi and Yu, Bo and Ling, Kong and Sun, Dongliang and Zhang, Wei and Jiao, Kaituo and Tao, Wenquan},
  journal={Int. J. Heat Mass Transf.},
  volume={208},
  pages={124039},
  year={2023},
  publisher={Elsevier}
}

@article{magaletti2020unraveling,
  title={Unraveling low nucleation temperatures in pool boiling through fluctuating hydrodynamics simulations},
  author={Magaletti, Francesco and Georgoulas, Anastasios and Marengo, Marco},
  journal = {Int. J. Multiph. Flow},
  volume={130},
  pages={103356},
  year={2020},
}

@article{su2020investigation,
  title={Investigation of flow boiling heat transfer and boiling crisis on a rough surface using infrared thermometry},
  author={Su, G-Y and Wang, C and Zhang, L and Seong, Jee Hyun and Kommajosyula, R and Phillips, B and Bucci, M},
  journal={Int. J. Heat Mass Transf.},
  volume={160},
  pages={120134},
  year={2020},
}

@article{kim2009review,
  title={Review of nucleate pool boiling bubble heat transfer mechanisms},
  author={Kim, Jungho},
  journal = {Int. J. Multiph. Flow},
  volume={35},
  number={12},
  pages={1067--1076},
  year={2009},
  publisher={Elsevier}
}

@article{scapin2022mass,
  title={A mass-conserving pressure-based method for two-phase flows with phase change},
  author={Scapin, N. and Shahmardi, A. and Chan, W. H. R. and Jain, S. S. and Mirjalili, S. and Pelanti, M. and Brandt, L.},
  journal={Center for Turbulence Research Proceedings of the Summer Program},
  year={2022},
  pages={195--204},
}

@article{roccon2024boiling,
  title={Boiling heat transfer by phase-field method},
  author={Roccon, A.},
  journal={Acta Mech.},
  volume={236},
  pages={5623–-5638},
  year={2025},
  publisher={Springer}
}

@article{dodd2014fast,
  title={A fast pressure-correction method for incompressible two-fluid flows},
  author={Dodd, Michael S and Ferrante, Antonino},
  journal={J. Comput. Phys.},
  volume={273},
  pages={416--434},
  year={2014},
  publisher={Elsevier}
}

@article{chen2024review,
  title={Review on numerical simulation of boiling heat transfer from atomistic to mesoscopic and macroscopic scales},
  author={Chen, Yujie and Yu, Bo and Lu, Wei and Wang, Bohong and Sun, Dongliang and Jiao, Kaituo and Zhang, Wei and Tao, Wenquan},
  journal={Int. J. Heat Mass Transf.},
  volume={225},
  pages={125396},
  year={2024},
}

@article{soligo2021turbulent,
  title={Turbulent flows with drops and bubbles: What numerical simulations can tell us - {Freeman} scholar lecture},
  author={Soligo, G. and Roccon, A. and Soldati, A.},
  journal={J. Fluids Eng.},
  volume={143},
  number={8},
  pages={080801},
  year={2021},
}

@article{gallo2025complex,
  title={Complex transition pathways in boiling and cavitation},
  author={Gallo, M. and Occhioni, F. and Magaletti, F. and Casciola, C. M.},
  journal={J. Fluid Mech.},
  volume={1019},
  pages={A53},
  year={2025},
  publisher={Cambridge University Press}
}

@article{vachhani2025numerical,
  title={Numerical simulation of pool boiling on biphilic surfaces},
  author={Vachhani, S. and Riaz, A. and Balaras, E.},
  journal={Int. J. Heat Mass Transf.},
  volume={239},
  pages={126531},
  year={2025},
  publisher={Elsevier}
}

@article{youssoufi2025direct,
  title={Direct numerical simulations of subcooled pool boiling: Studies of heat transfer and bubble dynamics},
  author={Youssoufi, S. and Riaz, A. and Balaras, E.},
  journal={Phys.  Fluids},
  volume={37},
  number={2},
  year={2025},
  pages={023393},
  publisher={AIP Publishing}
}

@article{torres2024coupling,
  title={On the coupling between direct numerical simulation of nucleate boiling and a micro-region model at the contact line},
  author={Torres, L. and Urbano, A. and Colin, C. and Tanguy, S.},
  journal={J. Comput. Phys.},
  volume={497},
  pages={112602},
  year={2024},
  publisher={Elsevier}
}

@article{chiramell2024sharp,
  title={On sharp and diffuse interfaces-based level set methods for heat mass transfer induced phase change},
  author={Chiramell, S. and Weber, L. and Zamani, S. and Sharma, A. and Brandt, L. and Pelanti, M. and Solsvik, J.},
  journal={SSRN Preprint},
}

@article{garcia2025numerical,
  title={Numerical methods for multiphase flows},
  author={Garcia-Villalba, M. and Colonius, T. and Desjardins, O. and Lucas, D. and Mani, A. and Marchisio, D. and Matar, O. K. and Picano, F. and Zaleski, S.},
  journal={Int. J. Multiph. Flow},
  volume={191},
  pages={105285},
  year={2025},
  publisher={Elsevier}
}

@article{sielaff2022multiscale,
  title={The multiscale boiling investigation on-board the {International Space Station}: {A}n overview},
  author={Sielaff, A. and Mangini, D. and Kabov, O. and Raza, M. Q. and Garivalis, A. I. and Zupan{\v{c}}i{\v{c}}, M. and Dehaeck, S. and Evgenidis, S. and Jacobs, C. and Van Hoof, D. and others},
  journal={Appl. Therm. Eng.},
  volume={205},
  pages={117932},
  year={2022},
  publisher={Elsevier}
}

@article{raj2012pool,
    author = {Raj, R. and Kim, J. and McQuillen, J.},
    title = {Pool Boiling Heat Transfer on the {International Space Station}: Experimental Results and Model Verification},
    journal = {J. Heat Transf.},
    volume = {134},
    number = {10},
    pages = {101504},
    year = {2012},
    month = {08},
}

@article{mudawar2023heat,
  title={Heat transfer and interfacial flow physics of microgravity flow boiling in single-side-heated rectangular channel with subcooled inlet conditions -- {Experiments onboard the International Space Station}},
  author={Mudawar, Issam and Devahdhanush, VS and Darges, Steven J and Hasan, Mohammad M and Nahra, Henry K and Balasubramaniam, R and Mackey, Jeffrey R},
  journal={Int. J. Heat Mass Transf.},
  volume={207},
  pages={123998},
  year={2023},
  publisher={Elsevier},
}

@article{garivalis2021critical,
  title={Critical heat flux enhancement in microgravity conditions coupling microstructured surfaces and electrostatic field},
  author={Garivalis, A. I. and Manfredini, G. and Saccone, G. and Di Marco, P. and Kossolapov, A. and Bucci, M.},
  journal={{NPJ} Microgravity},
  volume={7},
  number={1},
  pages={37},
  year={2021},
}

@article{mudawar1999ultra,
  title={Ultra-high critical heat flux ({CHF}) for subcooled water flow boiling—{I}: {CHF} data and parametric effects for small diameter tubes},
  author={Mudawar, I. and Bowers, M. B.},
  journal={Int. J. Heat Mass Transf.},
  volume={42},
  number={8},
  pages={1405--1428},
  year={1999},
  publisher={Elsevier}
}

@article{ganesan2021universal,
  title={Universal critical heat flux ({CHF}) correlations for cryogenic flow boiling in uniformly heated tubes},
  author={Ganesan, Vishwanath and Patel, Raj and Hartwig, Jason and Mudawar, Issam},
  journal={Int. J. Heat Mass Transf.},
  volume={166},
  pages={120678},
  year={2021},
  publisher={Elsevier}
}

@article{dhir1998boiling,
  title={Boiling heat transfer},
  author={Dhir, V. K.},
  journal={Annu. Rev. Fluid Mech.},
  volume={30},
  number={1},
  pages={365--401},
  year={1998},
}

@article{dhir1991nucleate,
  title={Nucleate and transition boiling heat transfer under pool and external flow conditions},
  author={Dhir, V. K.},
  journal={Int. J. Heat Fluid Flow},
  volume={12},
  number={4},
  pages={290--314},
  year={1991},
  publisher={Elsevier}
}

@article{moghaddam2009physical1,
  title={Physical mechanisms of heat transfer during single bubble nucleate boiling of {FC-72} under saturation conditions. {I:} Experimental investigation},
  author={Moghaddam, S. and Kiger, K.},
  journal={Int. J. Heat Mass Transf.},
  volume={52},
  number={5-6},
  pages={1284--1294},
  year={2009},
  publisher={Elsevier}
}

@article{moghaddam2009physical2,
  title={Physical mechanisms of heat transfer during single bubble nucleate boiling of {FC-72} under saturation conditions. {II:} Theoretical analysis},
  author={Moghaddam, S. and Kiger, K.},
  journal={Int. J. Heat Mass Transf.},
  volume={52},
  number={5-6},
  pages={1295--1303},
  year={2009},
  publisher={Elsevier}
}

@article{nukiyama1966maximum,
  title={The maximum and minimum values of the heat {Q} transmitted from metal to boiling water under atmospheric pressure},
  author={Nukiyama, S.},
  journal={Int. J. Heat Mass Transf.},
  volume={9},
  number={12},
  pages={1419--1433},
  year={1966},
  publisher={Elsevier}
}

@article{cheung2014modelingp1,
  title={Modeling subcooled flow boiling in vertical channels at low pressures-{Part 1}: Assessment of empirical correlations},
  author={Cheung, S. C. P. and Vahaji, S. and Yeoh, G. H. and Tu, J. Y.},
  journal={Int. J. Heat Mass Transf.},
  volume={75},
  pages={736--753},
  year={2014},
  publisher={Elsevier}
}

@article{cheung2014modelingp2,
  title={Modeling subcooled flow boiling in vertical channels at low pressures-{Part 2}: Evaluation of mechanistic approach},
  author={Cheung, S. C. P. and Vahaji, S. and Yeoh, G. H. and Tu, J. Y.},
  journal={Int. J. Heat Mass Transf.},
  volume={75},
  pages={754-768},
  year={2014},
  publisher={Elsevier}
}

@article{darges2022assessment,
  title={Assessment and development of flow boiling critical heat flux correlations for partially heated rectangular channels in different gravitational environments},
  author={Darges, S. J. and Devahdhanush, V. S. and Mudawar, I.},
  journal={Int. J. Heat Mass Transf.},
  volume={196},
  pages={123291},
  year={2022},
  publisher={Elsevier}
}

@article{wu2020molecular,
  title={Molecular dynamics study of rapid boiling of thin liquid water film on smooth copper surface under different wettability conditions},
  author={Wu, Nini and Zeng, Liangcai and Fu, Ting and Wang, Zhaohui and Lu, Chang},
  journal={Int. J. Heat Mass Transf.},
  volume={147},
  pages={118905},
  year={2020},
  publisher={Elsevier}
}

@article{lin2024recent,
  title={Recent advances of molecular dynamics simulation on bubble nucleation and boiling heat transfer: A state-of-the-art review},
  author={Lin, Xiang-Wei and Wu, Wei-Tao and Li, Yu-Bai and Jing, Deng-Wei and Chen, Bin and Zhou, Zhi-Fu},
  journal={Adv. Colloid Interface Sci},
  volume={334},
  pages={103312},
  year={2024},
  publisher={Elsevier}
}

@article{wang2020cfd,
  title={{CFD} simulation of flow and heat transfer characteristics in a 5$\times$ 5 fuel rod bundles with spacer grids of advanced {PWR}},
  author={Wang, Yingjie and Wang, Mingjun and Ju, Haoran and Zhao, Minfu and Zhang, Dalin and Tian, Wenxi and Liu, Tiancai and Qiu, Suizheng and Su, GH},
  journal={Nucl. Eng. Technol.},
  volume={52},
  number={7},
  pages={1386--1395},
  year={2020},
  publisher={Elsevier}
}

@article{karalis2021deciphering,
  title={Deciphering the molecular mechanism of water boiling at heterogeneous interfaces},
  author={Karalis, K. and Zahn, D. and Prasianakis, N. I. and Ni{\v{c}}eno, B. and Churakov, S. V.},
  journal={Sci. Rep.},
  volume={11},
  number={1},
  pages={19858},
  year={2021},
  publisher={Nature Publishing Group UK London}
}

@article{gu2024three,
  title={Three-dimensional simulation of film boiling on a horizontal surface with magnetic field},
  author={Gu, H.-T. and Sahu, k. Chandra and Zhang, J. and Ni, M.-J.},
  journal={J. Fluid Mech.},
  volume={999},
  pages={A74},
  year={2024},
}

@article{hao2025experimental,
  title={Experimental investigation of the electro-de-wetting effect on the regulation of pool boiling behavior},
  author={Hao, Xue-Ying and Wang, Yan and Liu, Jian-Shu and Li, Xiao-Bin and Zhang, Hong-Na and Li, Feng-Chen},
  journal = {Appl. Therm. Eng.},
  pages={129610},
  year={2025},
  publisher={Elsevier}
}

@article{warrier2015nucleate,
  title={Nucleate pool boiling experiment {(NPBX)} in microgravity: {International Space Station}},
  author={Warrier, G. R. and Dhir, V. K. and Chao, D. F.},
  journal={Int. J. Heat Mass Transf.},
  volume={83},
  pages={781--798},
  year={2015},
  publisher={Elsevier}
}

@article{dhir2012nucleate,
  title={Nucleate pool boiling experiments {(NPBX)} on the  {International Space Station}},
  author={Dhir, V. K. and Warrier, G. R. and Aktinol, E. and Chao, D. and Eggers, J. and Sheredy, W. and Booth, W.},
  journal={Microgravity Sci. Technol.},
  volume={24},
  number={5},
  pages={307--325},
  year={2012},
  publisher={Springer}
}

@article{urbano2018direct,
  title={Direct numerical simulation of nucleate boiling in micro-layer regime},
  author={Urbano, Annafederica and Tanguy, S{\'e}bastien and Huber, Gr{\'e}gory and Colin, Catherine},
  journal={Int. J. Heat Mass Transf.},
  volume={123},
  pages={1128--1137},
  year={2018},
  publisher={Elsevier}
}

@article{jung2014experimental,
  title={An experimental method to simultaneously measure the dynamics and heat transfer associated with a single bubble during nucleate boiling on a horizontal surface},
  author={Jung, Satbyoul and Kim, Hyungdae},
  journal={Int. J. Heat Mass Transf.},
  volume={73},
  pages={365--375},
  year={2014},
  publisher={Elsevier}
}

@article{chen2020measurement,
  title={Measurement of the microlayer characteristics in the whole range of nucleate boiling for water by laser interferometry},
  author={Chen, Zhihao and Hu, Xiaocheng and Hu, Kang and Utaka, Yoshio and Mori, Shoji},
  journal={Int. J. Heat Mass Transf.},
  volume={146},
  pages={118856},
  year={2020},
  publisher={Elsevier}
}

@article{gilman2017self,
  title={A self-consistent, physics-based boiling heat transfer modeling framework for use in computational fluid dynamics},
  author={Gilman, Lindsey and Baglietto, Emilio},
journal = {Int. J. Multiph. Flow},
  volume={95},
  pages={35--53},
  year={2017},
  publisher={Elsevier}
}

@article{yuan2024assessment,
  title={Assessment of advanced {RANS} models to predict the heat transfer for supercritical fluids in vertical tubes},
  author={Yuan, Baoqiang and Wang, Wei and Du, Wenjing},
  journal={Int. J. Heat Mass Transf.},
  volume={219},
  pages={124829},
  year={2024},
  publisher={Elsevier}
}

@article{garbin2025bubbles,
  title={Bubbles and bubbly flows},
  author={Garbin, Valeria and Bothe, Dieter and Brenn, G{\"u}nter and Casciola, Carlo Massimo and Colin, Catherine and Marengo, Marco and Risso, Fr{\'e}d{\'e}ric and Tryggvason, Gretar and Lohse, Detlef},
journal = {Int. J. Multiph. Flow},
  volume={190},
  pages={105240},
  year={2025},
  publisher={Elsevier}
}

@article{liang2018pool1,
  title={Pool boiling critical heat flux ({CHF})--{Part 2}: Assessment of models and correlations},
  author={Liang, Gangtao and Mudawar, Issam},
  journal={Int. J. Heat Mass Transf.},
  volume={117},
  pages={1368--1383},
  year={2018},
}

@article{liang2018pool2,
  title={Pool boiling critical heat flux ({CHF})--{Part 1}: Review of mechanisms, models, and correlations},
  author={Liang, Gangtao and Mudawar, Issam},
  journal={Int. J. Heat Mass Transf.},
  volume={117},
  pages={1352--1367},
  year={2018},
}

@article{kam2020heat,
  title={A heat transfer model development for {CHF} prediction with consideration of dry patch characteristics},
  author={Kam, D. H. and Jeong, Y. H. and No, H. C.},
  journal={Int. J. Heat Mass Transf.},
  volume={148},
  pages={118812},
  year={2020},
  publisher={Elsevier}
}

@article{shang2025comprehensive,
  title={Comprehensive review of enhancement techniques and mechanisms for flow boiling in micro/mini-channels},
  author={Shang, H. and Xia, G. and Cheng, L. and Miao, S.},
  journal={Appl. Therm. Eng.},
  volume={258},
  pages={124783},
  year={2025},
  publisher={Elsevier}
}

@article{yabuki2016microscale,
  title={Microscale wall heat transfer and bubble growth in single bubble subcooled boiling of water},
  author={Yabuki, T. and Nakabeppu, O.},
  journal={Int. J. Heat Mass Transf.},
  volume={100},
  pages={851--860},
  year={2016},
  publisher={Elsevier}
}

@article{jung2015experimental,
  title={An experimental study on heat transfer mechanisms in the microlayer using integrated total reflection, laser interferometry and infrared thermometry technique},
  author={Jung, Satbyoul and Kim, Hyungdae},
  journal={Heat Transf. Eng.},
  volume={36},
  number={12},
  pages={1002--1012},
  year={2015},
  publisher={Taylor \& Francis}
}

@article{baer1986two,
  title={A two-phase mixture theory for the deflagration-to-detonation transition ({DDT}) in reactive granular materials},
  author={Baer, Melvin R and Nunziato, Jace W},
journal = {Int. J. Multiph. Flow},
  volume={12},
  number={6},
  pages={861--889},
  year={1986},
  publisher={Elsevier}
}

@article{linga2019hierarchy,
  title={A hierarchy of non-equilibrium two-phase flow models},
  author={Linga, Gaute and Fl{\aa}tten, Tore},
  journal={{ESAIM}, Proc. surv.},
  volume={66},
  pages={109--143},
  year={2019},
  publisher={EDP Sciences}
}

@article{saurel2009simple,
  title={Simple and efficient relaxation methods for interfaces separating compressible fluids, cavitating flows and shocks in multiphase mixtures},
  author={Saurel, Richard and Petitpas, Fabien and Berry, Ray A},
	Journal = {J. Comput. Phys.},
  volume={228},
  number={5},
  pages={1678--1712},
  year={2009},
  publisher={Elsevier}
}

@article{pelanti2014mixture,
  title={A mixture-energy-consistent six-equation two-phase numerical model for fluids with interfaces, cavitation and evaporation waves},
  author={Pelanti, M. and Shyue, K.-M.},
	Journal = {J. Comput. Phys.},
  volume={259},
  pages={331--357},
  year={2014},
  publisher={Elsevier}
}

@article{kapila2001two,
  title={Two-phase modeling of deflagration-to-detonation transition in granular materials: Reduced equations},
  author={Kapila, A. K. and Menikoff, R. and Bdzil, J. B. and Son, S. F. and Stewart, D. S.},
  journal={Phys. Fluids},
  volume={13},
  number={10},
  pages={3002--3024},
  year={2001},
}

@article{allaire2002five,
  title={A five-equation model for the simulation of interfaces between compressible fluids},
  author={Allaire, G. and Clerc, S. and Kokh, S.},
	Journal = {J. Comput. Phys.},
  volume={181},
  number={2},
  pages={577--616},
  year={2002},
  publisher={Elsevier}
}

@article{perigaud2005compressible,
  title={A compressible flow model with capillary effects},
  author={Perigaud, G. and Saurel, R.},
	Journal = {J. Comput. Phys.},
  volume={209},
  number={1},
  pages={139--178},
  year={2005},
  publisher={Elsevier}
}

@article{bryngelson2021mfc,
  title={{MFC}: An open-source high-order multi-component, multi-phase, and multi-scale compressible flow solver},
  author={Bryngelson, S. H. and Schmidmayer, K. and Coralic, V. and Meng, J. C. and Maeda, K. and Colonius, T.},
  journal={Comp. Phys. Comm.},
  volume={266},
  pages={107396},
  year={2021},
  publisher={Elsevier}
}

@article{pelanti2022arbitrary,
  title={Arbitrary-rate relaxation techniques for the numerical modeling of compressible two-phase flows with heat and mass transfer},
  author={Pelanti, M.},
  journal = {Int. J. Multiph. Flow},
  volume={153},
  pages={104097},
  year={2022},
}

@article{saurel2018diffuse,
  title={Diffuse-interface capturing methods for compressible two-phase flows},
  author={Saurel, R. and Pantano, C.},
  journal = {Annu. Rev. Fluid Mech.},
  volume={50},
  pages={105--130},
  year={2018},
}

@article{saurel2008modelling,
  title={Modelling phase transition in metastable liquids: application to cavitating and flashing flows},
  author={Saurel, R. and Petitpas, F. and Abgrall, R.},
  journal={J. Fluid Mech.},
  volume={607},
  pages={313--350},
  year={2008},
}

@article{saurel1999multiphase,
  title={A multiphase Godunov method for compressible multifluid and multiphase flows},
  author={Saurel, R. and Abgrall, R.},
  journal = {J. Comput. Phys.},
  volume={150},
  number={2},
  pages={425--467},
  year={1999},
  publisher={Elsevier}
}

@article{lemartelot2013liquid,
  title={Liquid and liquid-gas flows at all speeds},
  author={Le Martelot, S. and Nkonga, B. and Saurel, R.},
  journal = {J. Comput. Phys.},
  volume={255},
  pages={53--82},
  year={2013},
}

@article{le2014towards,
  title={Towards the direct numerical simulation of nucleate boiling flows},
  author={Le Martelot, S. and Saurel, R. and Nkonga, B.},
  journal = {Int. J. Multiph. Flow},
  volume={66},
  pages={62--78},
  year={2014},
}

@article{sirianni2025mixture,
  title={Mixture-conservative temperature-based {Baer--Nunziato} solver for efficient full-disequilibrium simulations of real fluids},
  author={Sirianni, G. and Re, B. and Abgrall, R.},
  journal={Comput. Fluids},
  volume={300},
  pages={106761},
  year={2025},
}

@article{flaatten2011relaxation,
  title={Relaxation two-phase flow models and the subcharacteristic condition},
  author={Fl{\aa}tten, T. and Lund, H.},
  journal={Math. Models Methods Appl. Sci},
  volume={21},
  number={12},
  pages={2379--2407},
  year={2011},
  publisher={World Scientific}
}

@article{saurel2003multiphase,
  title={A multiphase model with internal degrees of freedom: application to shock--bubble interaction},
  author={Saurel, R. and Gavrilyuk, S. and Renaud, F.},
  journal={J. Fluid Mech.},
  volume={495},
  pages={283--321},
  year={2003},
}

@article{perrier2021derivation,
  title={Derivation and closure of {Baer and Nunziato} type multiphase models by averaging a simple stochastic model},
  author={Perrier, V. and Guti{\'e}rrez, E.},
  journal={Multsicale Model Simul.},
  volume={19},
  number={1},
  pages={401--439},
  year={2021},
  publisher={SIAM}
}

@article{schmidmayer2020assessment,
  title={An assessment of multicomponent flow models and interface capturing schemes for spherical bubble dynamics},
  author={Schmidmayer, Kevin and Bryngelson, Spencer H and Colonius, Tim},
  journal = {J. Comput. Phys.},
  volume={402},
  pages={109080},
  year={2020},
  publisher={Elsevier}
}

@article{massoni2002some,
  title={Some models and Eulerian methods for interface problems between compressible fluids with heat transfer.},
  author={Massoni, J. and Saurel, R. and Nkonga, B. and Abgrall, R.},
  journal={Int. J. Heat Mass Transf.},
  volume={45},
  number={6},
  pages={1287--1307},
  year={2002}
}

@article{saurel2007shock,
  title={Shock jump relations for multiphase mixtures with stiff mechanical relaxation},
  author={Saurel, R. and Le Metayer, O. and Massoni, J. and Gavrilyuk, S.},
  journal={Shock waves},
  volume={16},
  number={3},
  pages={209--232},
  year={2007},
  publisher={Springer}
}

@book{wood1930textbook,
  title={A textbook of sound},
  author={Wood, A. B.},
  year={1930},
  publisher={Macmillan}
}

@book{wallis2020one,
  title={One-dimensional two-phase flow},
  author={Wallis, Graham B},
  year={2020},
  publisher={Courier Dover Publications}
}

@article{coralic2014finite,
  title={Finite-volume {WENO} scheme for viscous compressible multicomponent flows},
  author={Coralic, Vedran and Colonius, Tim},
  Journal = {J. Comput. Phys.},
  volume={274},
  pages={95--121},
  year={2014},
}

@article{beig2015maintaining,
  title={Maintaining interface equilibrium conditions in compressible multiphase flows using interface capturing},
  author={Beig, Shahaboddin Alahyari and Johnsen, Eric},
  Journal = {J. Comput. Phys.},
  volume={302},
  pages={548--566},
  year={2015},
}

@article{meng2018numerical,
  title={Numerical simulation of the aerobreakup of a water droplet},
  author={Meng, Jomela C and Colonius, Tim},
  journal={J. Fluid Mech.},
  volume={835},
  pages={1108--1135},
  year={2018},
  publisher={Cambridge University Press}
}

@article{rasthofer2017large,
  title={Large scale simulation of cloud cavitation collapse},
  author={Rasthofer, U. and Wermelinger, F. and Hadjidoukas, P. and Koumoutsakos, P.},
  journal={Procedia Comput. Sci.},
  volume={108},
  pages={1763--1772},
  year={2017},
  publisher={Elsevier}
}

@article{hejazialhosseini2013vortex,
  title={Vortex dynamics in {3D} shock-bubble interaction},
  author={Hejazialhosseini, B. and Rossinelli, D. and Koumoutsakos, P.},
  journal={Phys. Fluids},
  volume={25},
  number={11},
  pages={110816},
  year={2013},
}

@article{lund2012hierarchy,
  title={A hierarchy of relaxation models for two-phase flow},
  author={Lund, H.},
  journal={SIAM J. Appl. Math.},
  volume={72},
  number={6},
  pages={1713--1741},
  year={2012},
  publisher={SIAM}
}

@article{adebayo2025review,
  title={A review of diffuse interface-capturing methods for compressible multiphase flows},
  author={Adebayo, E. M. and Tsoutsanis, P. and Jenkins, K. W.},
  journal={Fluids},
  volume={10},
  number={4},
  pages={93},
  year={2025},
  publisher={MDPI}
}

@inproceedings{romenski2015multiphase,
  title={Multiphase flow modeling based on the hyperbolic thermodynamically compatible systems theory},
  author={Romenski, E},
  booktitle={AIP Conference Proceedings},
  volume={1648},
  pages={030004},
  year={2015},
  organization={AIP Publishing LLC}
}

@article{tiwari2013diffuse,
  title={A diffuse interface model with immiscibility preservation},
  author={Tiwari, Arpit and Freund, Jonathan B and Pantano, Carlos},
  Journal = {J. Comput. Phys.},
  volume={252},
  pages={290--309},
  year={2013},
  publisher={Elsevier}
}

@article{saurel2016general,
  title={A general formulation for cavitating, boiling and evaporating flows},
  author={Saurel, R. and Boivin, P. and Le M{\'e}tayer, O.},
  Journal = {Comput. Fluids},
  volume={128},
  number={10},
  pages={53--64},
  year={2016},
  publisher={Elsevier}
}

@article{darshan2024numerical,
  title={Numerical modelling of flow boiling inside microchannels: {A} critical review of methods and applications},
  author={Darshan, M. B. and Magnini, M. and Matar, O. K.},
  journal={Appl. Therm. Eng.},
  volume={257},
  pages={124464},
  year={2024},
}

@article{ten2017acoustic,
  title={An acoustic-convective splitting-based approach for the {Kapila} two-phase flow model},
  author={ten Eikelder, M. F. P. and Daude, Fr{\'e}d{\'e}ric and Koren, B. and Tijsseling, A. S.},
  Journal = {J. Comput. Phys.},
  volume={331},
  pages={188--208},
  year={2017},
  publisher={Elsevier}
}

@article{koumoutsakos2024roads,
  title={On roads less travelled between {AI} and computational science},
  author={Koumoutsakos, Petros},
  journal={Nat. Rev. Phys.},
  volume={6},
  number={6},
  pages={342--344},
  year={2024},
  publisher={Nature Publishing Group UK London}
}

@article{bhoriya2026physical,
  title={Physical Constraint Preserving Alternative Finite-Difference {WENO} Scheme for Hyperbolic Systems with non-conservative Products},
  author={Bhoriya, Deepak and Balsara, Dinshaw S and Chandrashekar, Praveen and Shu, Chi-Wang},
  journal={J. Sci. Comput.},
  volume={107},
  number={1},
  pages={1},
  year={2026},
  publisher={Springer}
}

@article{dumbser2016new,
  title={A new efficient formulation of the {HLLEM} Riemann solver for general conservative and non-conservative hyperbolic systems},
  author={Dumbser, Michael and Balsara, Dinshaw S},
  Journal = {J. Comput. Phys.},
  volume={304},
  pages={275--319},
  year={2016},
  publisher={Elsevier}
}

@article{tokareva2010hllc,
  title={{HLLC-type Riemann} solver for the {Baer--Nunziato} equations of compressible two-phase flow},
  author={Tokareva, Svetlana A and Toro, Eleuterio F},
  Journal = {J. Comput. Phys.},
  volume={229},
  number={10},
  pages={3573--3604},
  year={2010},
  publisher={Elsevier}
}

@book{das2023vapor,
  title={Vapor liquid two phase flow and phase change},
  author={Das, Sarit Kumar and Chatterjee, Dhiman},
  year={2023},
  publisher={Springer}
}

@article{dhruv2021investigation,
  title={An investigation of the gravity effects on pool boiling heat transfer via high-fidelity simulations},
  author={Dhruv, Akash and Balaras, Elias and Riaz, Amir and Kim, Jungho},
  journal={Int. J. Heat Mass Transf.},
  volume={180},
  pages={121826},
  year={2021},
  publisher={Elsevier}
}

@article{fei2020mesoscopic,
  title={Mesoscopic simulation of three-dimensional pool boiling based on a phase-change cascaded lattice {Boltzmann} method},
  author={Fei, Linlin and Yang, Jiapei and Chen, Yiran and Mo, Huangrui and Luo, Kai H},
  journal={Phys. Fluids},
  volume={32},
  number={10},
  year={2020},
  pages={103312},
  publisher={AIP Publishing}
}

@article{Eikelder2026, 
title={Compressible N-phase fluid mixture models}, 
volume={1032}, 
journal={J. Fluid Mech.}, 
author={ten Eikelder, Marco F.P. and van Brummelen, E. Harald and Schillinger, Dominik}, 
year={2026}, 
pages={A25}
}

@article{sato2026interface,
  title={Interface tracking simulation for subcooled flow boiling of water at 10 bar},
  author={Sato, Y. and Kossolapov, A. and Ni{\v{c}}eno, B. and Bucci, M.},
journal={J. Fluid Mech.}, 
  volume={1030},
  pages={A40},
  year={2026},
  publisher={Cambridge University Press}
}

@article{biferale2013simulations,
  title={Simulations of boiling systems using a lattice {Boltzmann} method},
  author={Biferale, L and Perlekar, P and Sbragaglia, M and Toschi, F},
  journal={Commun. Comput. Phys.},
  volume={13},
  number={3},
  pages={696--705},
  year={2013},
  publisher={Cambridge University Press}
}

@article{wang2025mesoscopic,
  title={Mesoscopic insights into effects of electric field on pool boiling for leaky dielectric fluids},
  author={Wang, Geng and Yang, Junyu and Lei, Timan and Fei, Linlin and Zhao, Xiao and Zhao, Jianfu and Li, Kai and Luo, Kai H},
  journal={Commun. Phys.},
  volume={8},
  number={1},
  pages={188},
  year={2025},
  publisher={Nature Publishing Group UK London}
}

@article{weber2026consistent,
  title={A consistent and scalable framework suitable for boiling flows using the conservative diffuse interface method},
  author={Weber, L. and Mukherjee, A. and Class, A. G. and Brandt, L.},
  journal = {J. Comput. Phys.},
  volume={551},
  pages={114680},
  year={2026},
}

@article{esmaeeli2004computationsp2,
  title={Computations of film boiling. {Part II}: multi-mode film boiling},
  author={Esmaeeli, A. and Tryggvason, G.},
  journal={Int. J. Heat Mass Transf.},
  volume={47},
  number={25},
  pages={5463--5476},
  year={2004},
  publisher={Elsevier}
}

@article{marek2001analysis,
  title={Analysis of the evaporation coefficient and the condensation coefficient of water},
  author={Marek, R and Straub, J},
  journal={Int. J. Heat Mass Transf.},
  volume={44},
  number={1},
  pages={39--53},
  year={2001},
  publisher={Elsevier}
}

@article{zanutto2026modelling,
title = {Modelling interfacial phase change in liquid–vapour systems: A geometric Volume-of-Fluid approach coupled with a temperature/mass fraction gradient-based phase change model},
journal = {Int. J. Heat Mass Transf.},
volume = {268},
pages = {129033},
year = {2026},
author = {Zanutto, C. P. and Ambrose, S. and Magnini, M. and Omar, R. and Eastwick, C.},
}

@article{sahut2021numerical,
  title={Numerical simulation of boiling on unstructured grids},
  author={Sahut, G. and Ghigliotti, G. and Balarac, G. and Bernard, M. and Moureau, V. and Marty, P.},
  Journal = {J. Comput. Phys.},
  volume={432},
  pages={110161},
  year={2021},
  publisher={Elsevier}
}

@article{sato2015depletable,
  title={A depletable micro-layer model for nucleate pool boiling},
  author={Sato, Y. and Ni{\v{c}}eno, B.},
  Journal = {J. Comput. Phys.},
  volume={300},
  pages={20--52},
  year={2015},
  publisher={Elsevier}
}

@article{nikolayev1999boiling,
  title={Boiling crisis and non-equilibrium drying transition},
  author={Nikolayev, V. S. and Beysens, D. A.},
  journal={Europhys. Lett.},
  volume={47},
  number={3},
  pages={345--351},
  year={1999}
}

@article{nikolayev2006experimental,
  title={Experimental evidence of the vapor recoil mechanism in the boiling crisis},
  author={Nikolayev, V. S. and Chatain, D. and Garrabos, Y. and Beysens, D.},
  journal={Phys. Rev. Lett.},
  volume={97},
  number={18},
  pages={184503},
  year={2006},
  publisher={APS}
}

@article{long2025direct,
  title={Direct numerical simulation of nucleate boiling with a resolved microlayer and conjugate heat transfer},
  author={Long, T. and Pan, J. and Cipriano, E. and Bucci, M. and Zaleski, S.},
  journal={J. Fluid Mech.},
  volume={1020},
  pages={A30},
  year={2025},
}

@article{chu2012structured,
  title={Structured surfaces for enhanced pool boiling heat transfer},
  author={Chu, K.-H. and Enright, R. and Wang, E. N.},
  journal={Appl. Phys. Lett.},
  volume={100},
  pages={241603},
  year={2012},
}

@article{kim2015review,
  title={Review of boiling heat transfer enhancement on micro/nanostructured surfaces},
  author={Kim, D. E. and Yu, D. In and Jerng, D. W. and Kim, M. H. and Ahn, H. S.},
  journal={Exp. Therm. Fluid Sci.},
  volume={66},
  pages={173--196},
  year={2015},
}

@article{liang2019review,
  title={Review of pool boiling enhancement by surface modification},
  author={Liang, G. and Mudawar, I.},
  journal={Int. J. Heat Mass Trans.},
  volume={128},
  pages={892--933},
  year={2019},
  publisher={Elsevier}
}

@article{dhillon2015critical,
  title={Critical heat flux maxima during boiling crisis on textured surfaces},
  author={Dhillon, N. S. and Buongiorno, J. and Varanasi, K. K.},
  journal={Nature Comm.},
  volume={6},
  number={1},
  pages={8247},
  year={2015},
  publisher={Nature Publishing Group UK London}
}

@article{kim2015enhanced,
  title={Enhanced critical heat flux by capillary driven liquid flow on the well-designed surface},
  author={Kim, D. E. and Park, S. Cheong and Yu, D. In and Kim, M. H. and Ahn, H. S.},
  journal={Appl. Phys. Lett.},
  volume={107},
  pages={023903},
  year={2015},
  publisher={AIP Publishing}
}

@article{yadigaroglu2014cmfd,
  title={{CMFD} and the critical-heat-flux grand challenge in nuclear thermal--hydraulics--{A} letter to the {Editor} of this special issue},
  author={Yadigaroglu, G.},
  journal={Int. J. Multiph. Flow},
  volume={67},
  pages={3--12},
  year={2014},
  publisher={Elsevier}
}

@article{mudawar2002assessment,
  title={Assessment of high-heat-flux thermal management schemes},
  author={Mudawar, I.},
  journal={IEEE Trans. Compon. Packag. Technol.},
  volume={24},
  number={2},
  pages={122--141},
  year={2001},
}

@article{giustini2016evaporative,
  title={Evaporative thermal resistance and its influence on microscopic bubble growth},
  author={Giustini, G. and Jung, S. and Kim, H. and Walker, S. P.},
  journal={Int. J. Heat Mass Trans.},
  volume={101},
  pages={733--741},
  year={2016},
  publisher={Elsevier}
}

@article{sirianni2025efficient,
  title={Efficient solution of mechanical and thermo-chemical finite-rate relaxation for compressible liquid-vapor flows under arbitrary equations of state},
  author={Sirianni, G. and Re, B.},
  journal={J. Comput. Phys.},
  volume={545},
  pages={114467},
  year={2025},
}

@article{magnini2026evaporation,
  title={Evaporation microlayer shape, thickness and heat transfer upon nucleate boiling of water},
  author={Magnini, M. and Bucci, M. and Bucci, M.},
  journal={Int. J. Heat Mass Trans.},
  volume={267},
  pages={128926},
  year={2026},
}

@article{tsui2014phase,
  title={Phase change calculations for film boiling flows},
  author={Tsui, Yeng-Yung and Lin, Shi-Wen and Lai, Yin-Nan and Wu, Feng-Chi},
  journal={Int. J. Heat Mass Trans.},
  volume={70},
  pages={745--757},
  year={2014},
}

@phdthesis{minozzi2025influence,
  title={Influence of wettability and scale-up in nucleate boiling heat transfer: a {DNS} study},
  author={Minozzi, G.},
  school={The University of Edinburgh},
  year={2025}
}

@article{sun2012development,
  title={Development of a vapor-liquid phase change model for volume-of-fluid method in {FLUENT}},
  author={Sun, D. L. and Xu, J. L. and Wang, L.},
  journal={Int. Commun. Heat Mass Transfer},
  volume={39},
  number={8},
  pages={1101--1106},
  year={2012},
  publisher={Elsevier}
}

@book{knudsen1950kinetic,
  title={The kinetic theory of gases: some modern aspects},
  author={Knudsen, M.},
  volume={3},
  year={1950},
  publisher={Methuen}
}

@article{wang2007characteristics,
  title={Characteristics of an evaporating thin film in a microchannel},
  author={Wang, H. and Garimella, S. V. and Murthy, J. Y.},
  journal={Int. J. Heat Mass Trans.},
  volume={50},
  number={19-20},
  pages={3933--3942},
  year={2007},
  publisher={Elsevier}
}

@article{kharangate2015computational,
  title={Computational modeling of turbulent evaporating falling films},
  author={Kharangate, C. R. and Lee, H. and Mudawar, I.},
  journal={Int. J. Heat Mass Trans.},
  volume={81},
  pages={52--62},
  year={2015},
  publisher={Elsevier}
}

@article{rattner2014simple,
  title={Simple mechanistically consistent formulation for volume-of-fluid based computations of condensing flows},
  author={Rattner, A. S. and Garimella, S.},
  journal={J. Heat Transf.},
  volume={136},
  number={7},
  pages={071501},
  year={2014},
}

@article{pan2016saturated,
  title={A saturated-interface-volume phase change model for simulating flow boiling},
  author={Pan, Z. and Weibel, J. A. and Garimella, S. V.},
  journal={Int. J. Heat Mass Trans.},
  volume={93},
  pages={945--956},
  year={2016},
}

@article{bahreini2015numerical,
  title={Numerical simulation of bubble behavior in subcooled flow boiling under velocity and temperature gradient},
  author={Bahreini, M. and Ramiar, A. and Ranjbar, A. A.},
  journal={ Nucl. Eng. Des.},
  volume={293},
  pages={238--248},
  year={2015},
  publisher={Elsevier}
}

@article{salimi2024robin,
  title   = {A {Volume-of-Fluid} method for multicomponent droplet evaporation with {Robin} boundary conditions},
  author  = {Salimi, S. Z.  and Scapin, N. and Popescu, E.-R. and Costa, P. and Brandt, L.},
  journal = {J. Comput. Phys.},
  volume  = {513},
  pages   = {112955},
  year    = {2024},
}

@article{scapin2020,
  title   = {A volume-of-fluid method for interface-resolved simulations of phase-changing two-fluid flows},
  author  = {Scapin, N. and Costa, P. and Brandt, L.},
  journal = {J. Comput. Phys.},
  volume  = {407},
  pages   = {109251},
  year    = {2020},
}

@article{zhao2022,
  title   = {Boiling and evaporation model for liquid-gas flows: {A} sharp and conservative method based on the geometrical {VOF} approach},
  author  = {Zhao, S. and Zhang, J. and Ni, M.-J.},
  journal = {J. Comput. Phys.},
  volume  = {452},
  pages   = {110908},
  year    = {2022},
}

@article{zhang2018direct,
  title={Direct numerical simulations of incompressible multiphase magnetohydrodynamics with phase change},
  author={Zhang, J. and Ni, M.-J.},
  journal={J. Comput. Phys.},
  volume={375},
  pages={1104--1122},
  year={2018},
}

@article{chai2020,
  title   = {A finite difference discretization method for heat and mass transfer with {Robin} boundary conditions on irregular domains},
  author  = {Chai, M. and Luo, K. and Shao, C. and Wang, H. and Fan, J.},
  journal = {J. Comput. Phys.},
  volume  = {400},
  pages   = {108890},
  year    = {2020},
}

@article{burevs2022comprehensive,
  title={Comprehensive simulations of boiling with a resolved microlayer: validation and sensitivity study},
  author={Bure{\v{s}}, L. and Sato, Y.},
  journal={J. Fluid Mech.},
  volume={933},
  pages={A54},
  year={2022},
}

@article{zhong2025front,
  title={A front-tracking immersed-boundary framework for simulating Lagrangian melting problems},
  author={Zhong, Kevin and Howland, Christopher J and Lohse, Detlef and Verzicco, Roberto},
  journal={J. Comput. Phys.},
  volume={525},
  pages={113762},
  year={2025},
  publisher={Elsevier}
}

@article{tryggvason2005direct,
  title={Direct numerical simulations of flows with phase change},
  author={Tryggvason, Gretar and Esmaeeli, A and Al-Rawahi, N},
  journal={Comput. Struct.},
  volume={83},
  number={6-7},
  pages={445--453},
  year={2005},
}

@article{jin2024direct,
  title={Direct numerical simulation of bubble collision, bounce and coalescence in bubble-induced turbulence},
  author={Jin, Y. and Weiland, C. and Hoffmann, M. and Schl{\"u}ter, M.},
  journal={Chem. Eng. Sci.},
  volume={284},
  pages={119502},
  year={2024},
}

@article{gennari2025marching,
  title={A marching cubes based method for topology changes in three-dimensional two-phase flows with front tracking},
  author={Gennari, G. and Gorges, C. and Denner, F. and Wachem, B.},
  journal={J. Comput. Phys.},
  volume={540},
  pages={114284},
  year={2025},
  publisher={Elsevier}
}

@article{singh2007three,
  title={Three-dimensional adaptive Cartesian grid method with conservative interface restructuring and reconstruction},
  author={Singh, R. and Shyy, W.},
  journal={J. Comput. Phys.},
  volume={224},
  number={1},
  pages={150--167},
  year={2007},
  publisher={Elsevier}
}

@article{li2016lattice,
  title={Lattice Boltzmann methods for multiphase flow and phase-change heat transfer},
  author={Li, Q. and Luo, K. H. and Kang, Q. J. and He, Y. L. and Chen, Q. and Liu, Q.},
  journal={Prog. Energy Combust. Sci.},
  volume={52},
  pages={62--105},
  year={2016},
  publisher={Elsevier}
}

@article{chen1998lattice,
  title={Lattice {Boltzmann} method for fluid flows},
  author={Chen, S. and Doolen, G. D.},
  journal={Annu. Rev. Fluid Mech.},
  volume={30},
  number={1},
  pages={329--364},
  year={1998},
}

@article{aidun2010lattice,
  title={Lattice-{Boltzmann} method for complex flows},
  author={Aidun, C. K. and Clausen, J. R.},
   Journal = {Annu. Rev. Fluid Mech.},
  volume={42},
  number={1},
  pages={439--472},
  year={2010},
}

@article{timm2016lattice,
  title={The lattice {Boltzmann} method: principles and practice},
  author={Timm, Kr{\"u}ger and Kusumaatmaja, Halim and Kuzmin, Alexandr and Shardt, O and Silva, G and Viggen, E},
  journal={Cham, Switzerland: Springer International Publishing AG},
  year={2016}
}

@book{succi2001lattice,
  title={The lattice {Boltzmann} equation: for fluid dynamics and beyond},
  author={Succi, Sauro},
  year={2001},
  publisher={Oxford university press}
}

@article{qian1992lattice,
  title={Lattice {BGK} models for {Navier-Stokes} equation},
  author={Qian, Yue-Hong and d'Humi{\`e}res, Dominique and Lallemand, Pierre},
  journal={Europhys. Lett.},
  volume={17},
  number={6},
  pages={479--484},
  year={1992}
}

@article{luo2024enhanced,
  title={Enhanced boiling heat transfer on structured surfaces with linear and staggered arrangements of hydrophilic and hydrophobic micro-pillars},
  author={Luo, Chao and Tagawa, Toshio},
  journal={Int. J. Heat Mass Trans.},
  volume={225},
  pages={125394},
  year={2024},
  publisher={Elsevier}
}

@article{johnsen2009numerical,
  title={Numerical simulations of non-spherical bubble collapse},
  author={Johnsen, E. and Colonius, T.},
  journal={J. Fluid Mech.},
  volume={629},
  pages={231--262},
  year={2009},
  publisher={Cambridge University Press}
}

@article{johnsen2006implementation,
  title={Implementation of {WENO} schemes in compressible multicomponent flow problems},
  author={Johnsen, E. and Colonius, T.},
  journal={J. Comput. Phys.},
  volume={219},
  number={2},
  pages={715--732},
  year={2006},
  publisher={Elsevier}
}

@article{ubbink1999method,
  title={A method for capturing sharp fluid interfaces on arbitrary meshes},
  author={Ubbink, Onno and Issa, RI1703648},
  journal={J. Comput. Phys.},
  volume={153},
  number={1},
  pages={26--50},
  year={1999},
  publisher={Elsevier}
}

@article{pirozzoli2019algebraic,
  title={On algebraic {TVD-VOF} methods for tracking material interfaces},
  author={Pirozzoli, Sergio and Di Giorgio, Simone and Iafrati, Alessandro},
  journal={Comput. Fluids},
  volume={189},
  pages={73--81},
  year={2019},
  publisher={Elsevier}
}

@article{xiao2011revisit,
  title={Revisit to the {THINC} scheme: a simple algebraic {VOF} algorithm},
  author={Xiao, Feng and Ii, Satoshi and Chen, Chungang},
  journal={J. Comput. Phys.},
  volume={230},
  number={19},
  pages={7086--7092},
  year={2011},
}

@article{hirt1981volume,
  title={Volume of fluid {(VOF)} method for the dynamics of free boundaries},
  author={Hirt, Cyril W and Nichols, Billy D},
  journal={J. Comput. Phys.},
  volume={39},
  number={1},
  pages={201--225},
  year={1981},
}

@article{scardovelli2003interface,
  title={Interface reconstruction with least-square fit and split {Eulerian--Lagrangian} advection},
  author={Scardovelli, R. and Zaleski, S.},
  journal={Int. J. Numer. Methods Fluids},
  volume={41},
  number={3},
  pages={251--274},
  year={2003},
}

@article{pilliod2004second,
  title={Second-order accurate volume-of-fluid algorithms for tracking material interfaces},
  author={Pilliod Jr, J. E. and Puckett, E. G.},
  journal={J. Comput. Phys.},
  volume={199},
  number={2},
  pages={465--502},
  year={2004},
}

@article{youngs1982time,
  title={Time-dependent multi-material flow with large fluid distortion},
  author={Youngs, David L},
  journal={Numerical methods for fluid dynamics},
  year={1982},
  publisher={Academic press}
}

@article{ii2012interface,
  title={An interface capturing method with a continuous function: the {THINC} method with multi-dimensional reconstruction},
  author={Ii, Satoshi and Sugiyama, Kazuyasu and Takeuchi, Shintaro and Takagi, Shu and Matsumoto, Yoichiro and Xiao, Feng},
  journal={J. Comput. Phys.},
  volume={231},
  number={5},
  pages={2328--2358},
  year={2012},
  publisher={Elsevier}
}

@article{gennari2022phase,
  title={A phase-change model for diffusion-driven mass transfer problems in incompressible two-phase flows},
  author={Gennari, Gabriele and Jefferson-Loveday, Richard and Pickering, Stephen J},
  journal={Chem. Eng. Sci.},
  volume={259},
  pages={117791},
  year={2022},
  publisher={Elsevier}
}

@article{giustini2023modelling,
  title={Modelling of free bubble growth with interface capturing computational fluid dynamics},
  author={Giustini, Giovanni and Issa, Raad I},
  journal={Exp. Comput. Multiphase Flow},
  volume={5},
  number={4},
  pages={357--364},
  year={2023},
  publisher={Springer}
}

@article{sato2013sharp,
  title={A sharp-interface phase change model for a mass-conservative interface tracking method},
  author={Sato, Y. and Ni{\v{c}}eno, B.},
  journal={J. Comput. Phys.},
  volume={249},
  pages={127--161},
  year={2013},
  publisher={Elsevier}
}

@article{luo2019level,
  title={Level set method for atomization and evaporation simulations},
  author={Luo, Kun and Shao, Changxiao and Chai, Min and Fan, Jianren},
  journal={Prog. Energy Combust. Sci.},
  volume={73},
  pages={65--94},
  year={2019},
  publisher={Elsevier}
}

@article{luo2015mass,
  title={A mass conserving level set method for detailed numerical simulation of liquid atomization},
  author={Luo, Kun and Shao, Changxiao and Yang, Yue and Fan, Jianren},
  journal={J. Comput. Phys.},
  volume={298},
  pages={495--519},
  year={2015},
  publisher={Elsevier}
}

@article{ge2018efficient,
  title={An efficient mass-preserving interface-correction level set/ghost fluid method for droplet suspensions under depletion forces},
  author={Ge, Zhouyang and Loiseau, Jean-Christophe and Tammisola, Outi and Brandt, Luca},
  journal={J. Comput. Phys.},
  volume={353},
  pages={435--459},
  year={2018},
  publisher={Elsevier}
}

@article{sussman2000coupled,
  title={A coupled level set and volume-of-fluid method for computing {3D} and axisymmetric incompressible two-phase flows},
  author={Sussman, Mark and Puckett, Elbridge Gerry},
  journal={J. Comput. Phys.},
  volume={162},
  number={2},
  pages={301--337},
  year={2000},
  publisher={Elsevier}
}

@article{gabbana2025flow,
  title={Flow-driven hysteresis in the transition boiling regime},
  author={Gabbana, Alessandro and de Wit, Xander M and Fei, Linlin and Wang, Ziqi and Livescu, Daniel and Toschi, Federico},
  journal={arXiv preprint arXiv:2509.14042},
  year={2025}
}

@article{fedkiw1999non,
  title={A non-oscillatory Eulerian approach to interfaces in multimaterial flows (the ghost fluid method)},
  author={Fedkiw, Ronald P and Aslam, Tariq and Merriman, Barry and Osher, Stanley},
  journal={J. Comput. Phys.},
  volume={152},
  number={2},
  pages={457--492},
  year={1999},
  publisher={Elsevier}
}

@article{gong2012lattice,
  title={A lattice {Boltzmann} method for simulation of liquid--vapor phase-change heat transfer},
  author={Gong, Shuai and Cheng, Ping},
  journal={Int. J. Heat Mass Transf.},
  volume={55},
  number={17-18},
  pages={4923--4927},
  year={2012},
  publisher={Elsevier}
}

@article{zhang2024predicting,
  title={Predicting initial microlayer thickness in nucleate boiling using {Landau--Levich} theory},
  author={Zhang, Xiaolong and El Mellas, Ismail and Magnini, Mirco},
  journal={J. Fluid Mech.},
  volume={997},
  pages={A44},
  year={2024},
  publisher={Cambridge University Press}
}

@article{menard2007coupling,
  title={Coupling level set/{VOF}/ghost fluid methods: Validation and application to {3D} simulation of the primary break-up of a liquid jet},
  author={M{\'e}nard, Thibault and Tanguy, Sebastien and Berlemont, Alain},
  journal={Int. J. Multiph. Flow},
  volume={33},
  number={5},
  pages={510--524},
  year={2007},
  publisher={Elsevier}
}

@article{villegas2016ghost,
  title={A ghost fluid/level set method for boiling flows and liquid evaporation: application to the {Leidenfrost} effect},
  author={Villegas, Lucia Rueda and Alis, Romain and Lepilliez, Mathieu and Tanguy, S{\'e}bastien},
  journal={J. Comput. Phys.},
  volume={316},
  pages={789--813},
  year={2016},
  publisher={Elsevier}
}

@article{nguyen2001boundary,
  title={A boundary condition capturing method for incompressible flame discontinuities},
  author={Nguyen, Duc Q and Fedkiw, Ronald P and Kang, Myungjoo},
  journal={J. Comput. Phys.},
  volume={172},
  number={1},
  pages={71--98},
  year={2001},
  publisher={Elsevier}
}

@article{ma20193d,
  title={{3D} simulations of pool boiling above smooth horizontal heated surfaces by a phase-change lattice {Boltzmann} method},
  author={Ma, X. and Cheng, P.},
  journal={Int. J. Heat Mass Transf.},
  volume={131},
  pages={1095--1108},
  year={2019},
  publisher={Elsevier}
}

@article{li2021manipulating,
  title={Manipulating the heat transfer of pool boiling by tuning the bubble dynamics with mixed wettability surfaces},
  author={Li, Yuanyuan and Li, Yuting and Jiao, Wei and Chen, Xueqing and Lu, Gui},
  journal={Int. J. Heat Mass Transf.},
  volume={170},
  pages={120996},
  year={2021},
}

@article{jiang2023review,
  title={A review of numerical investigation on pool boiling},
  author={Jiang, H. and Liu, Y. and Chu, H.},
  journal={J. Thermal Anal. Calorim.},
  volume={148},
  number={17},
  pages={8697--8745},
  year={2023},
}

@article{hwang2026review,
  title={Review of the conservative {Allen-Cahn} equations},
  author={Hwang, Y. and Ma, J. and Nam, Y. and Kim, J.},
  journal={Eng. Anal. Bound. Elem.},
  volume={187},
  pages={106713},
  year={2026},
}

@article{son1997numerical,
    author = {Son, G. and Dhir, V. K.},
    title = {Numerical Simulation of Saturated Film Boiling on a Horizontal Surface},
    journal = {J. Heat Transf.},
    volume = {119},
    number = {3},
    pages = {525-533},
    year = {1997},
    month = {08},
}

@article{li2007numerical,
    author = {Li, Ding and Dhir, Vijay K.},
    title = {Numerical Study of Single Bubble Dynamics During Flow Boiling},
    journal = {J. Heat Transf.},
    volume = {129},
    number = {7},
    pages = {864-876},
    year = {2007},
    month = {01},
}

@article{urbano2022semi,
  title={A semi implicit compressible solver for two-phase flows of real fluids},
  author={Urbano, Annafederica and Bibal, Marie and Tanguy, S{\'e}bastien},
  journal={J. Comput. Phys.},
  volume={456},
  pages={111034},
  year={2022},
  publisher={Elsevier}
}

@article{son2007level,
  title={A level set method for analysis of film boiling on an immersed solid surface},
  author={Son, G. and Dhir, V. K},
  journal={Numer. Heat Transf. B.},
  volume={52},
  number={2},
  pages={153--177},
  year={2007},
  publisher={Taylor \& Francis}
}

@article{tecchio2024jfm,
  title={Microlayer in nucleate boiling seen as {Landau--Levich} film with dewetting and evaporation},
  author={Tecchio, C. and Zhang, X. and Cariteau, Benjamin and Zalczer, Gilbert and i Cabarrocas, P. R. and Bulkin, P. and Charliac, J. and Vassant, Simon and Nikolayev, V. S.},
  journal={J. Fluid Mech.},
  volume={989},
  pages={A4},
  year={2024},
}

@article{tecchio2024ijhmt,
  title={Microlayer evaporation during bubble growth in nucleate boiling},
  author={Tecchio, C. and Cariteau, B. and Le Houedec, C. and Bois, G. and Saikali, E. and Zalczer, G. and Vassant, S. and i Cabarrocas, P. R. and Bulkin, P. and Charliac, J. and Nikolayev, V. S.},
  journal={Int. J. Heat Mass Transf.},
  volume={231},
  pages={125860},
  year={2024},
  publisher={Elsevier}
}

@article{cooper1969microlayer,
  title={The microlayer in nucleate pool boiling},
  author={Cooper, M. G. and Lloyd, A. J. P.},
  journal={Int. J. Heat Mass Transf.},
  volume={12},
  number={8},
  pages={895--913},
  year={1969},
  publisher={Elsevier}
}

@article{le2016noble,
  title={The {Noble-Abel} stiffened-gas equation of state},
  author={Le M{\'e}tayer, O. and Saurel, R.},
  journal={Phys. Fluids},
  volume={28},
  number={4},
  pages={046102},
  year={2016},
  publisher={AIP Publishing}
}

@article{radulescu2019noble,
  title={On the {Noble-Abel} stiffened-gas equation of state},
  author={Radulescu, M. I.},
  journal={Phys. Fluids},
  volume={31},
  number={11},
  year={2019},
  publisher={AIP Publishing}
}

@article{harlow1968numerical,
  title={Numerical calculation of almost incompressible flow},
  author={Harlow, F. H. and Amsden, A. A.},
  journal={J. Comput. Phys.},
  volume={3},
  number={1},
  pages={80--93},
  year={1968},
  publisher={Elsevier}
}

@article{menikoff1989riemann,
  title={The Riemann problem for fluid flow of real materials},
  author={Menikoff, R. and Plohr, B. J},
  journal={Rev. Mod. Phys.},
  volume={61},
  number={1},
  pages={75},
  year={1989},
  publisher={APS}
}

@book{toro2013riemann,
  title={Riemann solvers and numerical methods for fluid dynamics: a practical introduction},
  author={Toro, E. F.},
  year={2013},
  publisher={Springer Science \& Business Media}
}

@article{prosperetti2017vapor,
  title={Vapor bubbles},
  author={Prosperetti, Andrea},
  journal={Annu. Rev. Fluid Mech.},
  volume={49},
  pages={221--248},
  year={2017},
  publisher={Annual Reviews}
}

@article{bibal2024compressible,
  title={A compressible solver for two phase-flows with phase change for bubble cavitation},
  author={Bibal, M. and Deferrez, M. and Tanguy, S. and Urbano, A.},
  journal={J. Comput. Phys.},
  volume={500},
  pages={112750},
  year={2024},
}

@article{mialhe2023extended,
  title={An extended model for the direct numerical simulation of droplet evaporation. Influence of the {Marangoni} convection on {Leidenfrost} droplet},
  author={Mialhe, Guillaume and Tanguy, S{\'e}bastien and Tranier, L{\'e}o and Popescu, E.-R. and Legendre, D.},
  journal={J. Comput. Phys.},
  volume={491},
  pages={112366},
  year={2023},
}

@article{bourdon2025direct,
  title={Direct Numerical Simulation of film boiling around a superheated sphere immersed in a subcooled liquid},
  author={Bourdon, G. and Tanguy, S. and Airiau, C.},
  journal={Int. J. Heat Mass Transf.},
  volume={239},
  pages={126520},
  year={2025},
}

@article{palmore2019volume,
  title={A volume of fluid framework for interface-resolved simulations of vaporizing liquid-gas flows},
  author={Palmore Jr, John and Desjardins, Olivier},
  journal={J. Comput. Phys.},
  volume={399},
  pages={108954},
  year={2019},
}

@article{salimi2026evaporation,
  title={Evaporation of finite-size ammonia and n-heptane droplets in weakly compressible turbulence: {An} interface-resolved {DNS} study},
  author={Salimi, Salar Zamani and Gruber, Andrea and Scapin, Nicol{\`o} and Brandt, Luca},
  journal={Int. J. Heat Mass Transf.},
  volume={256},
  pages={128188},
  year={2026},
  publisher={Elsevier}
}

@article{sun2010coupled,
  title={A coupled volume-of-fluid and level set {(VOSET)} method for computing incompressible two-phase flows},
  author={Sun, DL and Tao, WQ},
  journal={Int. J. Heat Mass Transf.},
  volume={53},
  number={4},
  pages={645--655},
  year={2010},
  publisher={Elsevier}
}

@article{lee2019experimental,
  title={Experimental and computational investigation on two-phase flow and heat transfer of highly subcooled flow boiling in vertical upflow},
  author={Lee, Jeongmin and O'Neill, Lucas E and Lee, Seunghyun and Mudawar, Issam},
  journal={Int. J. Heat Mass Transf.},
  volume={136},
  pages={1199--1216},
  year={2019},
  publisher={Elsevier}
}

@article{aslam2004partial,
  title={A partial differential equation approach to multidimensional extrapolation},
  author={Aslam, Tariq D},
  journal={J. Comput. Phys.},
  volume={193},
  number={1},
  pages={349--355},
  year={2004},
  publisher={Elsevier}
}

@incollection{osher2002extrapolation,
  title={Extrapolation in the normal direction},
  author={Osher, S. and Fedkiw, R.},
  booktitle={Level Set Methods and Dynamic Implicit Surfaces},
  pages={75--77},
  year={2002},
  publisher={Springer}
}

@article{chen20223d,
  title={{3-D} numerical study of subcooled flow boiling in a horizontal rectangular mini-channel by {VOSET}},
  author={Chen, Y.-J. and Ling, K. and Ding, H. and Wang, Y. and Jin, S.-Q. and Tao, W.-Q.},
  journal={Int. J. Heat Mass Transf.},
  volume={183},
  pages={122218},
  year={2022},
  publisher={Elsevier}
}

@article{chen2023numerical,
  title={Numerical investigation of critical heat flux during subcooled flow boiling in a vertical rectangular Mini-channel},
  author={Chen, Yu-Jie and Ling, Kong and Jin, Shu-Qi and Lu, Wei and Yu, Bo and Sun, Dongliang and Zhang, Wei and Tao, Wen-Quan},
  journal={App. Therm. Eng.},
  volume={221},
  pages={119862},
  year={2023},
  publisher={Elsevier}
}

@article{poblador2025momentum,
  title={A momentum balance correction to the non-conservative one-fluid formulation in boiling flows using volume-of-fluid},
  author={Poblador-Ibanez, Jordi and Valle, Nicol{\'a}s and Boersma, Bendiks Jan},
  journal={J. Comput. Phys.},
  volume={524},
  pages={113704},
  year={2025},
  publisher={Elsevier}
}

@article{yin2020experimental,
  title={Experimental investigation of pool boiling characteristics of surfactant solutions on bi-conductive surfaces},
  author={Yin, Jialun and Xiao, Xin and Feng, Longlong and Zhong, Ke and Jia, Hongwei},
  journal={Int. J. Heat Mass Transf.},
  volume={157},
  pages={119914},
  year={2020},
}

@article{liu2025phase,
  title={Phase field-lattice {Boltzmann} model for two-phase flows with near-contact interactions},
  author={Liu, Da and Hu, Yang and Zhang, Shiting and Zhu, Yuqi and He, Qiang and Li, Decai},
  journal={Phys. Fluids},
  volume={37},
  number={9},
  year={2025},
  pages={092123},
}

@article{montessori2019mesoscale,
  title={Mesoscale modelling of near-contact interactions for complex flowing interfaces},
  author={Montessori, Andrea and Lauricella, Marco and Tirelli, Nicola and Succi, Sauro},
  journal={J. Fluid Mech.},
  volume={872},
  pages={327--347},
  year={2019},
}

@article{hetsroni2004boiling,
  title={Boiling enhancement with environmentally acceptable surfactants},
  author={Hetsroni, Gurevich and Gurevich, M and Mosyak, A and Rozenblit, R and Segal, Z},
  journal={Int. J. Heat Fluid Flow},
  volume={25},
  number={5},
  pages={841--848},
  year={2004},
}

@article{moore1961measurement,
  title={The measurement of rapid surface temperature fluctuations during nucleate boiling of water},
  author={Moore, F. D. and Mesler, R. B.},
  journal={AIChE J.},
  volume={7},
  number={4},
  pages={620--624},
  year={1961},
}

@article{schweikert2019transition,
  title={On the transition between contact line evaporation and microlayer evaporation during the dewetting of a superheated wall},
  author={Schweikert, Kai and Sielaff, A and Stephan, P},
  journal={Int. J. Therm. Sci.},
  volume={145},
  pages={106025},
  year={2019},
}

@article{utaka2018measurement,
  title={Measurement of contribution of microlayer evaporation applying the microlayer volume change during nucleate pool boiling for water and ethanol},
  author={Utaka, Yoshio and Hu, Kang and Chen, Zhihao and Morokuma, Takayuki},
  journal={Int. J. Heat Mass Transf.},
  volume={125},
  pages={243--247},
  year={2018},
}

@article{stephan1994new,
  title={A new model for nucleate boiling heat transfer},
  author={Stephan, P and Hammer, J},
  journal={Heat Mass Transf.},
  volume={30},
  number={2},
  pages={119--125},
  year={1994},}

@article{burevs2021modelling,
  title={On the modelling of the transition between contact-line and microlayer evaporation regimes in nucleate boiling},
  author={Bure{\v{s}}, L. and Sato, Y.},
  journal={J. Fluid Mech.},
  volume={916},
  pages={A53},
  year={2021},
}

@article{wayner1976interline,
  title={The interline heat-transfer coefficient of an evaporating wetting film},
  author={Wayner Jr, P. C. and Kao, Y. K. and LaCroix, L. V.},
  journal={Int. J. Heat Mass Transf.},
  volume={19},
  number={5},
  pages={487--492},
  year={1976},
  publisher={Elsevier}
}

@article{utaka2014heat,
  title={Heat transfer characteristics based on microlayer structure in nucleate pool boiling for water and ethanol},
  author={Utaka, Yoshio and Kashiwabara, Yuki and Ozaki, Michio and Chen, Zhihao},
  journal={Int. J. Heat Mass Transf.},
  volume={68},
  pages={479--488},
  year={2014},
}

@article{chen2015heat,
  title={On heat transfer and evaporation characteristics in the growth process of a bubble with microlayer structure during nucleate boiling},
  author={Chen, Z. and Utaka, Y.},
  journal={Int. J. Heat Mass Transf.},
  volume={81},
  pages={750--759},
  year={2015},
}



\end{document}